\documentclass[twocolumn,aps,superscriptaddress,showpacs,nofootinbib,floatfix]{revtex4}
\usepackage{epsfig,bm,feynmf}
\usepackage{graphicx}
\usepackage{amsmath}
\usepackage{dcolumn}

\usepackage{graphicx}
\usepackage[usenames,dvipsnames,svgnames,table]{xcolor}

\usepackage{tikz-feynman,contour}
\tikzfeynmanset{compat=1.0.0}
\usepackage[normalem]{ulem}  

\renewcommand\sout{\bgroup \color{red} \ULdepth=-.5ex \ULset}

\newcommand{\com}[1]{{\color{black}#1}}

\newcommand{\GP}[3]{\Pi_{#1#2}(#3)}

\newcommand{\QP}[1]{\Lambda(#1)}

\begin{document}


\title{Charmonium production at energies available at the CERN Super Proton Synchrotron and Facility for Antiproton and Ion Research}


\author{Taesoo Song}\email{t.song@gsi.de}
\affiliation{GSI Helmholtzzentrum f\"{u}r Schwerionenforschung GmbH, Planckstrasse 1, 64291 Darmstadt, Germany}

\author{Jiaxing Zhao}\email{jzhao@itp.uni-frankfurt.de}
\affiliation{Institute for Theoretical Physics, Johann Wolfgang Goethe Universit\"{a}t, Frankfurt am Main, Germany}
\affiliation{Helmholtz Research Academy Hessen for FAIR (HFHF),GSI Helmholtz Center for Heavy Ion Research. Campus Frankfurt, 60438 Frankfurt, Germany}

\author{Joerg Aichelin}\email{aichelin@subatech.in2p3.fr}
\affiliation{SUBATECH UMR 6457 (IMT Atlantique,  Universit\'{e} de Nantes, IN2P3/CNRS), 4 Rue Alfred Kastler, F-44307 Nantes, France}
\affiliation{Frankfurt Institute for Advanced Studies, Ruth-Moufang-Strasse 1, 60438 Frankfurt am Main, Germany}

\author{Elena Bratkovskaya}\email{E.Bratkovskaya@gsi.de}
\affiliation{GSI Helmholtzzentrum f\"{u}r Schwerionenforschung GmbH, Planckstrasse 1, 64291 Darmstadt, Germany}
\affiliation{Institute for Theoretical Physics, Johann Wolfgang Goethe Universit\"{a}t, Frankfurt am Main, Germany}
\affiliation{Helmholtz Research Academy Hessen for FAIR (HFHF),GSI Helmholtz Center for Heavy Ion Research. Campus Frankfurt, 60438 Frankfurt, Germany}


\begin{abstract}
In this study we apply the Remler formalism to charmonium production at SPS and GSI/FAIR energies in order to investigate the effects of baryon-rich matter on charmonium production and dissociation in heavy-ion collisions within the Parton-Hadron-String Dynamics (PHSD). 
As a first step the Remler formalism is tested in p+p collisions and then applied to p+A collisions in order to extract the nuclear absorption cross section of charmonium, which is then utilized in heavy-ion collisions.
We find that the Remler formalism successfully describes charmonium production in heavy-ion collisions at SPS energies when an in-medium heavy quark potential is implemented, in which $J/\psi$ dissociates near $T_c$.
Finally the same formalism is applied to the low CERN/SPS energy and GSI/FAIR energies, where we estimate charmonium production.
\end{abstract}


\maketitle
\section{introduction}

Relativistic heavy-ion collisions provide a unique opportunity to create and study hot and dense strongly interacting matter under laboratory conditions. Lattice QCD calculations indicate that, with increasing temperature QCD matter undergoes a transition from hadronic stage to a deconfined quark--gluon plasma (QGP). In the temperature--baryon chemical potential ($T$--$\mu_B$) plane, this transition is established to be a crossover at small $\mu_B$. At larger $\mu_B$, a possible first-order phase transition is expected, which would be connected to the crossover region through a possible critical end point. However, the high-$\mu_B$ region remains challenging to access from first-principles lattice QCD calculations due to the fermion sign problem.

Among the proposed probes of the QGP, quarkonium occupies a central role. Matsui and Satz first suggested that the suppression of $J/\psi$ production could provide a signature of QGP formation~\cite{Matsui:1986dk}. In a deconfined medium, color screening reduces the attractive interaction between a heavy quark and antiquark, thereby weakening or even preventing the formation of bound quarkonium states.

In addition, quarkonium states can be dissociated through interactions with thermal quarks and gluons in the medium even after they have formed. This mechanism is referred to as thermal dissociation of quarkonium ~\cite{Park:2007zza,Song:2007gm}. In terms of the heavy-quark potential, color screening is related to the real part of the potential, while thermal dissociation is associated with its imaginary part~\cite{Laine:2006ns,Brambilla:2008cx}.

Quarkonium production remains of particular interest in the high-$\mu_B$ region. At relatively low collision energies, where baryon-rich matter is created, the production of charm--anticharm pairs is rare, and typically no more than one $c\bar{c}$ pair is produced per collision. 
In this case, each charm and anticharm quark can be traced without interference from additional charm pairs produced in the same event, which frequently occurs in high-energy collisions at RHIC and LHC.
This allows one to study the correlations between produced charm and anticharm quarks in heavy-ion collisions and provides valuable information on (anti)charm interactions in dense matter at high-$\mu_B$.

Several theoretical and phenomenological studies have investigated quarkonium properties in baryon-rich partonic and hadronic matter.
For example, in the hard thermal loop approach~\cite{Bellac:2011kqa,Zhao:2022cvl} increasing $\mu_B$ enhances the screening mass between a heavy quark pair, thereby weakening the binding of quarkonium.
It has also been suggested that the spectral function of charmonium~\cite{Klingl:1998sr,Hayashigaki:1998ey,Kim:2000kj,Friman:2002fs}, as well as those of open charm mesons~\cite{Hayashigaki:2000es,Digal:2001iu}, may be modified in nuclear matter.
According to QCD sum rule analyses, the $D$ meson mass is much more sensitive to chiral symmetry restoration in nuclear matter through the reduction of the light-quark condensate, whereas the medium modification of the charmonium mass arises only from changes in the gluon condensate.
Consequently, the mass shift of the $D$ meson in nuclear matter is about an order of magnitude larger than that of charmonium, which is only about 5-10 MeV at nuclear saturation density~\cite{Klingl:1998sr,Hayashigaki:2000es,Hayashigaki:1998ey,Kim:2000kj}.

In this study, we estimate charmonium production in heavy-ion collisions at the Super Proton Synchrotron (SPS) and Facility for Antiproton and Ion Research (FAIR) energies, which provide access to matter with relatively large $\mu_B$. For this study we neglect the possible changes of the properties of quarkonia in a purely hadronic environment  at large temperature and chemical potential.  For quarkonia in the environment of a QGP we investigate two scenarios, both discussed presently in the literature: We firstly study the case where the real part of the potential between charm and anti-charm quarks does not depend on the temperature of the QGP\cite{Bazavov:2023dci}, followed by investigating a heavy quark -  anti-quark potential whose strength decreases with increasing temperature of the QGP \cite{Kaczmarek:2003ph,Burnier:2014ssa,Gubler:2020hft} 

For this purpose, we adopt the Remler formalism~\cite{Remler:1975re,Remler:1975fm,Remler:1981du,Song:2017phm,Zhao:2024vqp,Song:2023ywt,Villar:2022sbv,Song:2023zma}, in which charmonium production is expressed as the projection of the charmonium Wigner function onto the phase-space distribution of charm and anticharm quarks.
This approach has already been applied to p+p collisions for both charmonium and bottomonium, yielding results consistent with experimental data~\cite{Song:2017phm,Zhao:2024vqp,Song:2023zma}.
Charmonium production is then investigated in p+A collisions at SPS energies in order to extract the nuclear absorption cross sections within the Parton-Hadron-String dynamics (PHSD) approach~\cite{Cassing:2008sv,Cassing:2009vt,Moreau:2019vhw}. 
Subsequently, the Remlar formalism is extended to heavy-ion collisions, including charmonium formation and interactions in the QGP.

PHSD is a non-equilibrium microscopic transport approach that describes strongly interacting hadronic and partonic matter produced in p+p, p+A, and heavy-ion collisions.
It has successfully described the production not only of bulk particles but also of electromagnetic probes~\cite{Bratkovskaya:2007jk,Linnyk:2015rco,Jorge:2025wwp} and heavy flavors~\cite{Cassing:2000vx,Bratkovskaya:2003ux,Song:2017phm,Zhao:2024vqp,Song:2023zma} in heavy-ion collisions.
Since the baryon chemical potential at SPS energies is not very large, the corresponding trajectory in the QCD phase diagram is not expected to passes through the critical end point.
In contrast, the GSI/FAIR program will probe a much larger value of $\mu_B$ and may approach the critical end point, providing valuable information about this region of the phase diagram.

This paper is organized as follows: The general theoretical framework for charm dynamics, including charm production in PHSD, the Remler formalism for charmonium formation in the QGP, and the hadronic interactions of charmonium, is outlined in Sec.~\ref{theory}.
The results for charmonium production in p+p, p+A, and heavy-ion collisions at SPS energies, as well as estimates for GSI/FAIR energies, are presented in Sec.~\ref{results}. Finally, the conclusions are summarized in Sec.~\ref{summary}.

\section{Charm dynamics}
\label{theory}

In this section, we outline the theoretical framework used to describe charmonium production.
First, we briefly review how heavy flavors are produced within the PHSD approach.
We then discuss the Remler formalism, which has been adopted to describe bottomonium production in p+p and heavy-ion collisions at the LHC~\cite{Song:2023zma}.
The Remler formalism for charmonium is applied up to the critical temperature $T_c$, where all charm quarks hadronize.
Subsequently, the hadronic interactions of charmonium are taken into acocunt.

\subsection{Charm quark production in Parton-Hadron-String Dynamics (PHSD)}

In the PHSD approach, heavy quarks are produced through hard nucleon-nucleon scatterings, which are simulated using the PYTHIA event generator~\cite{Sjostrand:2006za}.
PYTHIA is based on leading-order calculations in perturbative QCD (pQCD). To include higher order effects, the transverse momentum and rapidity of heavy quarks are rescaled such that their distributions reproduce those obtained from the Fixed-Order-Next-to-Leading-Logarithm (FONLL) calculations~\cite{Cacciari:1998it,Cacciari:2001td,Song:2015sfa,Song:2015ykw}.

\begin{figure}[h]
\centerline{
\includegraphics[width=9 cm]{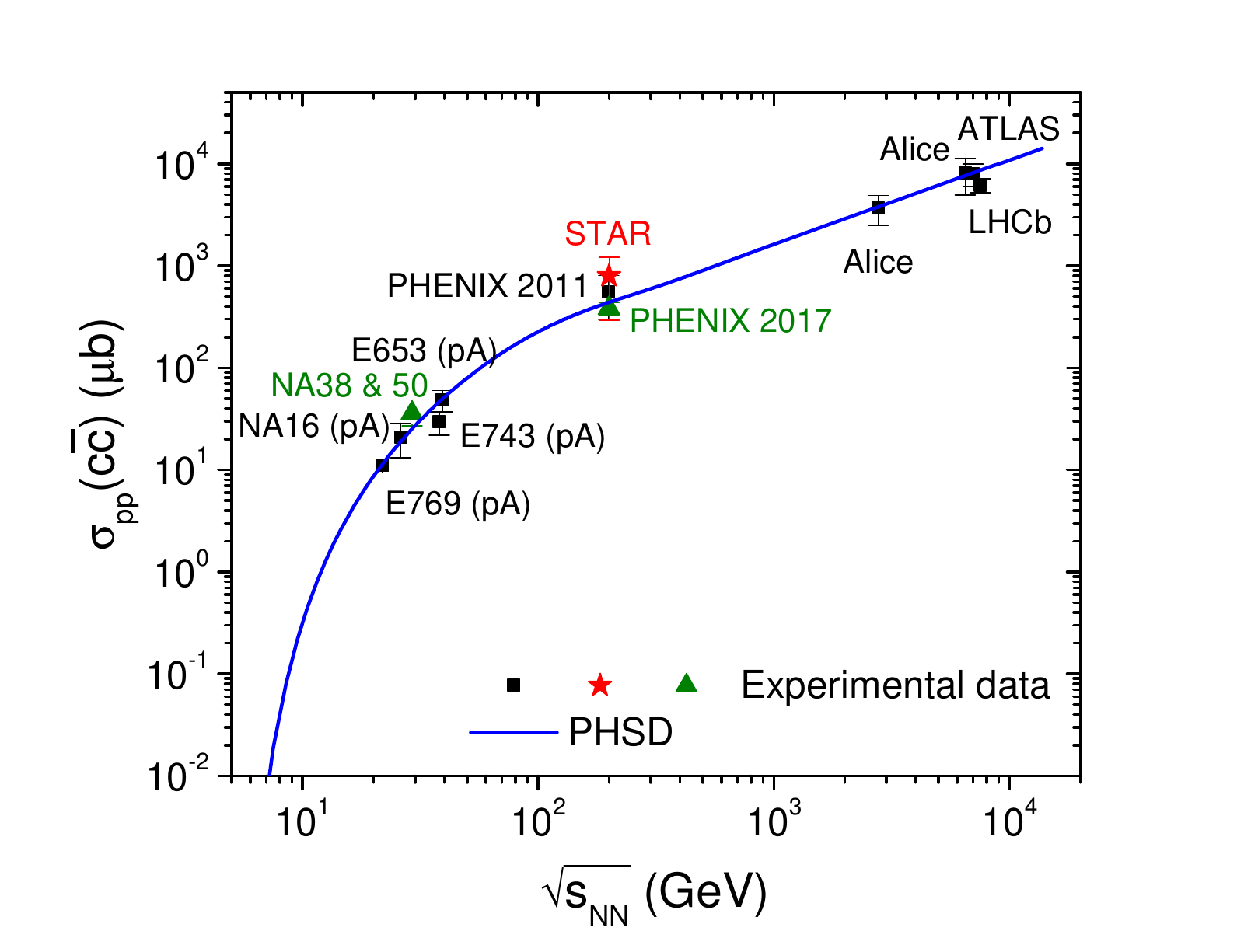}}
\caption{The parameterized cross section for charm production in p+p collisions as a function of center-of-mass energy, compared with various experimental data.}
\label{sigma}
\end{figure}

However, PYTHIA is not used to determine the total cross section for charm production in p+p collisions.
Instead, the total cross section is parameterized~
\cite{Cassing:2000vx} using experimental data covering a wide range of collision energies, shown in Fig.~\ref{sigma}.

In heavy-ion collisions, the parton distribution functions (PDFs) are modified inside nuclei due to multiple scatterings of partons, an effect known as (anti)shadowing.
Since heavy quarks are produced through parton-parton scattering, these modifications of PDFs, which affect charm production, are taken into account using the EPS09 nuclear parton distribution set~\cite{Eskola:2009uj}.
At relatively low energies such as at SPS and FAIR, charm production occurs predominantly in the anti-shadowing region. This enhances the charm yield.
In contrast, at LHC it takes place mainly in the shadowing region, leading to a suppression of charm production at mid-rapidity~\cite{Song:2015ykw}.

\subsection{Charm quark interactions in QGP}

The produced heavy quarks subsequently interact with massive off-shell quarks and gluons in the QGP described by the dynamical quasi-particle model (DQPM) \cite{Cassing:2007yg, Cassing:2007nb, Moreau:2019vhw, Soloveva:2019xph}.

We recall that the Dynamical QuasiParticle Model \cite{Cassing:2007yg, Cassing:2007nb, Moreau:2019vhw, Soloveva:2019xph} is a phenomenological effective model for describing non-perturbative QCD matter in terms of strongly interacting quasi-particles. It is formulated in a two-particle-irreducible (2PI) propagator representation, where dressed partonic propagators encode medium-dependent masses and widths.
 The model’s equation of state (EoS) is matched to lattice-QCD (lQCD) results, allowing the DQPM to describe the thermodynamic and transport properties of the strongly coupled quark-gluon plasma beyond perturbative QCD in the whole $(T,\mu_B)$ plane \cite{Moreau:2019vhw}.
 
\com{
The spectral functions are no longer $\delta$-functions, but have the form separately for quarks, antiquarks, and gluons ($j = q,\bar q,g$)~\cite{Linnyk:2015rco}
\begin{align}
  \rho_{j}(\omega,{\bf p}) &= \frac{\gamma_{j}}{\tilde{E}_j}
  \left(\frac{1}{(\omega-\tilde{E}_j)^2+\gamma^{2}_{j}}
  -\frac{1}{(\omega+\tilde{E}_j)^2+\gamma^{2}_{j}}\right) 
  \nonumber\\
  &\equiv \frac{4\omega\gamma_j}{\left( \omega^2 - \vec{p}^2 - M^2_j \right)^2 + 4\gamma^2_j \omega^2},
  \label{eq:spectral_function}
\end{align}
where $\tilde{E}_{j}^2({\bf p})={\bf p}^2+M_{j}^{2}-\gamma_{j}^{2}$ with the widths $\gamma_{j}$ and the masses $M_{j}$. The spectral function is antisymmetric in $\omega$ and normalized as
\begin{equation}
  \label{eq:spectral_function_norm}
  \int\limits_{-\infty}^{\infty}\frac{d\omega}{2\pi}\
  \omega \ \rho_{j}(\omega,{\bf p})=
  \int\limits_{0}^{\infty} d\omega \frac{\omega}{\pi}\ 
  \rho_{j}(\omega,{\bf p})=1.
\end{equation}

The dynamical quasiparticle mass $M$ is given by the HTL thermal mass in the asymptotic high-momentum regime, i.e., for gluons by \cite{Linnyk:2015rco}
\begin{equation}
  \label{eq:Mg}
  M^2_{g}(T,\mu_q)=\frac{g^2(T,\mu_q)}{6}\left(\left(N_{c}+\frac{1}{2}N_{f}\right)T^2
  +\frac{N_c}{2}\sum_{q}\frac{\mu^{2}_{q}}{\pi^2}\right),
\end{equation}
and for quarks (antiquarks) by
\begin{equation}
  \label{eq:Mq}
  M^2_{q(\bar q)}(T,\mu_q)=\frac{N^{2}_{c}-1}{8N_{c}}g^2(T,\mu_q)\left(T^2+\frac{\mu^{2}_{q}}{\pi^2}\right),
\end{equation}
where $N_{c}\ (=3)$ stands for the number of colors while $N_{f}\ (=3)$ denotes the number of light flavors. Eq.~(\ref{eq:Mq}) determines masses for the ($u,d$) quarks, the strange quark has a larger bare mass for controlling the strangeness ratio in the QGP. Empirically, we find $M_s(T,\mu_B)= M_{u/d}(T,\mu_B)+ \Delta M$, where $\Delta M$ = 30 MeV, which has been fixed once in comparison to experimental data. Furthermore, the effective quarks, antiquarks, and gluons in the DQPM have finite widths $\gamma$, which are taken in the form \cite{Linnyk:2015rco}
\begin{equation}
		\gamma_{j}(T,\mu_\mathrm{B}) = \frac{1}{3} C_j \frac{g^2(T,\mu_\mathrm{B})T}{8\pi}\ln\left(\frac{2c_m}{g^2(T,\mu_\mathrm{B})}+1\right).
	\label{eq:widths}
	\end{equation}
~\\
Here $c_m=14.4$ is related to a magnetic cutoff, which is an additional parameter of the DQPM, and $C_q = \dfrac{N_c^2 - 1}{2 N_c} = 4/3$ and $C_g = N_c = 3$ are the {QCD} color factors for quarks and for gluons, respectively. We also assume that all (anti-)quarks have the same width. With the choice of Eq. (\ref{eq:spectral_function}), the complex self-energies for gluons $\Pi = M_g^2-2i \omega \gamma_g$ and for (anti)quarks $\Sigma_{q} = M_{q}^2 - 2 i \omega \gamma_{q}$ are fully defined via Eqs. (\ref{eq:Mg}), (\ref{eq:Mq}) and (\ref{eq:widths}).

The strength of the interaction is defined by the coupling constant $g^2 = 4\pi\alpha_s$. This is an essential quantity which enters the definition of the DQPM thermal masses and widths and so determines all the microscopic properties as well as transport coefficients. In the DQPM this quantity can be extracted from lQCD in the following way. At vanishing chemical potential $g^2$ is obtained by a parametrization, introduced in Ref. \cite{Berrehrah:2016vzw}, of the lQCD results for the entropy density $s(T,\mu_B = 0)$ obtained by the BW Collaboration~ \cite{Borsanyi:2012cr}
\begin{equation}
  g^2(T,\mu_B=0) = d \left( (s(T,0)/s^{QCD}_{SB})^e - 1 \right)^f.
  \label{eq:g2mub0}
\end{equation}
Here $s^{QCD}_{SB} = 19/9 \pi^2T^3$ is the Stefan-Boltzmann entropy density and $d = 169.934, e = -0.178434$ and $f = 1.14631$ are the dimensionless parameters. 

The extension of the coupling constant to finite baryon chemical potential $\mu_B$ is realized using a scaling hypothesis \cite{Cassing:2008nn}, which works up to $\mu_B \approx 500$ MeV. It assumes that $g^2$ is a function of the ratio of the effective temperature $T^* = \sqrt{T^2 + \mu_q^2/\pi^2}$ and the $\mu_B$-dependent critical temperature $T_c(\mu_B)$ as
\begin{equation}
\label{eq:coupling}
  g^2(T/T_c,\mu_B) = g^2\left(\frac{T^*}{T_c(\mu_B)},\mu_B =0 \right),
\end{equation}
with $T_c(\mu_B) = T_c \sqrt{1 - \alpha\mu_B^2}$, where $T_c$ is the critical temperature at vanishing chemical potential ($\approx$ 0.158~GeV) and $\alpha \simeq 0.974$ GeV$^{-2}$ \cite{Berrehrah:2016vzw}. This extension of the coupling constant for finite $\mu_B$ increases the $\mu_B$ dependence of pole masses and widths, especially in the vicinity of $T_c$.

\begin{figure}[h!]
  \centering
  \includegraphics[width=0.8\columnwidth]{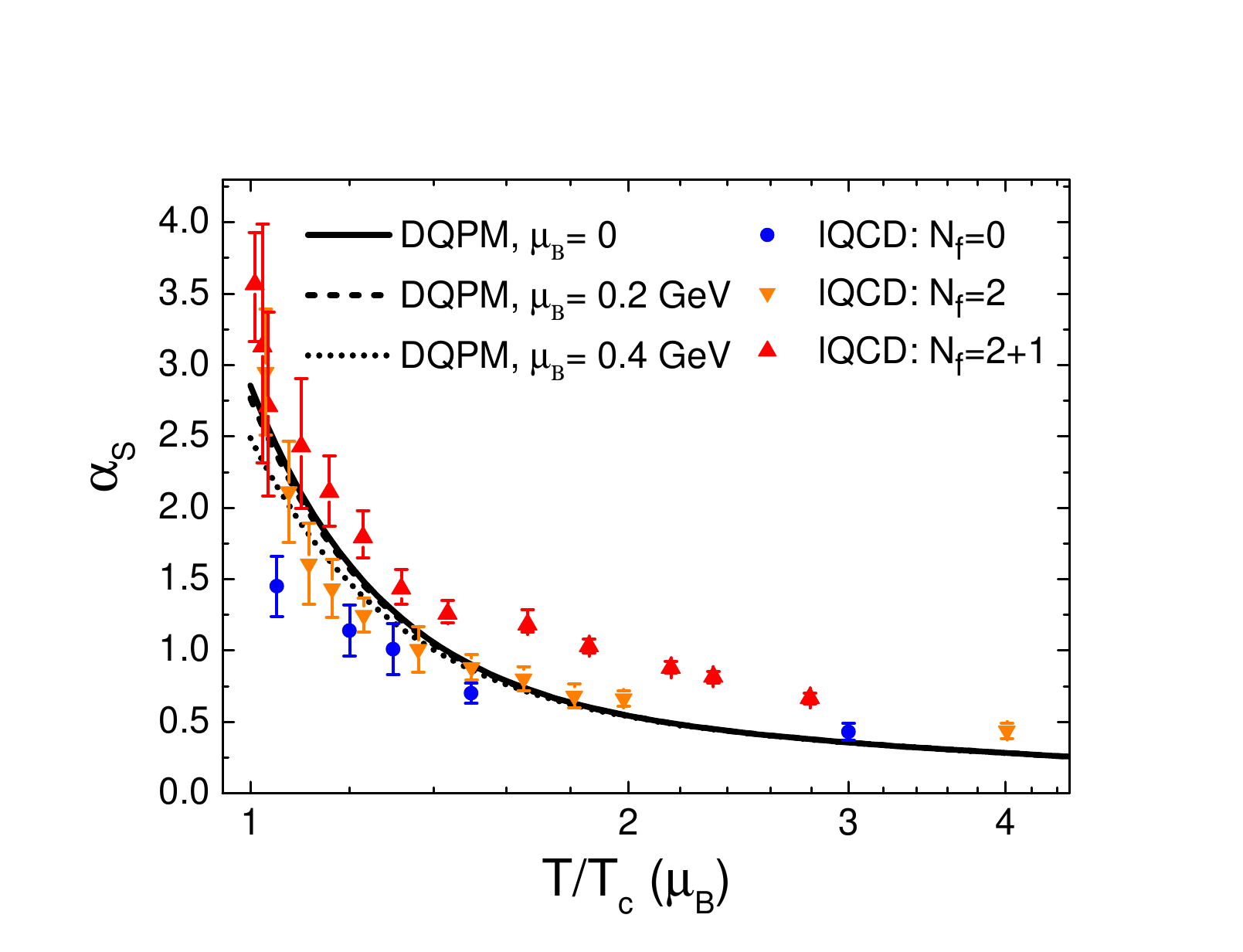}
  \caption{
    Running coupling constant $\alpha_s$ as a function of $T/T_c$ for $\mu_B=0, 0.2, 04$ MeV.  The lattice results for quenched QCD, $N_f = 0$ (blue circles), are taken from Ref. \cite{Kaczmarek:2004gv}, for $N_f = 2$ (orange inverted triangles) from Ref. \cite{Kaczmarek:2005ui}, and for $N_f =2 +1$ (red triangles) from Ref. \cite{Kaczmarek:2007pb}.
  }
  \label{fig:alphas}
\end{figure}

Figure \ref{fig:alphas} displays the running coupling constant $\alpha_s$ for $\mu_B=0$ as a function of $T/T_c$, represented by the black solid line in accordance with Eq. \eqref{eq:g2mub0}. The lattice results for quenched QCD, $N_f = 0$ (blue circles) are taken from Ref. \cite{Kaczmarek:2004gv}, for $N_f = 2$ (orange inverted triangles) from Ref. \cite{Kaczmarek:2005ui}, and for $N_f = 2+1$ (red triangles) from Ref. \cite{Kaczmarek:2007pb}. The fit obtained from the entropy density for the DQPM $\alpha_s$ shows a significant increase approaching $T_c$ comparable to the predictions from lQCD. 
}


\com{The scattering cross sections of charm quarks are evaluated from tree-level Feynman diagrams within the DQPM framework~\cite{Moreau:2019vhw,Grishmanovskii:2023gog,Song:2024hvv}. The partonic propagators entering these diagrams are dressed and thus effectively resum medium effects through temperature- and density-dependent quasiparticle masses, $M_q$ and $M_g$, and widths, $\gamma_q$ and $\gamma_g$. The quark propagator is written as
\begin{equation}
\label{eq:prop_q}
\com{\QP{q}= i \frac{\not{q} + M_q}{q^2 - M^2_q + 2i \gamma_q q_0},}
\end{equation}
\com{and the gluon propagator as} 
\begin{equation}
\label{eq:prop_g}
\com{\GP{\mu}{\nu}{q}= -i \frac{g^{\mu \nu} - q^\mu q^\nu / M^2_g}{q^2 - M^2_g + 2i \gamma_g q_0}.}
\end{equation}

As a consequence, the resulting scattering amplitudes incorporate nonperturbative medium effects beyond the conventional pQCD treatment with bare propagators. In particular, the finite quasiparticle masses and widths regularize the infrared behavior and lead to finite, divergence-free cross sections without introducing an additional external Debye screening mass; the dynamically generated gluon mass $M_g$ provides the relevant screening scale. We emphasize that the DQPM parameters are constrained by lattice-QCD thermodynamics and that the resulting transport coefficients of both light and heavy quarks in the QGP are consistent with available lattice-QCD results~\cite{Berrehrah:2014kba,Song:2019cqz,Grishmanovskii:2022tpb,Grishmanovskii:2024gag,Grishmanovskii:2025mnc}.
}

\begin{figure}[h]
\centerline{
\includegraphics[width=9 cm]{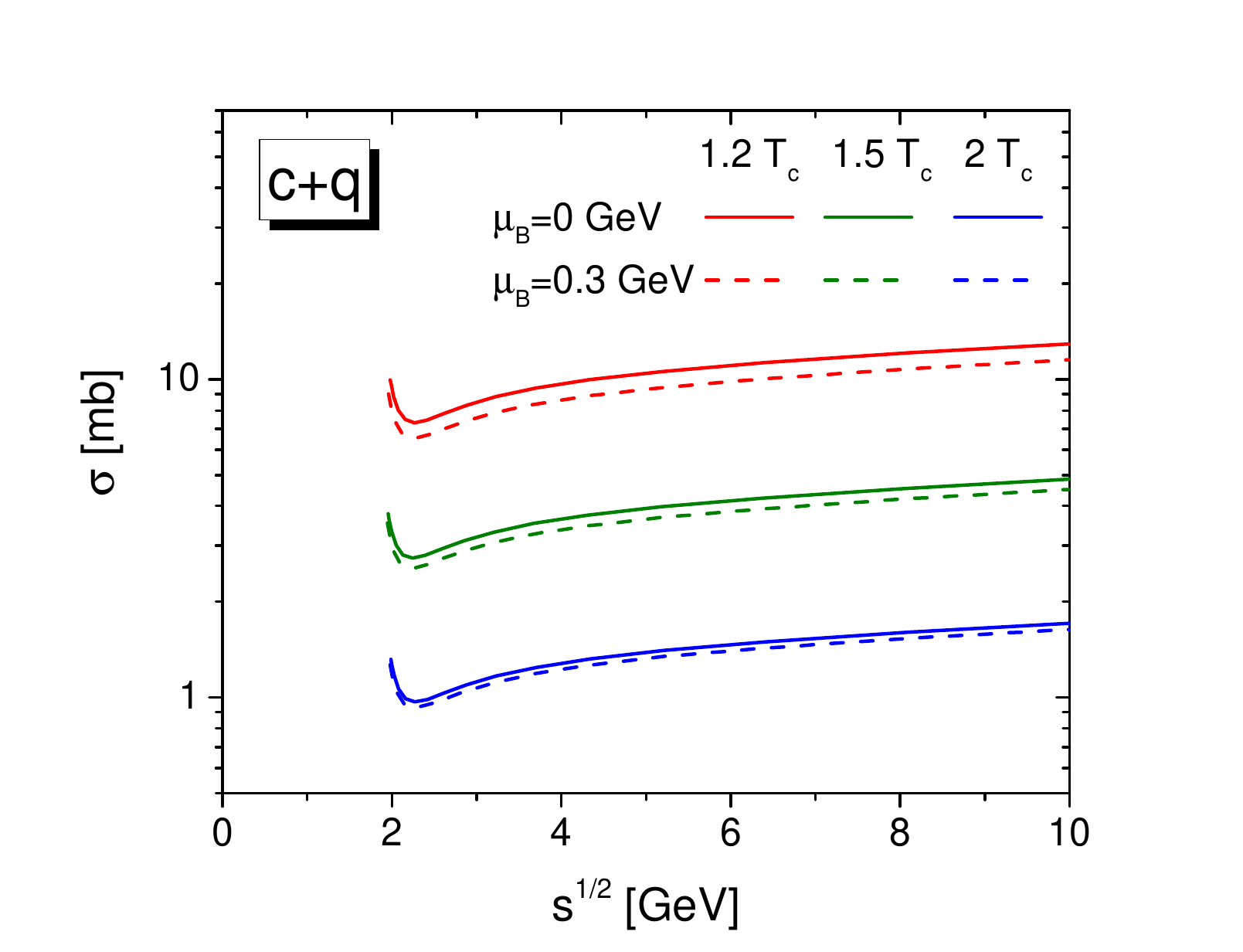}}
\centerline{
\includegraphics[width=9 cm]{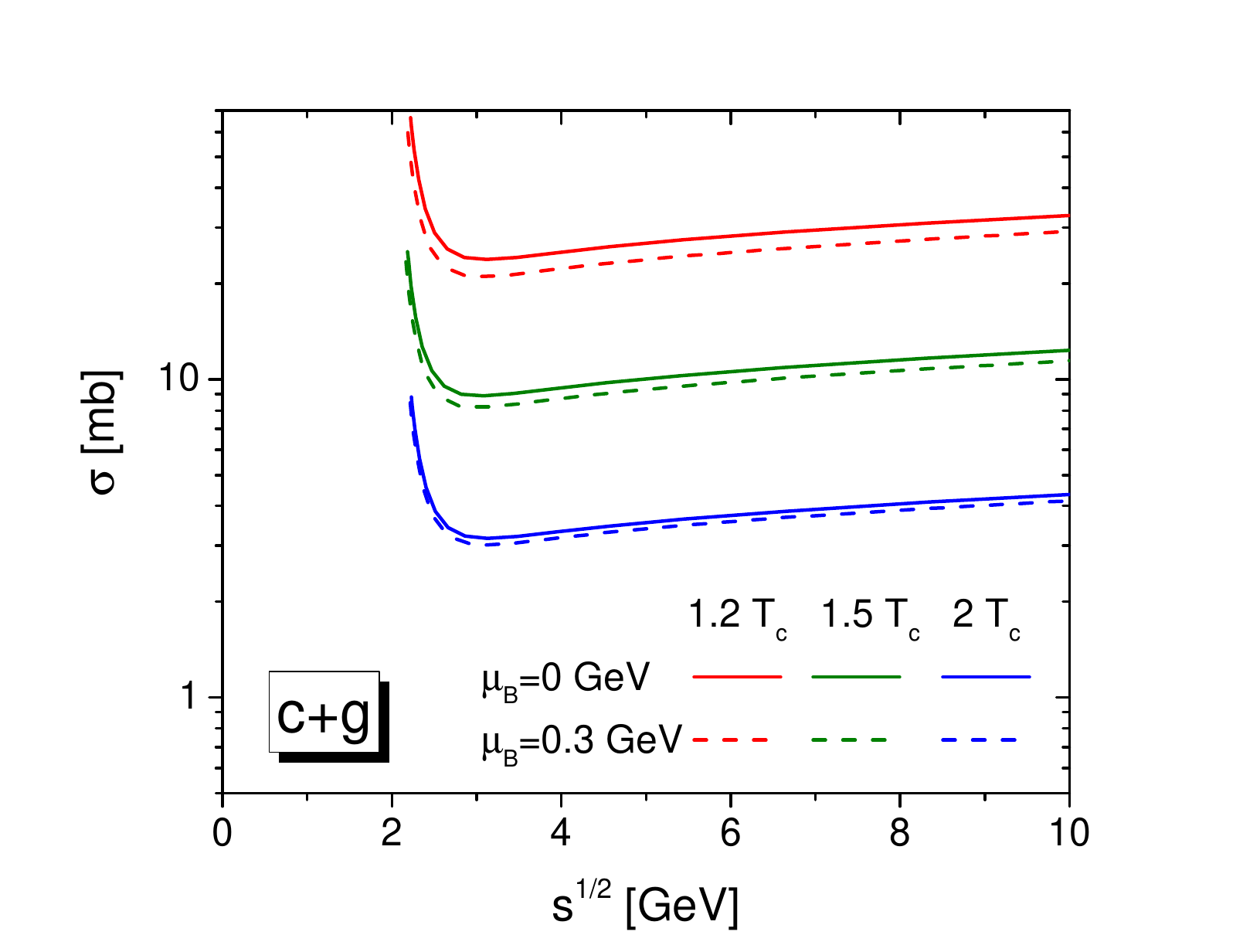}}
\caption{Charm quark scattering cross sections with (upper) light quarks and (lower) gluons, assuming pole masses, as a function of scattering energy at $T=$ 1.2, 1.5 and 2.0 $T_c$ for $\mu_B=$0 or 0.3 $\rm GeV$.}
\label{sigma-charm}
\end{figure}


\com{As an example, Fig.~\ref{sigma-charm} shows the charm-quark scattering cross sections with light quarks and gluons, calculated using pole masses, as functions of the center-of-mass energy at $T=1.2$, $1.5$, and $2.0,T_c$ for $\mu_B=0$, as well as at $\mu_B=0.3$ GeV~\cite{Berrehrah:2013mua}. Since the strong coupling increases as the temperature approaches $T_c$, the corresponding scattering cross sections become larger at lower temperatures. The figure also shows a slight suppression of the cross sections at finite $\mu_B$, reflecting the decrease of the effective strong coupling with increasing baryon chemical potential. We note that the calculation of charm-quark scattering cross sections within the DQPM has recently been extended to include radiative $2\to3$ processes~\cite{Grishmanovskii:2025mnc}.
}

\begin{figure}[h]
\centerline{
\includegraphics[width=9 cm]{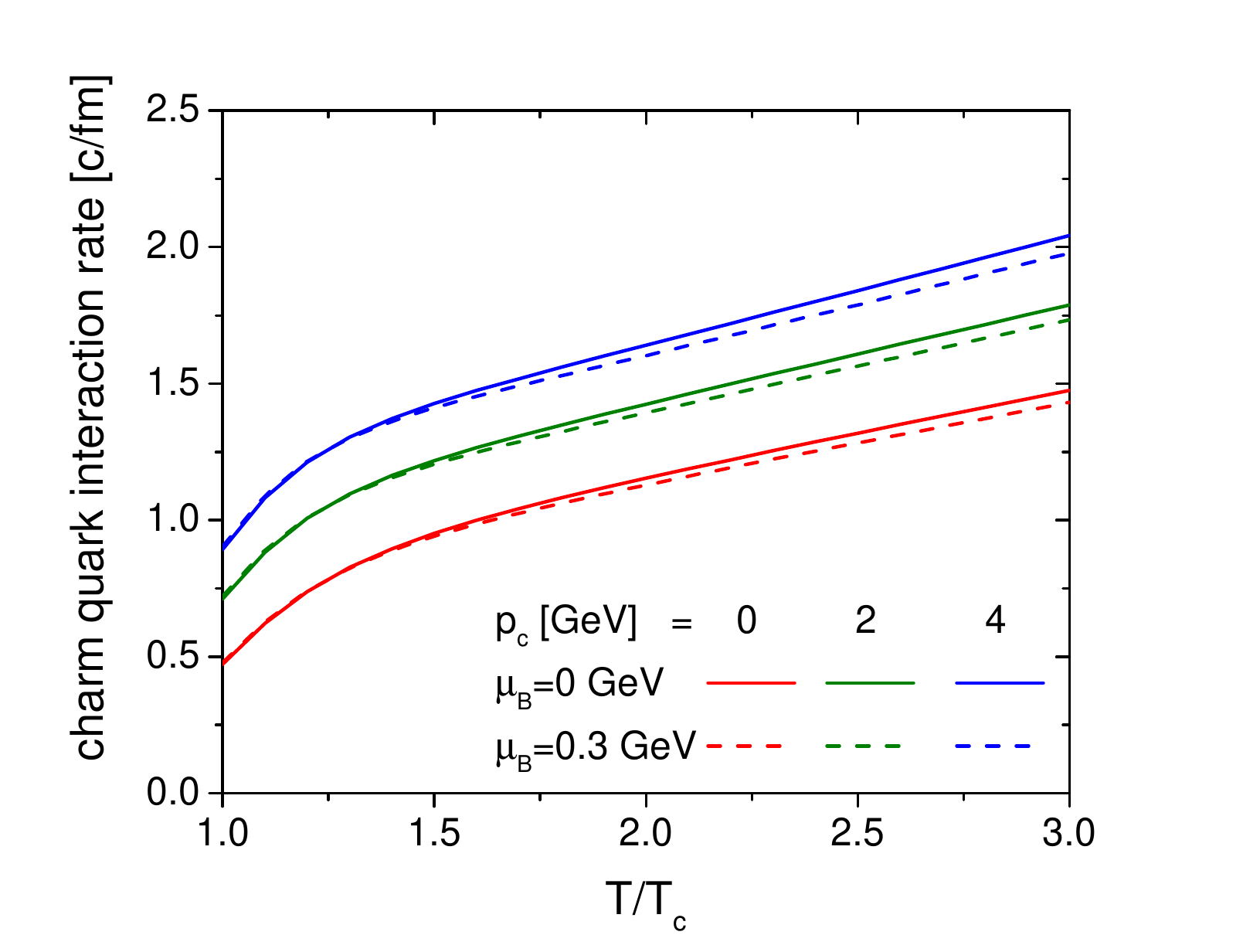}}
\caption{The interaction rate of charm quark in QGP as a function of temperature for charm quark momentum $p_c=$ 0, 2 and 4 GeV and for $\mu_B=$0 or 0.3 $\rm GeV$.}
\label{interaction-rate}
\end{figure}


\com{Additionally, Fig.~\ref{interaction-rate} shows the interaction rate of charm
quarks in the QGP as a function of temperature for different charm-quark
momenta. The interaction rate increases with both temperature and charm-quark
momentum and is typically of the order of $1$--$2~c/{\rm fm}$, implying that a
charm quark undergoes approximately one to two scatterings per $fm/c$ while
propagating through the QGP. Other heavy-quark transport coefficients,
including the drag and momentum-diffusion coefficients, have also been
studied extensively within the DQPM. In particular, the resulting spatial
diffusion coefficient is consistent with available lattice-QCD
calculations~\cite{Song:2019cqz,Grishmanovskii:2025mnc}. These results
demonstrate that the DQPM provides a nonperturbative description of
heavy-quark interactions in the QGP that is compatible with lattice-QCD
constraints.
}

\subsection{Remler formalism}

We start out with a discussion of the Remler formalism for quarkonium production in the QGP.
In the Remler formalism ~\cite{Remler:1975re,Remler:1975fm,Remler:1981du,Song:2017phm,Zhao:2024vqp,Song:2023ywt,Villar:2022sbv,Song:2023zma}, the expected number of quarkonia as a function of time is given by 
\begin{equation}
    N(t) = N(0)+\int_0^t \Gamma(t^\prime)dt^\prime,
\label{probability}    
\end{equation}
where $t$ denotes the time when the QGP phase ends and the rate $\Gamma(t)$ is expressed as
\begin{eqnarray}
\Gamma(t)
= \sum_{i,j}\sum_{\nu_{i(j)}}\frac{1}{(2\pi)^{3N}}  \int d^3r_1d^3p_1 ... d^3r_N d^3p_N\nonumber\\
\times W_0(\vec{r}_i-\vec{r}_j,\vec{p}_i-\vec{p}_j)
\bigg\{\delta\bigg(t-t^\nu_{i(j)}\bigg)\nonumber\\
-\delta\bigg(t-t^{\nu-1}_{i(j)}\bigg)\bigg\}W^{(N)}(t+\varepsilon),~~~
\label{new}
\end{eqnarray}
where $N$ is the total number of heavy quarks and heavy antiquarks.
The indices $i$ and $j$ in Eq.~(\ref{new}) run over all heavy quarks and heavy antiquarks, respectively.
$t^\nu_{i(j)}$ denotes the time of $\nu$ th scattering of the heavy quark $i$ (or heavy antiquark $j$) in the QGP. 
$W_0$ is the Wigner function of the quarkonium 
composed of heavy quark $i$ and heavy antiquark $j$, whereas $W^{(N)}(t)$ is the quantal density matrix of the $N$ heavy quarks and heavy antiquarks.
In practice, the latter is approximated by their classical phase-space density distribution,
\begin{eqnarray}
W^{(N)}(t)\approx \prod_{i=1}^N (2\pi)^{3N}\delta(r_i-r_i^*(t))\delta(p_i-p_i^*(t)),
\label{deltas}
\end{eqnarray} 
which satisfies the normalization condition:
\begin{eqnarray}
\prod_{i=1}^N \int \frac{d^3r_i d^3p_i}{(2\pi)^{3N}}W^{(N)}(t)=1.\nonumber
\end{eqnarray}

Eq.~(\ref{new}) implies that the Wigner projection at $t=t^{\nu-1}_{i(j)}$,
\begin{eqnarray}
W_0(\vec{r}_i-\vec{r}_j,\vec{p}_i-\vec{p}_j)W^{(N)}(t^{\nu-1}_{i(j)})
\end{eqnarray}
is replaced by the Wigner projection at $t=t^{\nu}_{i(j)}$,
\begin{eqnarray}
W_0(\vec{r}_i-\vec{r}_j,\vec{p}_i-\vec{p}_j)W^{(N)}(t^{\nu}_{i(j)})
\end{eqnarray}
through the $\nu$th scattering of heavy quark $i$ (or heavy antiquark $j$)~\cite{Song:2023ywt}.
The parameter $\varepsilon$ in Eq.~(\ref{new}) signifies that the Wigner density
has to be applied after the momentum change of the heavy (anti)quark due to a collision with the QGP. 
In terms of the thermal interaction of quarkonium with the medium, the first term in the curly bracket of Eq.~(\ref{new}) can be interpreted as quarkonium regeneration, while the second term corresponds to thermal dissociation.

Integrating the rate over time, the number of quarkonium becomes
\begin{eqnarray}
    N(t)=\sum_{i,j}\frac{1}{(2\pi)^{3N}}  \int d^3r_1d^3p_1 ... d^3r_N d^3p_N\nonumber\\
\times W_0(\vec{r}_i-\vec{r}_j,\vec{p}_i-\vec{p}_j)\bigg\{
W^{(N)}(t_0)\nonumber\\
+W^{(N)}(t^1_{i(j)})-W^{(N)}(t_0)+~...\nonumber\\
+W^{(N)}(t^f_{i(j)})-W^{(N)}(t^{f-1}_{i(j)})\bigg\}\nonumber\\
=\sum_{i,j}\frac{1}{(2\pi)^{3N}}  \int d^3r_1d^3p_1 ... d^3r_N d^3p_N\nonumber\\
\times W_0(\vec{r}_i-\vec{r}_j,\vec{p}_i-\vec{p}_j)W^{(N)}(t^f_{i(j)}),
\label{tinfinity}
\end{eqnarray}
where $t_0$ is the initial projection time, which will be discussed in the next section, and $t^f_{i(j)}$ denotes the last scattering time of heavy quark $i$ (or heavy antiquark $j$).
Thus, Eq.~(\ref{tinfinity}) corresponds to the Wigner projection of $ W_0(\vec{r},\vec{p})$ onto $W^{(N)}(t)$ for all heavy quark pair combinations at their final scattering times, $t=t^f_{i(j)}$.
If a charm quark does not scatter in QGP, what often happens in p+p and p+A collisions, only the initial Wigner projection is left.

This approach has been tested in both thermalized and thermalizing box simulations~\cite{Song:2023ywt} and has successfully described bottomonium production not only in p+p but also in heavy-ion collisions at LHC energies~\cite{Song:2023zma}.
Recently, it was shown that, in addition to heavy-quark scattering,  the heavy-quark potential should be included to describe the heavy-quark phase-space distribution more realistically~\cite{Song:2025zfy}.

In the present study, we apply this method to charmonium production at relatively low collision energies, as those at SPS and FAIR, which are relevant for studying high-$\mu_B$ partonic and nuclear matter.

\subsection{Hadronic interactions of charmonium}\label{hadronic-int}

After the hadronization of the QGP, the produced charmonia - formed according to the Remler formalism - interact with mesons and baryons. 
The inelastic interactions with mesons are commonly referred to as comover effects, and dominate the hadronic suppression of $J/\psi$ in heavy collisions where many mesons are produced.
Inelastic interactions with baryons are usually called nuclear absorption, since they constitute the dominant  suppression mechanism of charmonium in p+A collisions where only few mesons are produced.
Historically, an anomalous $J/\psi$ suppression was observed in Pb+Pb at SPS by the NA50 Collaboration~\cite{NA50:1996lag,NA50:2000brc}, and this observation was suggested to be a possible signal for the onset of a phase transition to the QGP~\cite{Blaizot:1996nq,Wong:2000if}.
However, alternative explanations that do not invoke QGP formation were later proposed, attributing the suppression instead to interactions of charmonium with comoving hadrons~\cite{Cassing:1996zb,Cassing:1997kw,Armesto:1997sa,Capella:2000zp,Sibirtsev:1999jr}.

To estimate the hadronic interactions of charmonium, it is necessary to know the scattering cross sections of charmonium with mesons and baryons.
However, the cross sections are quite uncertain and strongly depend on the theoretical model employed~\cite{Martins:1994hd,Lin:1999ad,Haglin:2000ar,Wong:2001td,Duraes:2002ux,Oh:2000qr,Oh:2001rm,Song:2005yd}.
For this reason the cross section for the charmonium-nucleon interaction can be extracted from experimental data in p+A collisions,  for simplicity assuming a constant cross section, regardless of scattering energy.  In principle, the cross section should depend on the scattering energy but data do not allow yet to determine $J/\psi+N$ energy dependence.

Once the nuclear absorption cross section has been determined from p+A collisions, one can investigate the comover interaction cross sections.
The cross sections for the comover dissociation of charmonium and for its regeneration from two open-charm mesons are not well known, although they have been discussed in many studies~\cite{Braun-Munzinger:2000csl,Braun-Munzinger:2000eyl,Thews:2000rj,Braun-Munzinger:2000uqj,Braun-Munzinger:1999mxu,Martins:1994hd,Wong:1999zb,Wong:2001td,Ko:1998fs,Oh:2001rm,Song:2005yd}. 
Following Refs.~\cite{Bratkovskaya:2003ux,Bratkovskaya:2004cq}, we introduce a
simple two-body transition amplitude $|M_0|^2$ as a single parameter, which also allows the implementation of the backward reactions through detailed balance for each individual channel of comover reaction.

Since charmonium-meson dissociation and the corresponding backward reactions typically
occur with small relative momenta (i.e., among comovers), the cross section for a 2-to-2 process can be written as
\begin{eqnarray}
\label{model}
 \sigma_{1+2 \rightarrow 3+4}(s) = 2^4 \frac{E_1 E_2 E_3 E_4}{s}\nonumber\\
 \times |\tilde M_i|^2 \left(\frac{m_3+m_4}{\sqrt{s}}\right)^6  \frac{p_{out}}{p_{in}},
\end{eqnarray}
where $E_k$ and $m_k$ denote the energy and mass of the incoming $(k=1,2)$ and outgoing $(k=3,4)$ hadron, respectively, and 
\begin{eqnarray}
p_{in}^2 = \{s-(m_1+m_2)^2\}\{s-(m_1-m_2)^2\}/(4s), \nonumber\\
p_{out}^2 = \{s-(m_3+m_4)^2\}\{s-(m_3-m_4)^2\}/(4s),
\label{moment}
\end{eqnarray}
which represent the initial and final three-momenta for the invariant energy $\sqrt{s}$.
The quantity $|\tilde M_i|^2$ ($i=J/\psi,~\chi_c^0,~\chi_c^1,~\chi_c^2,~\psi^\prime$) denotes the
effective squared matrix element averaged over the initial particle spins and summed over the final particle spins.

For example, using the effective Lagrangian approach for light and heavy mesons~\cite{Haglin:1999xs,Lin:1999ad,Haglin:2000ar}, in PHSD we consider the following reactions:
\begin{eqnarray}
&&\hspace*{-3mm}|\tilde M_i|^2 =|M_i|^2  \ \ {\rm for} \
    \ \rho(\omega)+(c\bar c)_i \to D+\bar{D}, \label{mod}\\
&&\hspace*{-3mm}|\tilde M_i|^2 = 3 |M_i|^2  \ \ {\rm for} \
\rho(\omega)+(c\bar c)_i \to D^*+\bar{D}^*,    \   \nonumber\\
&& \pi+(c\bar c)_i \to D^*+\bar{D}, D+\bar{D}^*. 
\end{eqnarray}
Note that the final state $D^*+\bar{D}^*$ is multiplied by a factor 3 rather than 9.
Including anomalous parity interactions, additional scattering channels such as $J/\psi+\rho \to D+\bar{D}^*$ are also possible~\cite{Oh:2001rm,Song:2005yd}, although they are not considered in the present study.

Furthermore, a suppression factor of 1/3 is introduced for all channels involving $s, \bar s$ quarks. 
Consequently,
\begin{eqnarray}
&&|\tilde M_i|^2 = \frac{1}{3} |M_i|^2 \ \ {\rm for~channels~such~as} \nonumber\\
   &&\ K^*+(c\bar c)_i \to D_s + \bar{D}, \nonumber\\
&&\bar{K}^*+(c\bar c)_i \to D+\bar{D}_s,  \nonumber \\
&&|\tilde M_i|^2 =  |M_i|^2  \ \ {\rm for} \
    \ K+(c\bar c)_i  \to D_s + \bar{D}^*(D^*_s + \bar{D}), \nonumber \\
&&~~ \bar{K}+(c\bar c)_i  \to D^*+\bar{D}_s(D+\bar{D}^*_s),\nonumber\\
&& K^*+(c\bar c)_i \to D_s^* + \bar{D}^*, \bar{K}^*+(c\bar c)_i \to D^* + \bar{D}_s^*.
\end{eqnarray}
The factor $\left(
{(m_3+m_4)}/{\sqrt{s}} \right)^6 $ in Eq.~(\ref{model}) accounts for the
suppression of binary channels with increasing $\sqrt{s}$ and has
been fitted to experimental data for the reactions $\pi + N
\rightarrow \rho+N, \omega+N, \phi+N, K^+ +\Lambda$ in Ref.
\cite{Cassing:2001ud}.

For simplicity, we use the same matrix elements for the
dissociation of all charmonium states $i$ ($i=J/\psi,~\chi_c^0,~\chi_c^1,~\chi_c^2,~
\psi^\prime$) with mesons,
\begin{eqnarray}
 |M_{J/\Psi}|^2 = |M_{\chi_c}|^2 = |M_{\Psi^\prime}|^2 = |M_0|^2,
\label{MatrElem}
\end{eqnarray}
where the parameter $|M_0|^2$ is determined by fitting experimental data from the NA50 and NA60 Collaborations~\cite{NA60:2006ikc,NA50:2006rdp}.
\com{We note that, despite using the same scattering amplitudes for $J/\psi$ and the excited charmonia, the dissociation cross sections and rates of the excited states are larger than those of $J/\psi$ due to the phase-space integration, as will be shown in Fig.~\ref{rate}.}

\begin{figure}[h!]
\centerline{
\includegraphics[width=9 cm]{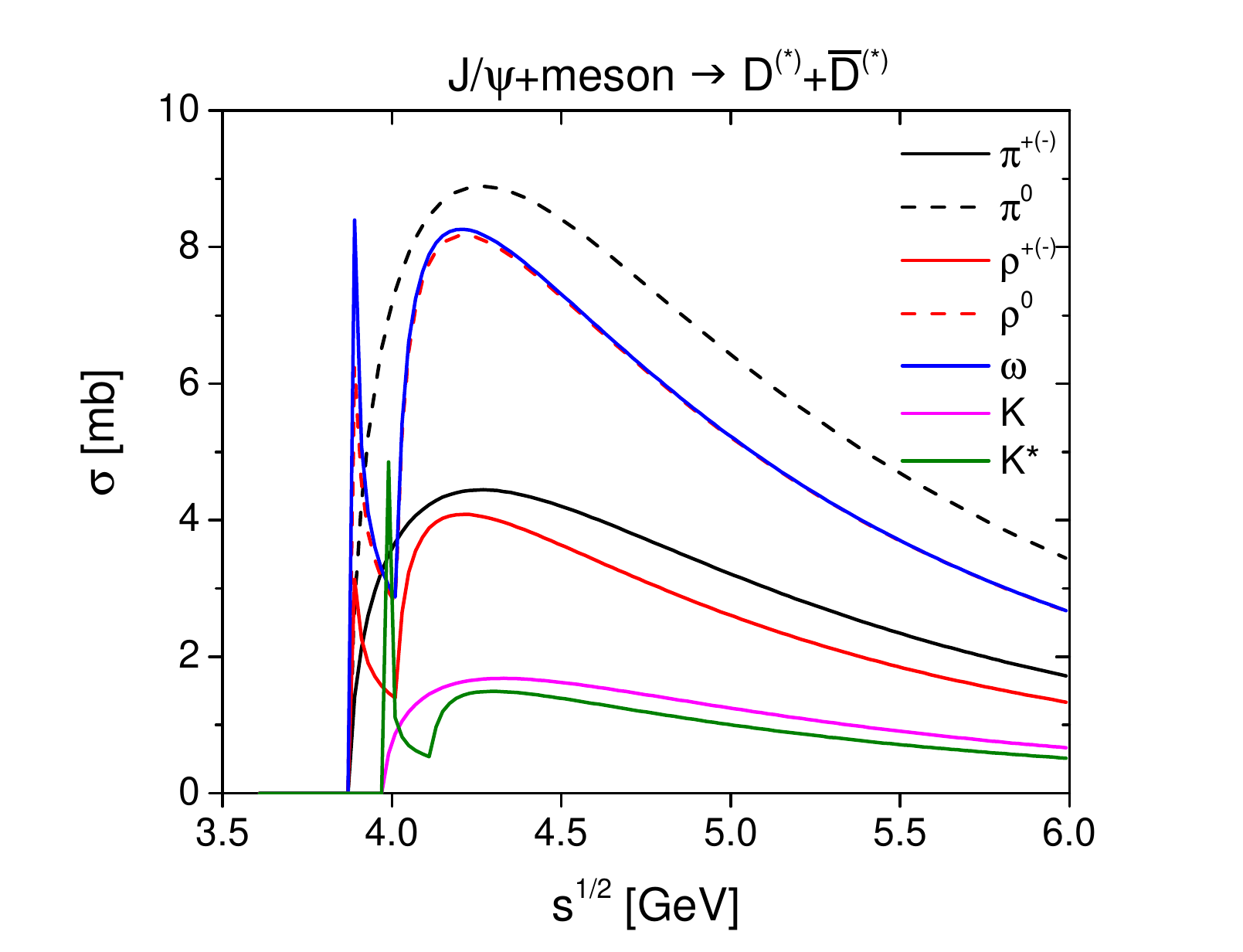}}
\centerline{
\includegraphics[width=9 cm]{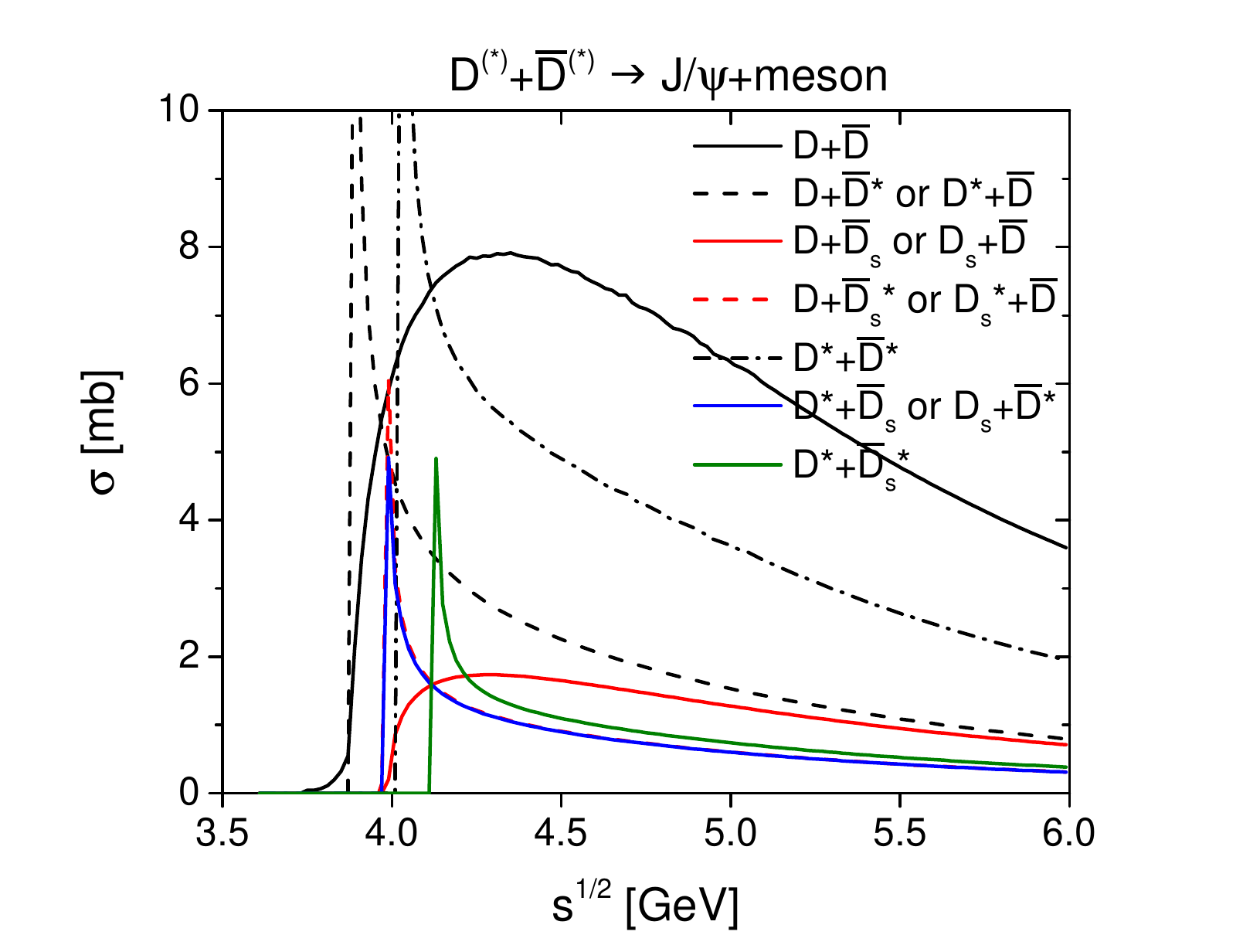}}
\caption{Cross sections for $J/\psi$ dissociation by light mesons ($\pi, \rho, \omega, K, K^*$) and regeneration from $D\bar{D}$ mesons as functions of the center-of-mass energy.}
\label{cs-jpsi}
\end{figure}

\begin{figure}[h!]
\centerline{
\includegraphics[width=9 cm]{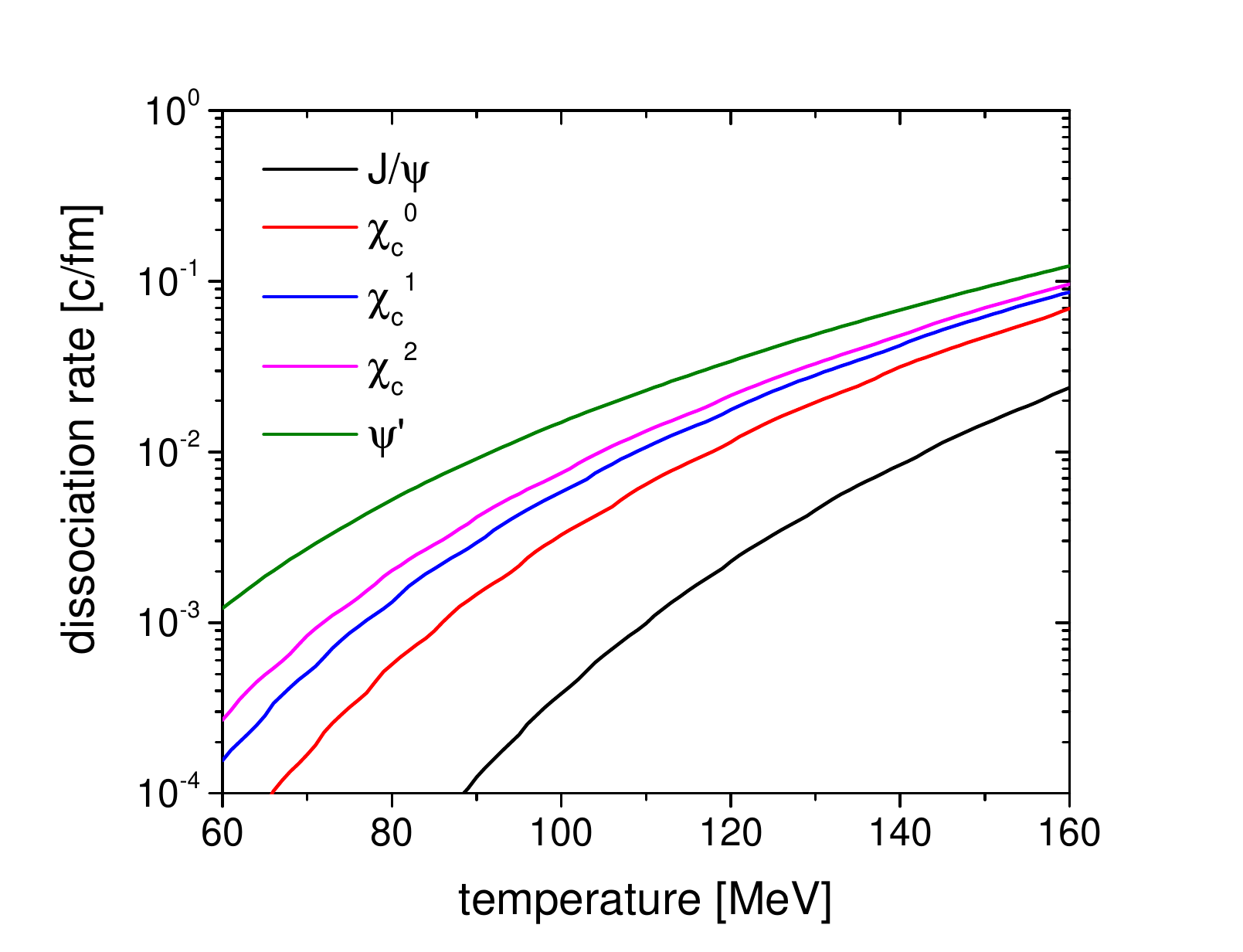}}
\caption{The dissociation rates of static charmonia in thermalized hadron gas as a function of temperature.}
\label{rate}
\end{figure}

In the present study, we adopt $|M_0|^2=0.11$~mb/GeV$^2$, which yields a maximum cross section of about 10 mb, as shown in Fig.~\ref{cs-jpsi}, except for divergent peaks near the threshold energy when the sum of the initial particle masses exceeds that of the final particles.
The impact of the value of $|M_0|^2$ on charmonium observables will be discussed in the next section.

An important advantage of this approach is
that the backward reactions - i.e., the formation of charmonia through $D+\bar{D}$ reactions - can be obtained without introducing any additional parameters, but instead through detailed balance for each individual channel,
\begin{eqnarray}
\!\!\sigma_{3+4 \rightarrow 1+2}(s) =
 \sigma_{1+2 \rightarrow 3+4}(s)\nonumber\\
 \times
\frac{(2S_1+1)(2S_2+1)}{(2S_3+1)(2S_4+1)} \ \frac{p_{in}^2}{p_{out}^2}, \
\label{balance}
\end{eqnarray}
where $S_j$
denotes the spin of particle $j$, and ${p_{in}^2}$ and ${p_{out}^2}$ are given in Eq.~(\ref{moment}).
The uncertainty of these cross sections
is comparable to that in effective Lagrangian approaches, since
the form factors at the interaction vertices are poorly known~\cite{Duraes:2002ux}. In PHSD, the comover
dissociation channels for charmonia are implemented with their
proper individual thresholds for each channel, in contrast to schematic comover absorption models~\cite{Armesto:1997sa,Armesto:1998rc,Capella:2000zp}.

\com{
Fig.~\ref{cs-jpsi} shows the cross sections for $J/\psi$ dissociation by
light mesons and for the inverse regeneration reactions from $D\bar D$
mesons. The dissociation cross sections for $J/\psi+\pi^0$ and
$J/\psi+\rho^0$ are approximately twice as large as those for the
corresponding charged mesons, $J/\psi+\pi^\pm$ and $J/\psi+\rho^\pm$,
because the neutral channels allow for more possible final states. For
example, $J/\psi+\rho^0$ can produce both $D^+D^-$ and
$D^0\bar D^0$, whereas $J/\psi+\rho^+$ can produce only
$D^+\bar D^0$.

The cross section for $J/\psi+\rho^0$ is also close to that for
$J/\psi+\omega$, reflecting the similar pole masses of the $\rho$ and
$\omega$ mesons. We further note that the reactions
$J/\psi+\rho(\omega)$ and $J/\psi+K^*$ receive contributions from two
different classes of final states, $D+\bar D$ and $D^*+\bar D^*$.
The former corresponds to an exothermic process, whereas the latter is
endothermic.

The corresponding $J/\psi$ regeneration cross sections are shown in the
lower panel of Fig.~\ref{cs-jpsi} for various combinations of $D$ and
$\bar D$ mesons. All regeneration cross sections are obtained from the
corresponding dissociation reactions using the detailed-balance relation
given in Eq.~(\ref{balance}).
}

\com{
Fig.~\ref{rate} shows the dissociation rates of static charmonia ($p=0$) in a thermalized hadron gas as a function of temperature:
\begin{eqnarray}
\Gamma_i=\sum_j \int \frac{d^3p}{(2\pi)^3}f_j(p)v_{ij}\sigma_{ij}(\sqrt{s}),
\end{eqnarray}
where $f_j(p)$, $v_{ij}$ and $\sigma_{ij}(\sqrt{s})$ denote the light meson distribution function, relative velocity, and dissociation cross section, respectively, with $i,~j$ labeling the charmonium state and light meson, respectively. 

One can see that, although the same scattering amplitude is used for all charmonium states in Eq.~(\ref{MatrElem}), their interaction rates differ significantly due to kinematic effects.
For example, the dissociation of $J/\psi$ requires a collision with a sufficiently energetic hadron to produce two open-charm mesons, whereas $\psi^\prime$ can be dissociated much more easily.
}

\section{results}
\label{results}

Now we describe in more detail how charmonium is produced in p+p, p+A, and heavy-ion collisions, and present the results for each case.

\subsection{Charmonium production in p+p collisions}
\label{pp}

\begin{figure*}[th!]
\centerline{
\includegraphics[width=8.6 cm]{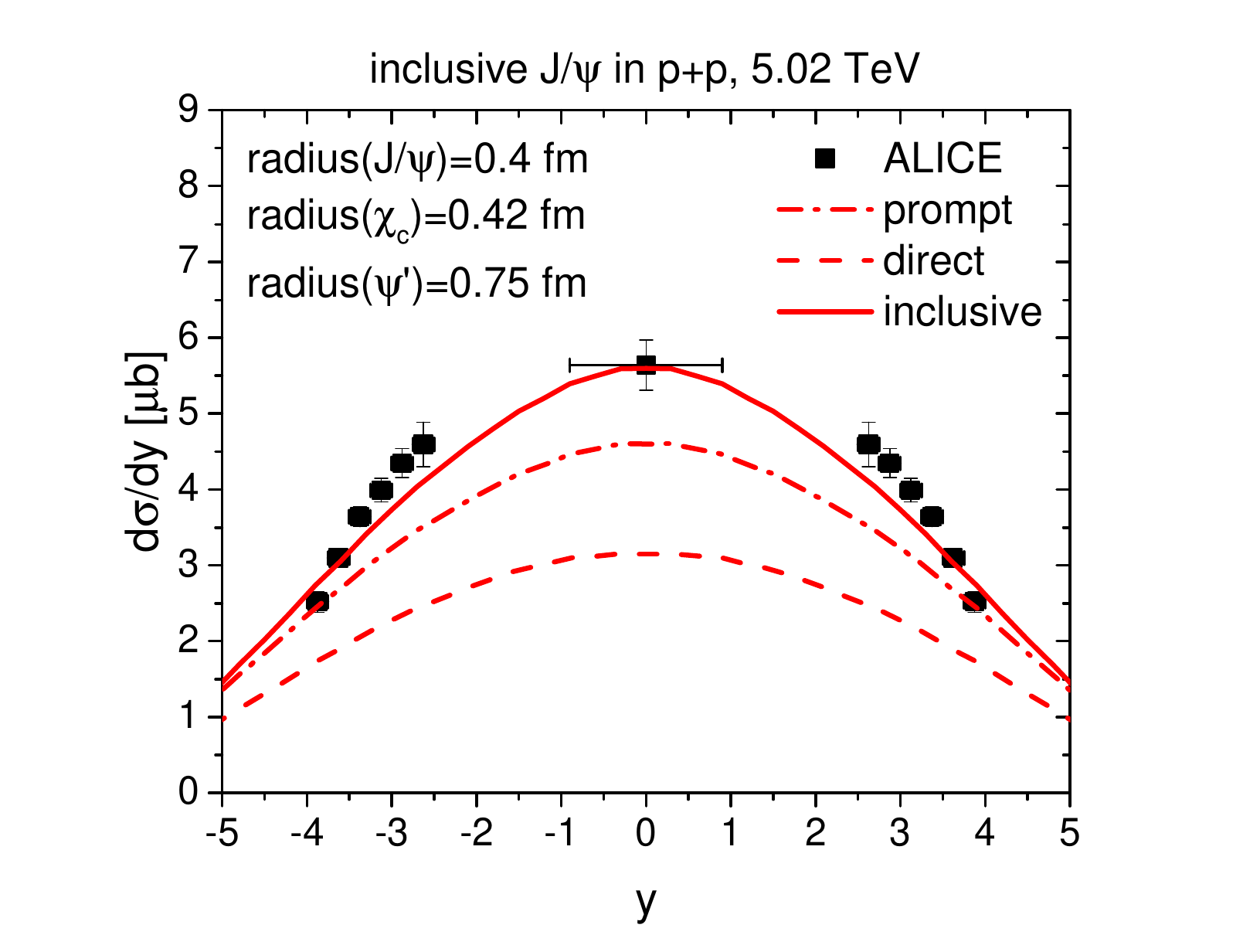}
\includegraphics[width=8.6 cm]{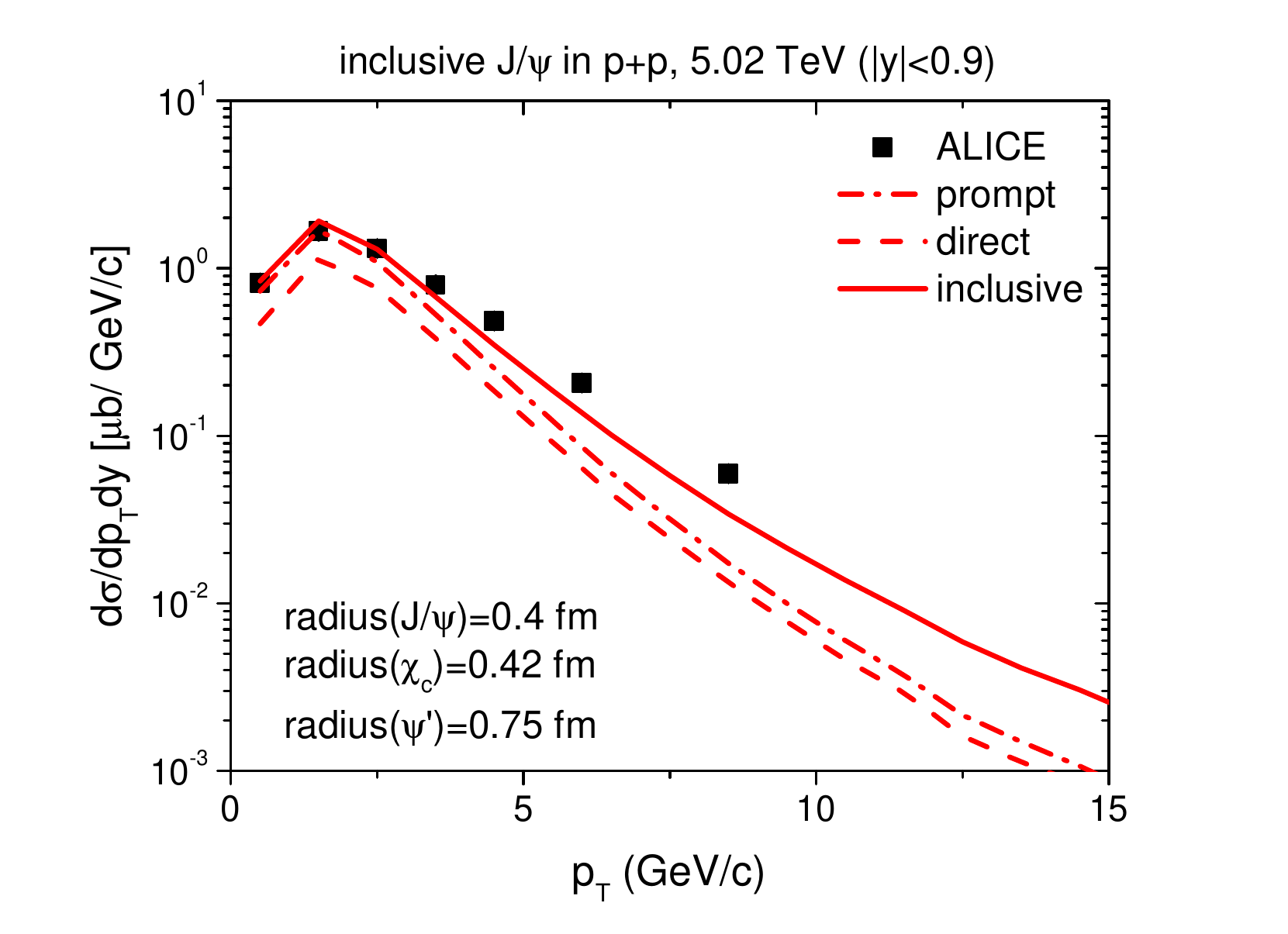}}
\centerline{
\includegraphics[width=8.6 cm]{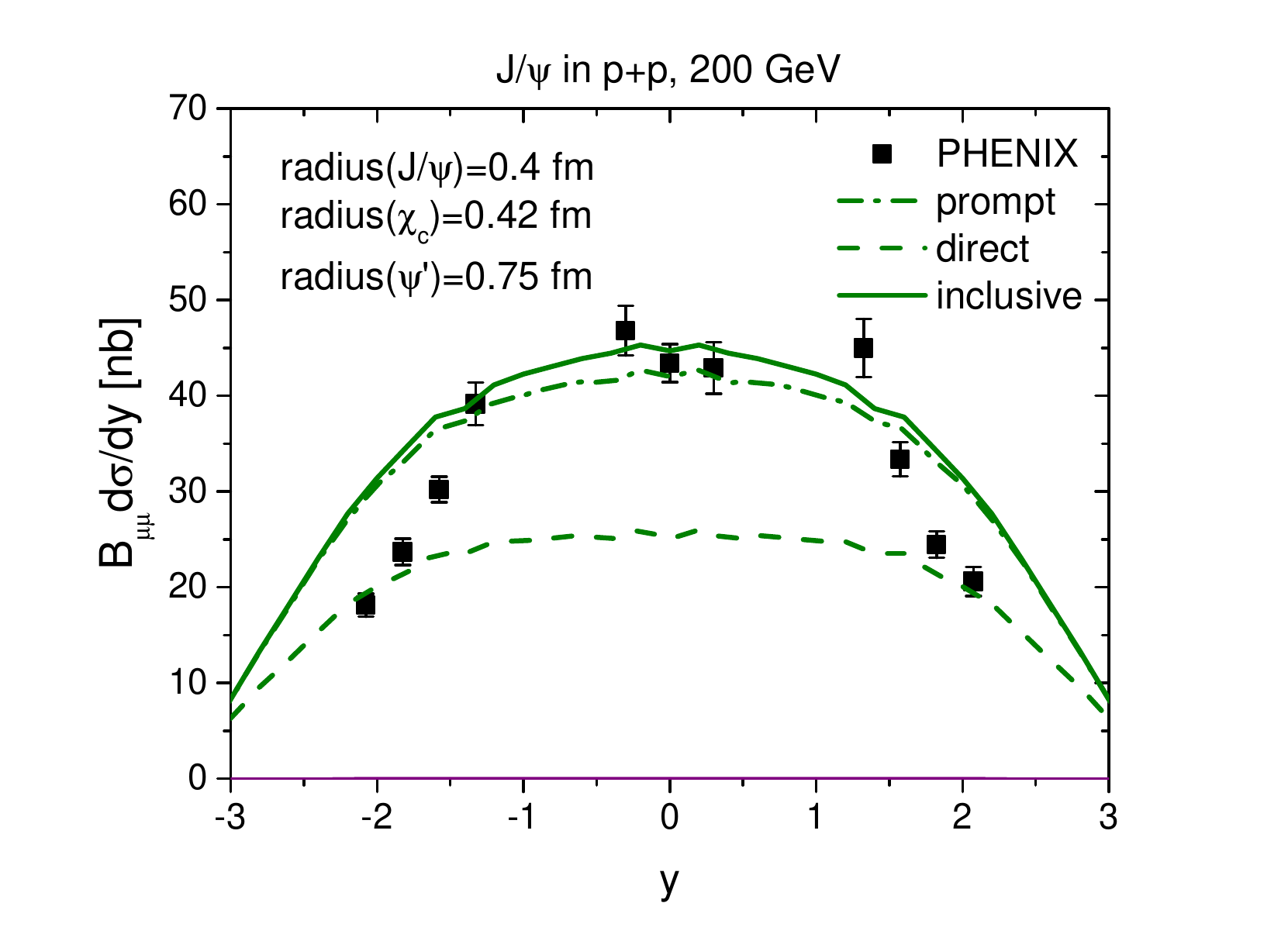}
\includegraphics[width=8.6 cm]{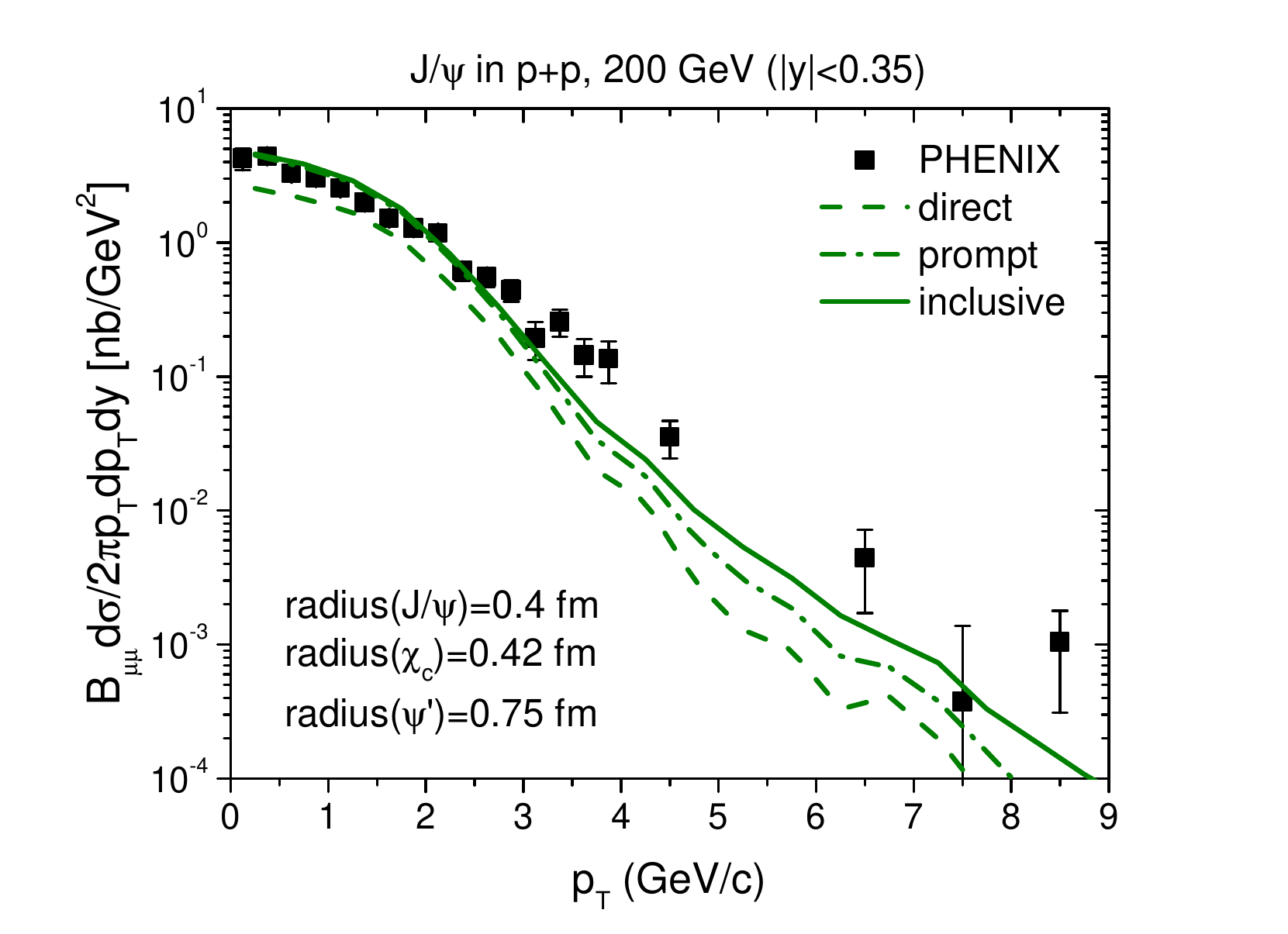}}
\centerline{
\includegraphics[width=8.6 cm]{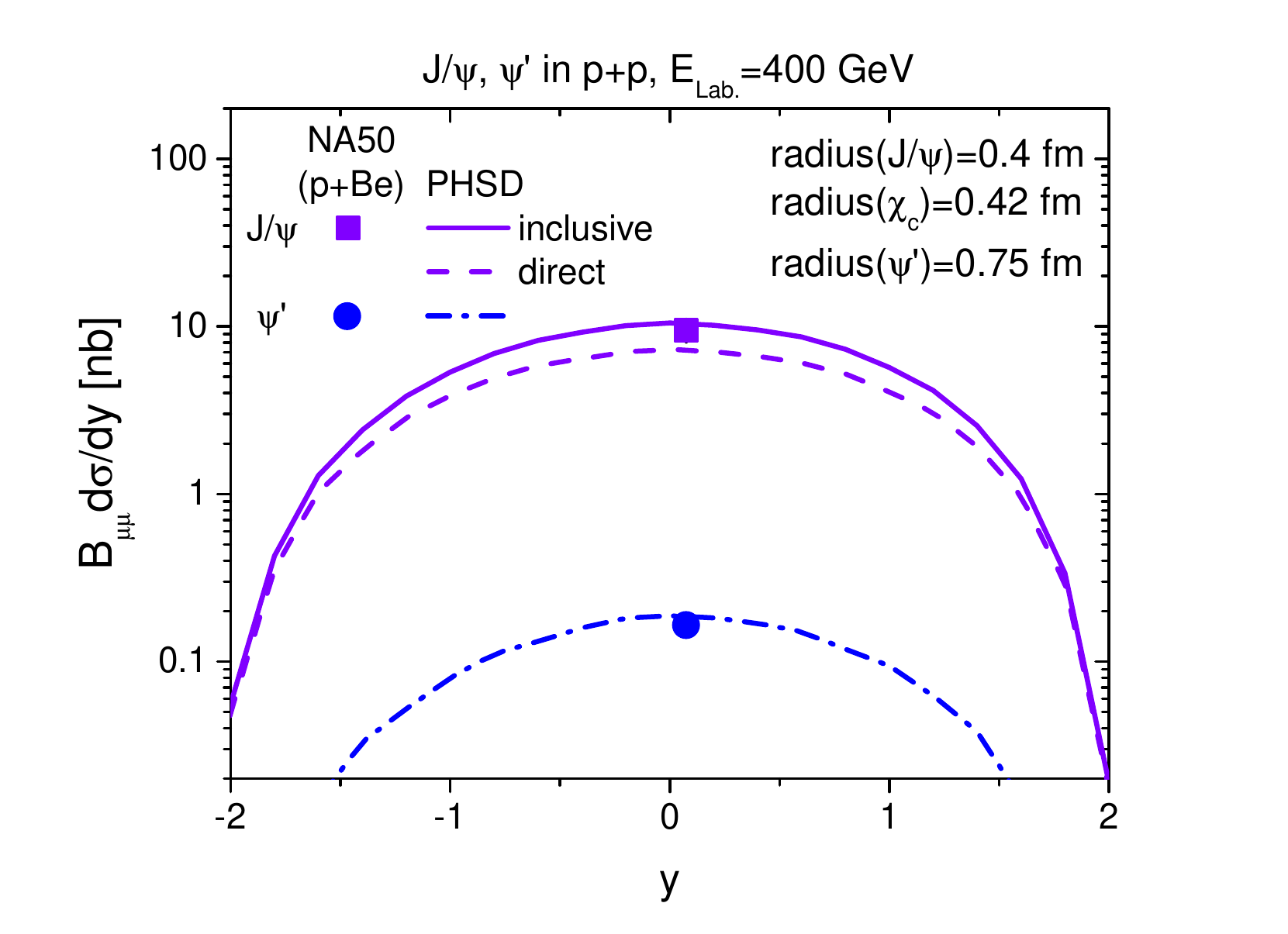}
\includegraphics[width=8.6 cm]{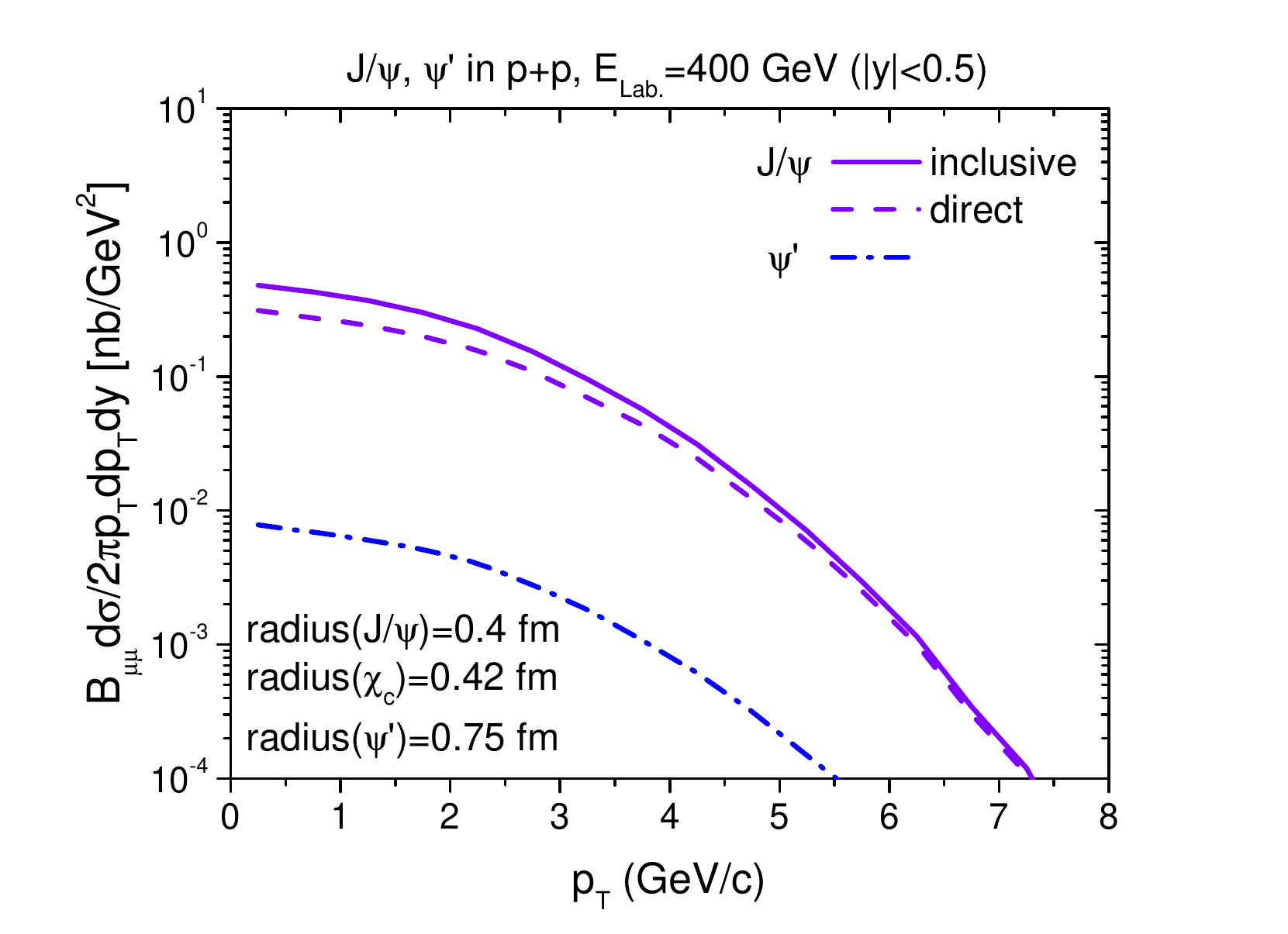}}
\caption{Differential cross sections for charmonium production, multiplied by the branching ratio to dimuon decay, as a function of rapidity and transverse momentum in p+p collisions at $\sqrt{s_{NN}}=$ 5.02 TeV, 200 GeV, and 27 GeV ($E_{kin}=$ 400 GeV), respectively. The experimental data are taken from the ALICE~\cite{ALICE:2019pid}, PHENIX~\cite{PHENIX:2006aub} and NA50~\cite{NA50:2006rdp} Collaborations.}
\label{sigma-pp}
\end{figure*}

From Eqs.~(\ref{probability}) and (\ref{new}), the initial Wigner projection is given by
\begin{eqnarray}
N(t_0)
= \sum_{i,j}\frac{1}{(2\pi)^{3N}}  \int d^3r_1d^3p_1 ... d^3r_N d^3p_N\nonumber\\
\times W_0(\vec{r}_i-\vec{r}_j,\vec{p}_i-\vec{p}_j)
W^{(N)}(t_0),
\label{begin}
\end{eqnarray}
where $t_0$ is the initial Wigner projection time of quarkonium. For example, $t_0$ may correspond to the time when the quarkonium begins to exist below its dissociation temperature $T_d$.
Since the temperature decreases with time, one may consider the situation where the temperature at $t=t^{\nu-1}_{i(j)}$ is above 
$T_d$, while the temperature at $t=t^\nu_{i(j)}$ is below $T_d$.
In this case, the loss term vanishes and 
only the gain term survives in Eq.~(\ref{tinfinity}):
\begin{eqnarray}
    N(t)=\sum_{i,j}\frac{1}{(2\pi)^{3N}}  \int d^3r_1d^3p_1 ... d^3r_N d^3p_N\nonumber\\
\times W_0(\vec{r}_i-\vec{r}_j,\vec{p}_i-\vec{p}_j)\bigg\{
W^{(N)}(t_0)\nonumber\\
+W^{(N)}(t^1_{i(j)})-W^{(N)}(t_0)+~...\nonumber\\
+W^{(N)}(t^{\nu}_{i(j)})-W^{(N)}(t^{\nu-1}_{i(j)})\nonumber\\
+W^{(N)}(t^{\nu+1}_{i(j)})-W^{(N)}(t^{\nu}_{i(j)})+~...\nonumber\\
+W^{(N)}(t^f_{i(j)})-W^{(N)}(t^{f-1}_{i(j)})\bigg\}\nonumber\\
=\sum_{i,j}\frac{1}{(2\pi)^{3N}}  \int d^3r_1d^3p_1 ... d^3r_N d^3p_N\nonumber\\
\times W_0(\vec{r}_i-\vec{r}_j,\vec{p}_i-\vec{p}_j)\bigg\{
W^{(N)}(t^{\nu}_{i(j)})\nonumber\\
+W^{(N)}(t^{\nu+1}_{i(j)})-W^{(N)}(t^{\nu}_{i(j)})+~...\nonumber\\
+W^{(N)}(t^f_{i(j)})-W^{(N)}(t^{f-1}_{i(j)})\bigg\}.
\label{example1}
\end{eqnarray}
The terms $W^{(N)}(t_0)+~ ...~ -W^{(N)}(t^{\nu-1}_{i(j)})$ are removed because the temperature before $t=t^{\nu-1}_{i(j)}$ is higher than the dissociation temperature of quarkonium. Therefore neither a bound state wave function nor its Wigner function 
exists.

Each line in Eq.~(\ref{example1}) represents the effect of a scattering.
If no scattering occurs at all, Eq. ~(\ref{example1}) becomes equivalent to Eq.~(\ref{begin}).
Even if a QGP is formed, it is concentrated in a small volume and exists for a short time only.
Therefore Eq.~(\ref{begin}) provides a suitable expression for charmonium production in p+p collisions.

Since typically only one charm-anticharm pair is produced in p+p collisions, the expression can be further simplified to 
\begin{eqnarray}
N(t_0)
= \frac{1}{(2\pi)^{6}}  \int d^3r_cd^3p_cd^3r_{\bar{c}} d^3p_{\bar{c}}\nonumber\\
\times W_0(\vec{r},\vec{p})
W^{(2)}(t_0),
\label{begin2}
\end{eqnarray}
where $\vec{r}=\vec{r}_c-\vec{r}_{\bar{c}}$ and $\vec{p}=(\vec{p}_c-\vec{p}_{\bar{c}})/2$.

Using the wavefunction of a 3-dimensional harmonic oscillator \com{instead of real wavefunction~\cite{Zhao:2025cnp,Song:2025zfy}},
the Wigner densities of a $S-$state and a $P-$state are given by
\begin{eqnarray}
W_0^{\rm S}({\bf r, p})&=&8\frac{D}{d_1 d_2}\exp\bigg[-\frac{r^2}{\sigma^2}-\sigma^2p^2\bigg],\label{wigner1}\\
W_0^{\rm P}({\bf r, p})&=&\frac{16}{3}\frac{D}{d_1 d_2}\bigg(\frac{r^2}{\sigma^2}-\frac{3}{2}+\sigma^2p^2\bigg)\nonumber\\
&&\times\exp\bigg[-\frac{r^2}{\sigma^2}-\sigma^2p^2\bigg].
\label{wigner2}
\end{eqnarray}
The only parameter $\sigma$ is related to the quarkonium Root Mean Square (RMS) through $\sigma^2=2/3\langle r^2\rangle$ for an $S-$state and $\sigma^2=2/5\langle r^2\rangle$ for a $P-$state, where $\sqrt{\langle r^2\rangle}$ denotes the root-mean-square (rms) distance.
The factors $D$, $d_1$, and $d_2$ represent the color-spin degeneracies of charmonium, the charm quark, and the charm antiquark, respectively.
The Wigner function for a $P-$state can locally become negative; following our previous study~\cite{Song:2017phm}, we set such values to zero.
The Wigner function for a $2S$ state is more complicated~\cite{Cho:2014xha,Kordell:2021prk,Zhao:2023dvk}, but for simplicity we use the form of Eq.~(\ref{wigner1}).

For the momenta of the charm and anticharm quarks we employ the PYTHIA event generator~\cite{Sjostrand:2006za,Song:2015sfa} (see section II.A).
However, PYTHIA does not provide information on the spatial separation of the charm-anticharm pair.
Therefore we assume that their separation follows a Gaussian distribution with the mean-square radius inversely proportional to the charm quark mass,
\begin{eqnarray}
W^{(2)}(\mathbf{r,p})
\sim r^2 \exp\bigg(-\frac{r^2}{2\delta^2}\bigg),
\label{separation}
\end{eqnarray}
where $\delta^2=\langle r^2\rangle/3=4/(3m_c^2)$ such that $\sqrt{\langle r^2\rangle}/2=1/m_c$.

Fig.~\ref{sigma-pp} shows the differential cross section for charmonium production, multiplied by the branching ratio to dimuon decay, as a function of rapidity and transverse momentum in p+p collisions from SPS to LHC energies, assuming radii ($\sqrt{\langle r^2\rangle}/2$) of $J/\psi$, $\chi_c$ and $\psi^\prime$ of 0.4, 0.42, and 0.75 fm, respectively ~\cite{Song:2017phm,Zhao:2023dvk}.
These values are somewhat larger than the radii obtaind from the Schr\"odinger equation solved with the heavy-quark potential in vacuum, which is typically modeled by the string potential~\cite{Satz:2005hx}. 
However, if additional particles are produced in p+p collisions besides the charm pair, the environment for charmonium formation may differ from that in vacuum. 

In Fig.~\ref{sigma-pp}, direct $J/\psi$ refers to the initial production of $J/\psi$, whereas prompt $J/\psi$ includes feed-down contributions from excited states such as $\chi_c$ and $\psi^\prime$.
The inclusive $J/\psi$ yield additionally contains non-prompt $J/\psi$ originating from the weak decay of beauty hadrons.
Since no experimental data are available in p+p collisions at SPS energies, we compare the PHSD results with p+Be collision data at $E_{kin}$=400 GeV from the NA50 Collaboration~\cite{NA50:2006rdp}, divided by the mass number of Beryllium. We note that the PHSD results are divided by two to imitate the experimental condition $-0.5<\cos\theta_{CS}<0.5$, where $\theta_{CS}$ is the polar angle in the Collins-Soper frame (the rest frame of the lepton pair).
We simply assume that this condition reduces the measured cross section by approximately a factor of two.
It is interesting to observe that the simple approach of Eq.~(\ref{begin2}) reproduces charmonium production in p+p collisions over a wide range of collision energies.

\subsection{Charmonium production in p+A collisions}

The Remler formalism for charmonium production is applied only in the partonic phase. 
Once the partons hadronize around $T_c$, the charmonia produced through the Remler formalism undergo hadronic interactions, which can, as shown in Sec.~\ref{hadronic-int}, be categorized into nuclear absorption and comover effects.
Nuclear absorption corresponds to the inelastic scattering of charmonium off nucleon (or baryon) and represents the dominant suppression mechanism for charmonium in p+A collisions. 
In principle, the cross section for nuclear absorption depends on the scattering energy~\cite{Oh:2001rm,Song:2005yd}. For simplicity, however, it is taken to be constant for each p+A collision energy and is extracted from experimental data~\cite{NA50:2006rdp,NA60:2010wey}.

\begin{figure}[h]
\centerline{
\includegraphics[width=9 cm]{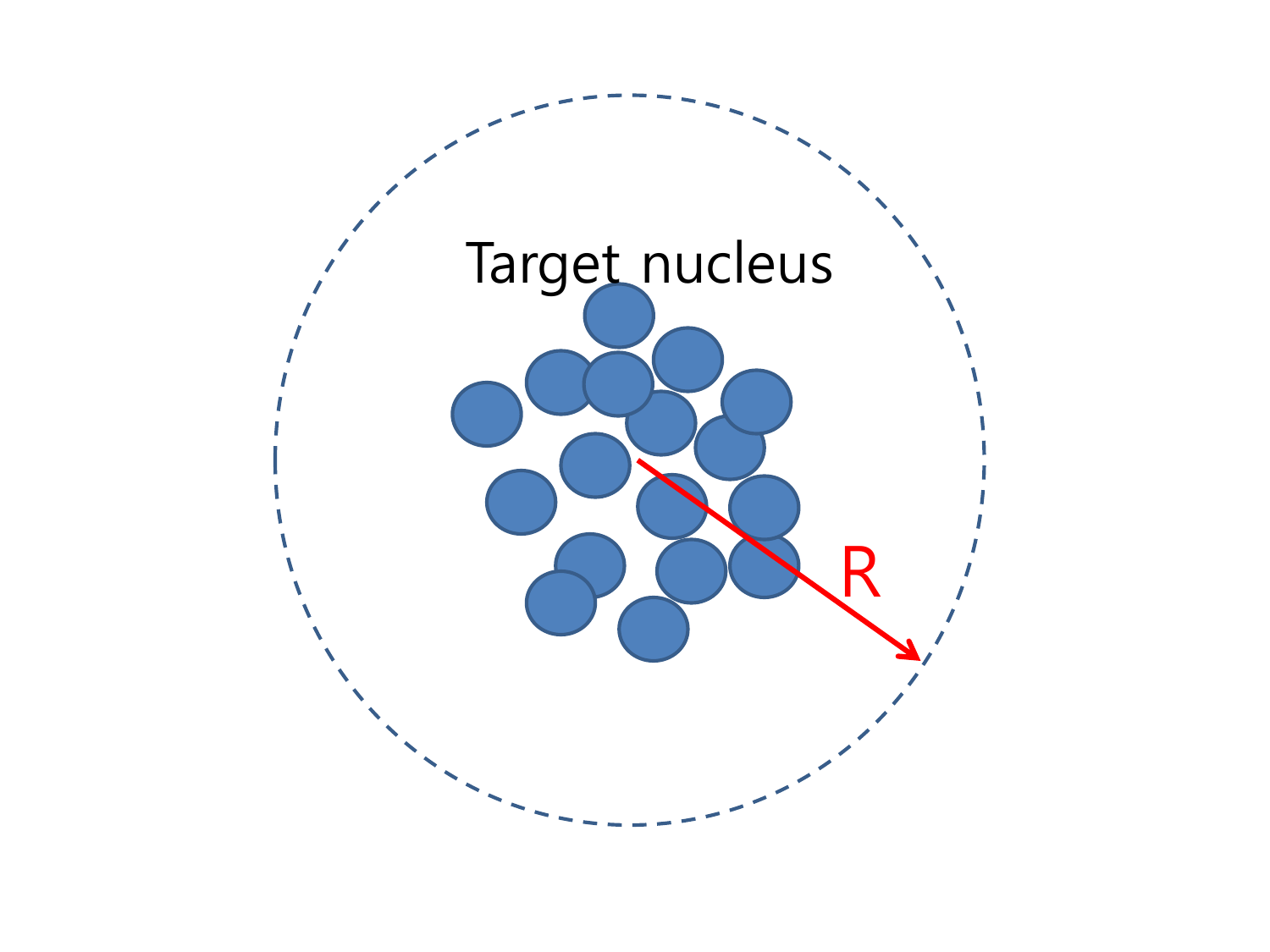}}
\caption{A schematic illustration of the simulation setup for p+A collisions}
\label{fig-sigma}
\end{figure}

Fig.~\ref{fig-sigma} schematically illustrates how charmonium production in p+A collisions is simulated within PHSD.
In the first step, the target nucleus is constructed according to the Glauber model using a Monte-Carlo method.
The circular area with radius $R$ is then uniformly bombarded with a proton beam.
The cross section for $J/\psi$ production is given by
\begin{eqnarray}
\sigma_{J/\psi}^{pA}=\pi R^2\frac{N_{J/\psi}}{N_p},
\end{eqnarray}
where $N_p$ is the number of incident proton projectiles and $N_{J/\psi}$ is the number of produced $J/\psi$ mesons.
We note that the cross section does not depend on $R$, because increasing $R$ also increases the number of proton projectiles that miss the target nucleus and therefore do not contribute to $J/\psi$ production.
In other words, there is a compensation between 
$\pi R^2$ and $N_{J/\psi}/N_p$.

\begin{figure}[h]
\centerline{
\includegraphics[width=9 cm]{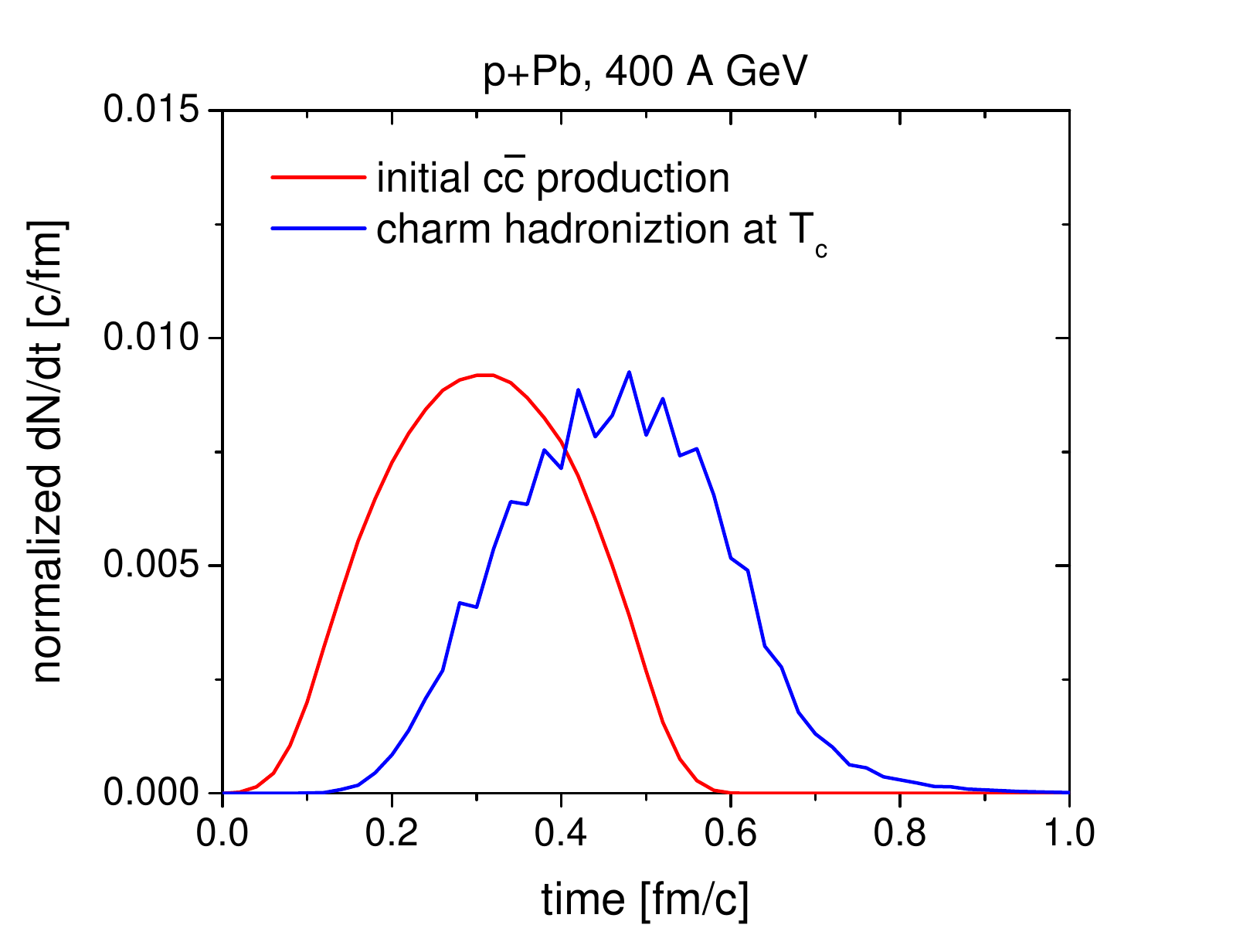}}
\caption{Distributions of the charm production time and charm hadronization time at $T_c$ in p+Pb collisions at $E_{kin}$=400 GeV.} 
\label{dndt-charm}
\end{figure}

Fig.~\ref{dndt-charm} shows the distribution of the charm production times and of the charm hadronization times both from PHSD calculations for p+Pb collisions at $E_{kin}$= 400 GeV.
The production times are estimated from the Glauber model and hadronization happens at the local energy density between 0.65 and 0.4 $\rm GeV/fm^3$~\cite{Song:2015sfa}.
As explained in the previous section for p+p collisions, the initial Wigner projection is carried out at the red line when a charm quark pair is produced.
If either the charm or anticharm quark scatters in the QGP during the evolution from the red line to the blue line, the Wigner projection is updated.
However, the time difference between the red and blue lines, which corresponds to the lifetime of the QGP in p+A collisions at SPS energies, is very short (on average about 0.15-0.17 fm/c).

Furthermore, partons in PHSD have a formation time given by $E/m_T^2$, where $m_T$ is the transverse mass, and they do not interact before this time.
As a result, the produced charm quarks hardly interact before reaching $T_c$ in p+A collisions at SPS energies, and the initial Wigner projection remains essentially unchanged until $T_c$. 
The produced charmonia subsequently  have nuclear absorption and comover interactions.

\begin{figure}[h]
\centerline{
\includegraphics[width=9 cm]{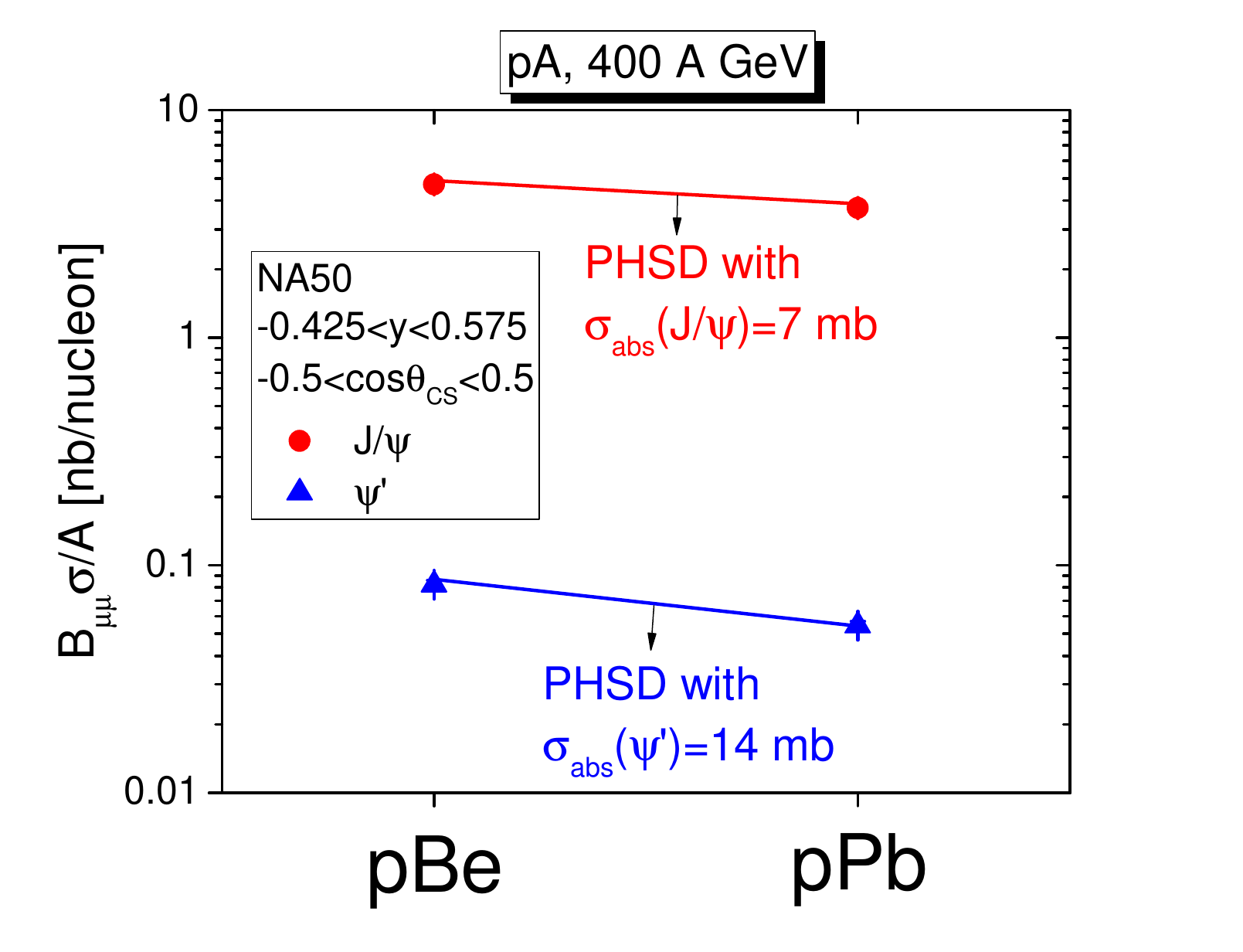}}
\caption{Production cross sections of $J/\psi$ and $\psi'$, multiplied by the branching ratio to dimuons and divided by the target mass number, in p+Be and p+Pb collisions at $E_{kin}$=400 GeV from the NA50 Collaboration~\cite{NA50:2006rdp}, compared with PHSD results.}
\label{pA400}
\end{figure}

Fig.~\ref{pA400} presents the PHSD  results for the production cross sections of $J/\psi$ and $\psi'$, multiplied by the branching ratio to dimuons and divided by the target mass number, in p+Be and p+Pb collisions at $E_{kin}$=400 A GeV. The results are compared with experimental data from the NA50 Collaboration~\cite{NA50:2006rdp}.
The dimuon branching ratios $B_{\mu\mu}$ are taken to be 6 \% for $J/\psi$ and 0.8 \% for $\psi^\prime$ according to the Particle Data Group~\cite{ParticleDataGroup:2024cfk}.
The angle $\theta_{CS}$ denotes the polar angle in the Collins-Soper frame, which corresponds to the rest frame of the lepton pair.
Since the mass number of Be target is 9, nuclear absorption has little impact on charmonium production, whereas much stronger effects are expected for a Pb target.
Indeed, the NA50 data show smaller production cross sections of $J/\psi$ and $\psi^\prime$ per nucleon in p+Pb collisions compared with p+Be collisions.
The figure demonstrates that the PHSD results are consistent with the NA50 data in p+Be collisions, as already indicated in Fig.~\ref{sigma-pp}.
Furthermore, nuclear absorption cross sections of 7 mb for direct $J/\psi$ and 14 mb for direct $\psi^\prime$ reproduce the measured charmonium production cross sections in p+Pb collisions~\cite{NA50:2006rdp}.
Since no experimental data are available for the nuclear absorption of $\chi_c$, we assume that its absorption cross section is the same as that of $J/\psi$. 
\com{We also note that charmonium absorption by mesonic comovers is included in the PHSD calculations, but its effect is rather small due to the relatively low meson density in p+A reactions.}
\begin{figure}[h]
\centerline{
\includegraphics[width=9 cm]{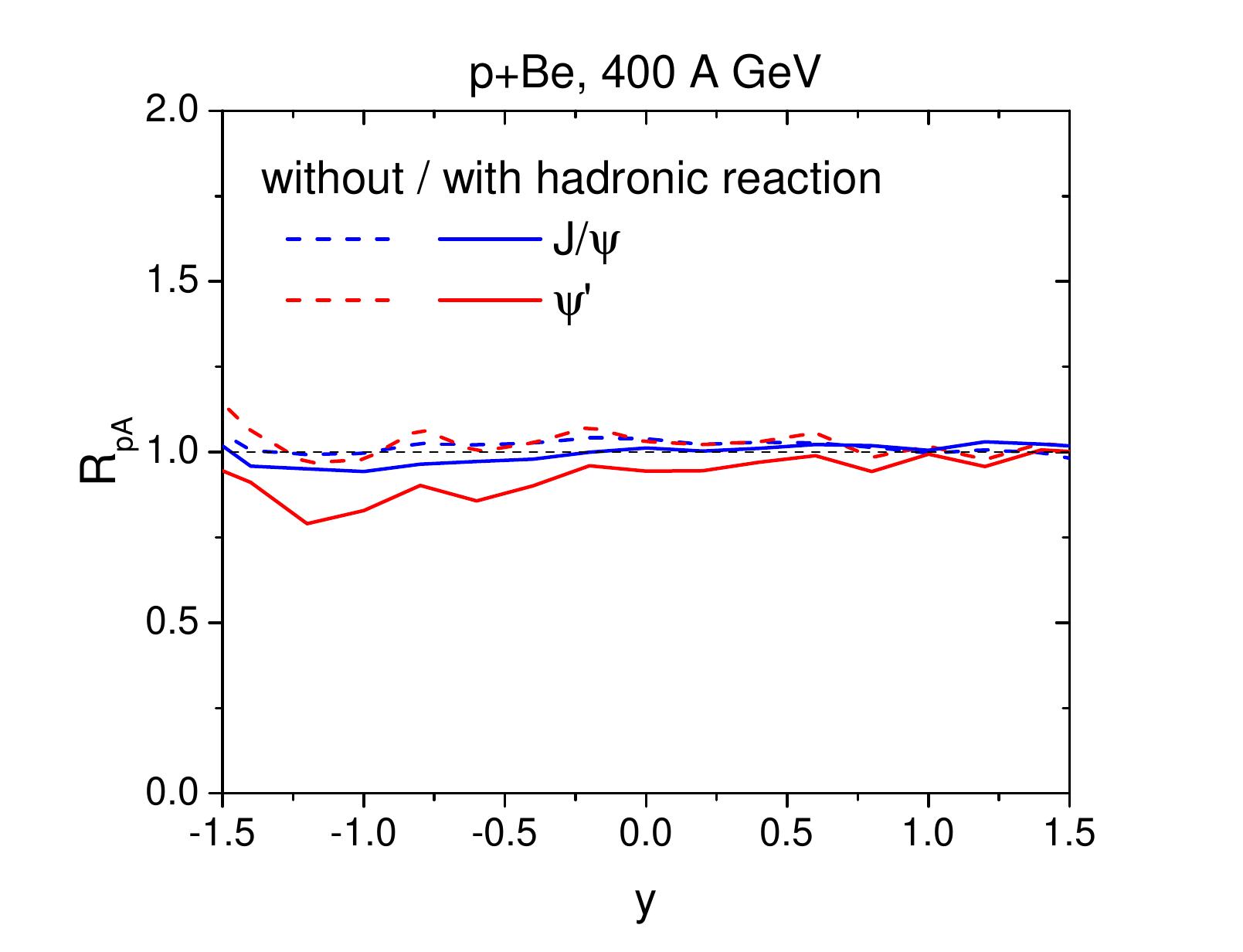}}
\centerline{
\includegraphics[width=9 cm]{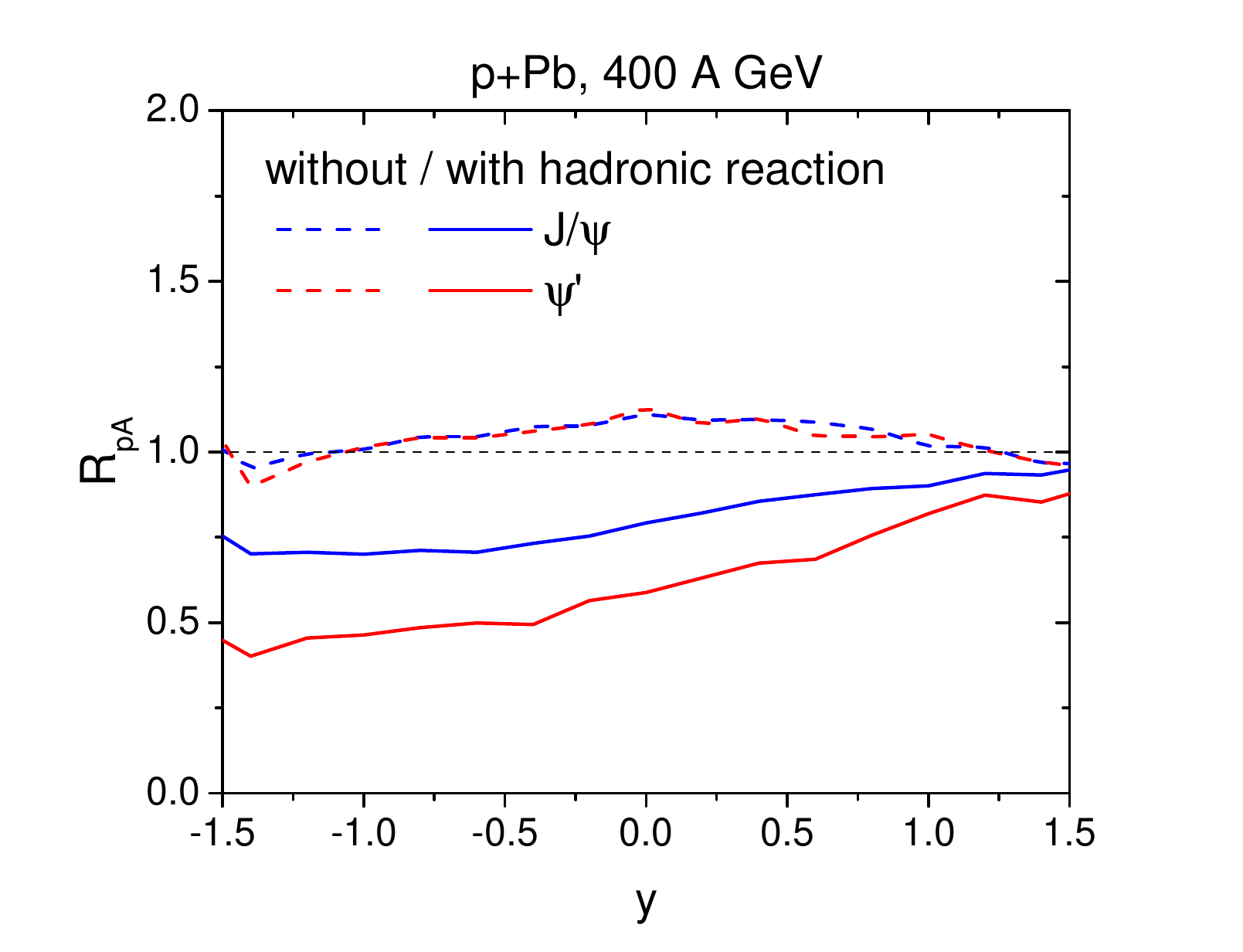}}
\caption{Nuclear modification factor $R_{pA}$ of (blue) $J/\psi$ and (red) $\psi'$ in p+Be and p+Pb collisions at $E_{kin}$=400 GeV, calculated (dashed lines) without and (solid lines) with nuclear reactions after production.}
\label{RpA400}
\end{figure}

Fig.~\ref{RpA400} shows the nuclear modification factors $R_{pA}$ of $J/\psi$ and $\psi^\prime$ in p+Be and p+Pb collisions.
Positive rapidity corresponds to the direction of the proton projectile, while negative corresponds to the target nucleus.
The dashed lines represent $R_{pA}$ without hadronic reactions such as nuclear absorption (charmonium-nucleon interactions) or 
charmonium-meson interactions.
In other words, the dashed lines correspond to the results obtained purely from the Remler formalism at $T_c$.
Since the number of nucleons is much larger than the number of produced mesons in p+A collisions at SPS energies, the dominant suppression of charmonium originates from nuclear absorption.
One can see that $R_{pA}$ without nuclear reactions in p+Be collisions is close to unity, while in p+Pb collisions it is slightly above unity due to antishadowing effects, which enhance charmonium production at SPS energies~\cite{Arnaldi:2009it,Lourenco:2008sk, Cassing:1997kw}. 
However, nuclear reactions significantly suppress charmonium production, especially in p+Pb collisions.
Because the target nucleus moves in the negative direction in the center-of-mass frame, a stronger suppression is observed in the negative rapidity region.
Even in p+Be collisions a slight suppression is visible due to nuclear reactions.

The nuclear absorption cross section should in principle depend on collision energy~\cite{Oh:2001rm,Song:2005yd}, as a scattering cross section of $J/\psi+N$ changes with scattering energy~\cite{Lin:1999ad,Wong:1999zb,Haglin:2000ar,Oh:2000qr}.
By simply assuming that the target nucleus moves opposite to the beam direction in the center-of-mass frame, one can estimate the scattering energy of the produced charmonium and nucleon in p+A collisions~\cite{Lourenco:2008sk}. 
Since the scattering energy depends on charmonium rapidity and transverse momentum, the $R_{pA}$ in Fig.~\ref{RpA400} will be modified if a more realistic absorption cross section is adopted.

\begin{figure}[h]
\centerline{
\includegraphics[width=9 cm]{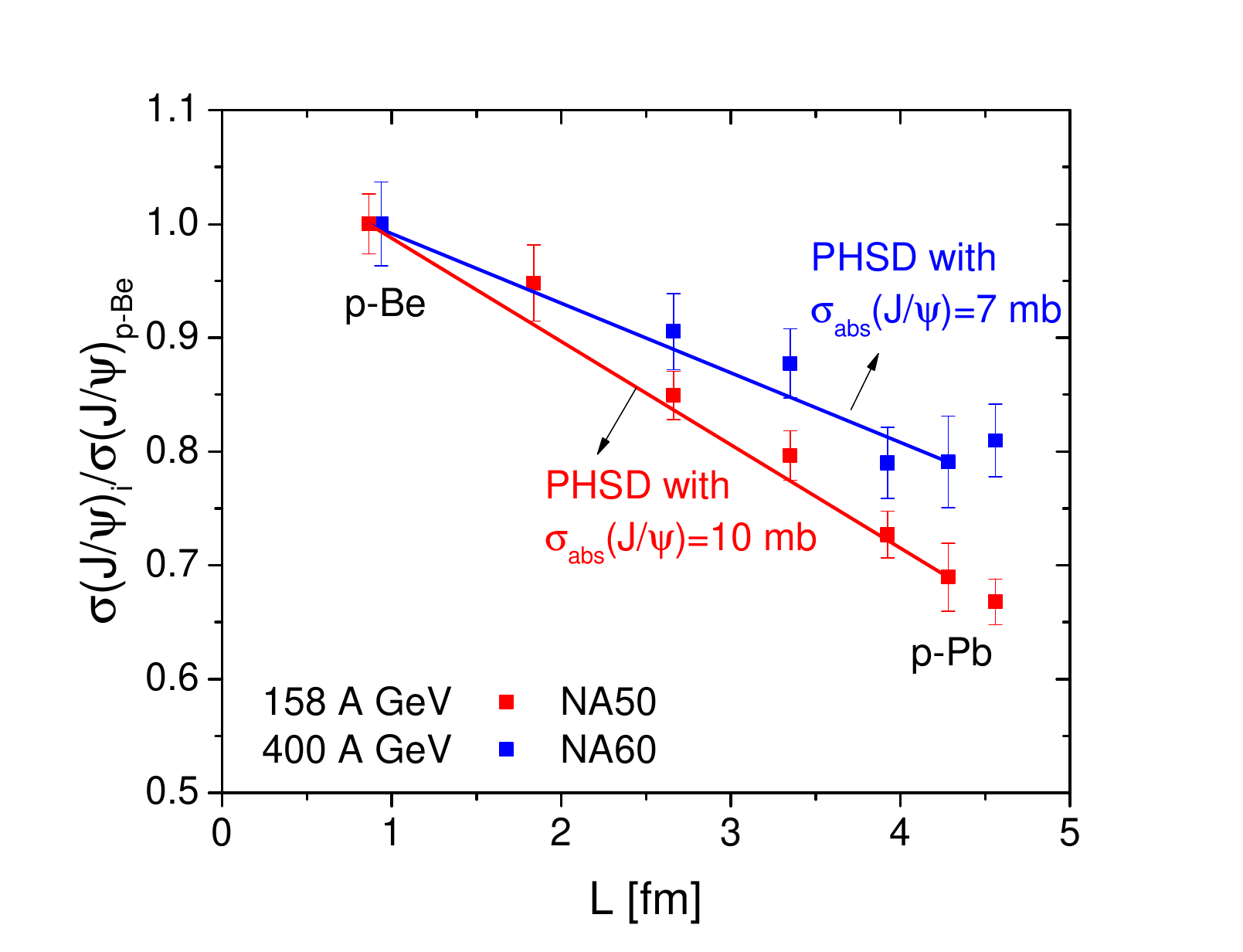}}
\caption{Ratio of $J/\psi$ production cross sections in p+A collisions to those in p+Be collisions as a function of path length at $E_{kin}$=400 and 158 GeV. Experimental data from the NA50 and NA60 Collaborations~\cite{NA60:2010wey} are compared with PHSD results, which are calculated only for p+Be and p+Pb collisions.
}
\label{pA}
\end{figure}

Fig.~\ref{pA} shows the ratio of the $J/\psi$ production cross section per nucleon in p+A collisions at $E_{kin}$= 400 and 158 GeV to that in p+Be collisions, plotted as a function of the average path length of nuclear matter traversed by the produced $J/\psi$ in p+A collisions.
In our calculations, the ratio is evaluated only for p+Pb collisions ($L\approx 4.3$), and straight lines are drawn to qualitatively compare the results with experimental data from the NA50 and NA60 Collaborations~\cite{NA60:2010wey}.
We find that an absorption cross section of 7~mb for $J/\psi$ is required at $E_{kin}$= 400 GeV, while a value of 10 mb is needed at $E_{kin}$= 158 GeV. This indicates that a larger absorption cross section is required at lower collision energies in order to reproduce the experimental data.
For $\psi^\prime$, we assume that the absorption cross section at $E_{kin}$= 158 GeV is \com{20 mb,} twice that of the $J/\psi$, based on the results shown in Fig.~\ref{pA400}.

Since charmonium production in p+A collisions is successfully described within PHSD using the Remler formalism and the absorption cross sections have been extracted from experimental data, we are now ready to extend the study to heavy-ion collisions.

\subsection{Charmonium production in heavy-ion collisions}\label{HIC}

The potential between the heavy quark and heavy anti-quark may depend on the temperature. Therefore the wave function of the relative motion as well as its Wigner density  may vary with temperature.
However, the heavy quark potential in the QGP has not yet been fully established. Since the temperature of the QGP at SPS energies is relatively moderate, we assume that the potential does not differ significantly from that in p+p collisions.
Furthermore, recent studies of the heavy quark potential suggest that the real part of the potential is rather insensitive to the temperature~\cite{Bazavov:2023dci}.
Therefore, as a first approximation, we use the same radii of charmonia as in p+p and p+A collisions.

In heavy-ion collisions, the initial Wigner projection is carried out according to Eq.~(\ref{begin2}) as in p+p and p+A collisions.
Subsequently, the Wigner projection is updated whenever a charm or anticharm quark undergoes a scattering in the QGP.
Within PHSD, (anti)charm quarks scatter in the QGP  with partons described by  the Dynamical Quasi-Particle Model (DQPM). This approach reproduces the spatial diffusion coefficient from lattice QCD calculations~\cite{Banerjee:2011ra,Berrehrah:2014kba,Song:2019cqz,Grishmanovskii:2025mnc}. 

However, if a charm-anticharm pair forms a bound state, its scattering cross section should be smaller than twice the charm-quark scattering cross section, because the pair is in a color-singlet (color-transparent) configuration.
Indeed, it has been shown that if the bottomonium cross section is about 10 \% of twice bottom-quark cross section, the nuclear modification factor $R_{AA}$ of bottomonium in Pb+Pb collisions at LHC energies can be reproduced~\cite{Song:2023zma}.
In the present study we adopt the same approach for charmonium.
Since  charmonium is less tightly bound than bottomonium, the ratio of the charmonium interaction rate to that of two open-charm quarks in the QGP is expected to be larger than that for bottomonium.
In the PHSD simulations, whether the Wigner projection is updated after a scattering of a (anti)charm quark is determined by a parameter defined as the ratio of the charmonium scattering cross section to twice charm-quark scattering cross section in the QGP.

\begin{figure}[h]
\centerline{
\includegraphics[width=9 cm]{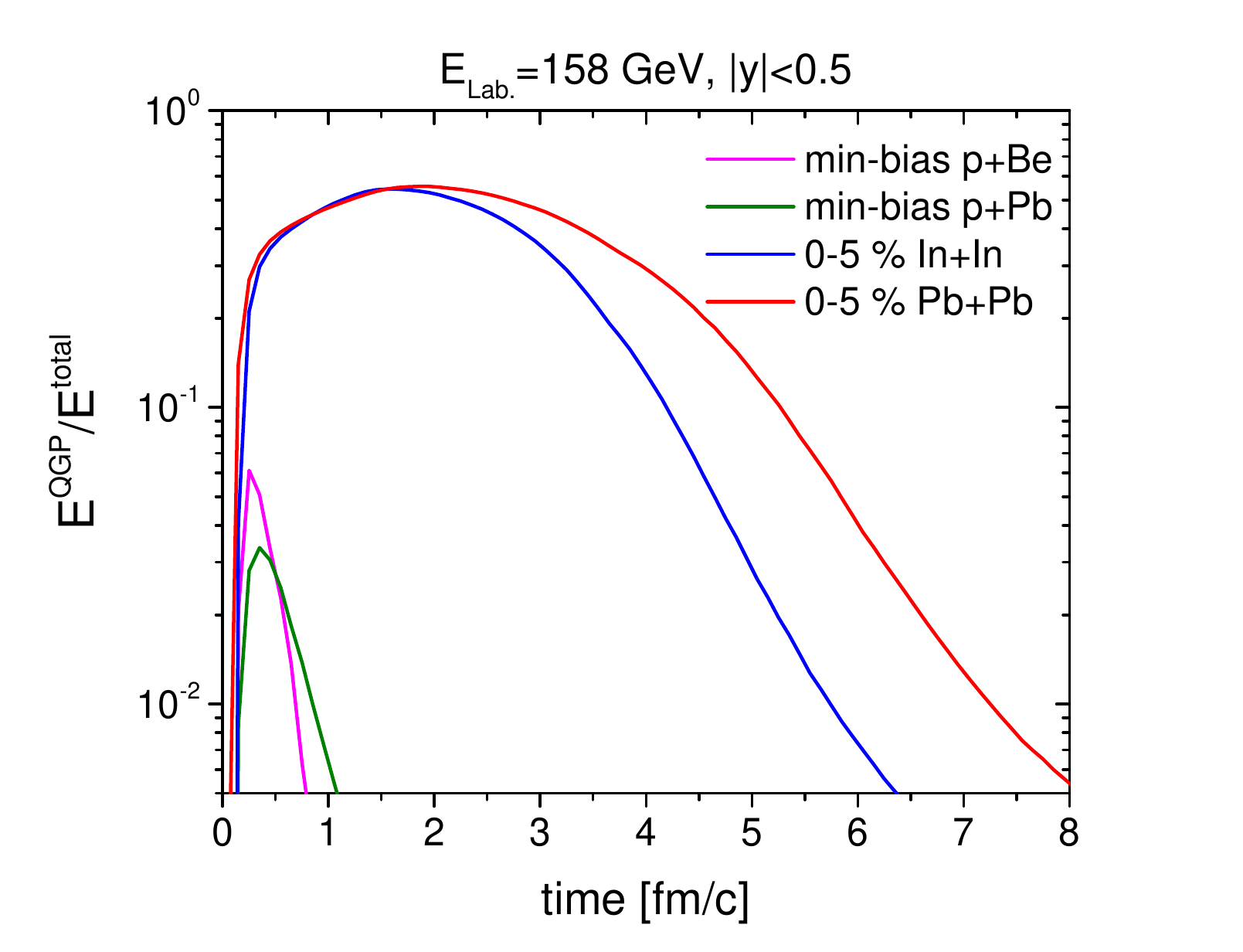}}
\caption{Fraction of QGP (parton) energy relative to the total (parton and hadron) energy at midrapidity as a function of time in minimum-bias p+Be, p+Pb and 0-10 \% central In+In and Pb+Pb collisions at $E_{kin}=$ 158 AGeV.}
\label{eden}
\end{figure}

The Remler formalism is limited to the QGP phase. In Fig.~\ref{eden} we show the fraction of QGP energy relative to the total energy at mid-rapidity as a function of time for p+A and heavy-ion collisions at $E_{kin}$=158 GeV.
One can see that the QGP fraction in p+A collisions remains below 10 \% and rapidly disappears with time.
In practice, the QGP in p+A collisions corresponds to a small droplet of partons.
The maximum QGP fraction in p+Pb collisions is slightly smaller but decreases slower than in p+Be collisions.
Because the lifetime of this small QGP system in  p+A collisions is extremely short, there is almost no opportunity for the Wigner projection to be updated through (anti)charm scattering.
In contrast, the QGP volume and lifetime in heavy-ion collisions are significantly larger.
The maximum QGP fraction reaches about 55 \%, and the QGP phase persists for approximately 4.2 and 5.3 fm/c in central In+In and Pb+Pb collisions, respectively, until the QGP energy fraction decreases to 10 \% .

\begin{figure}[h]
\centerline{
\includegraphics[width=9 cm]{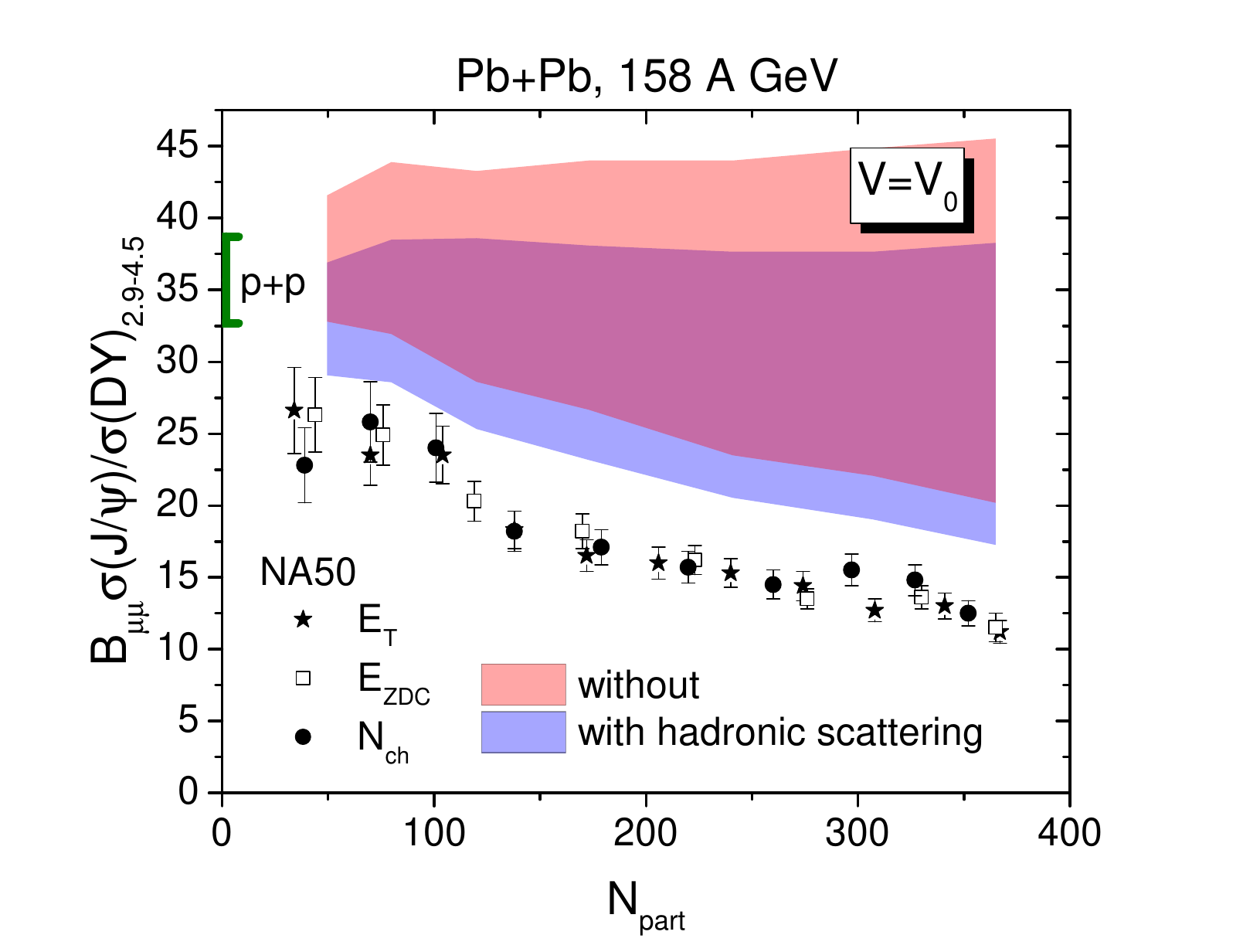}}
\centerline{
\includegraphics[width=9 cm]{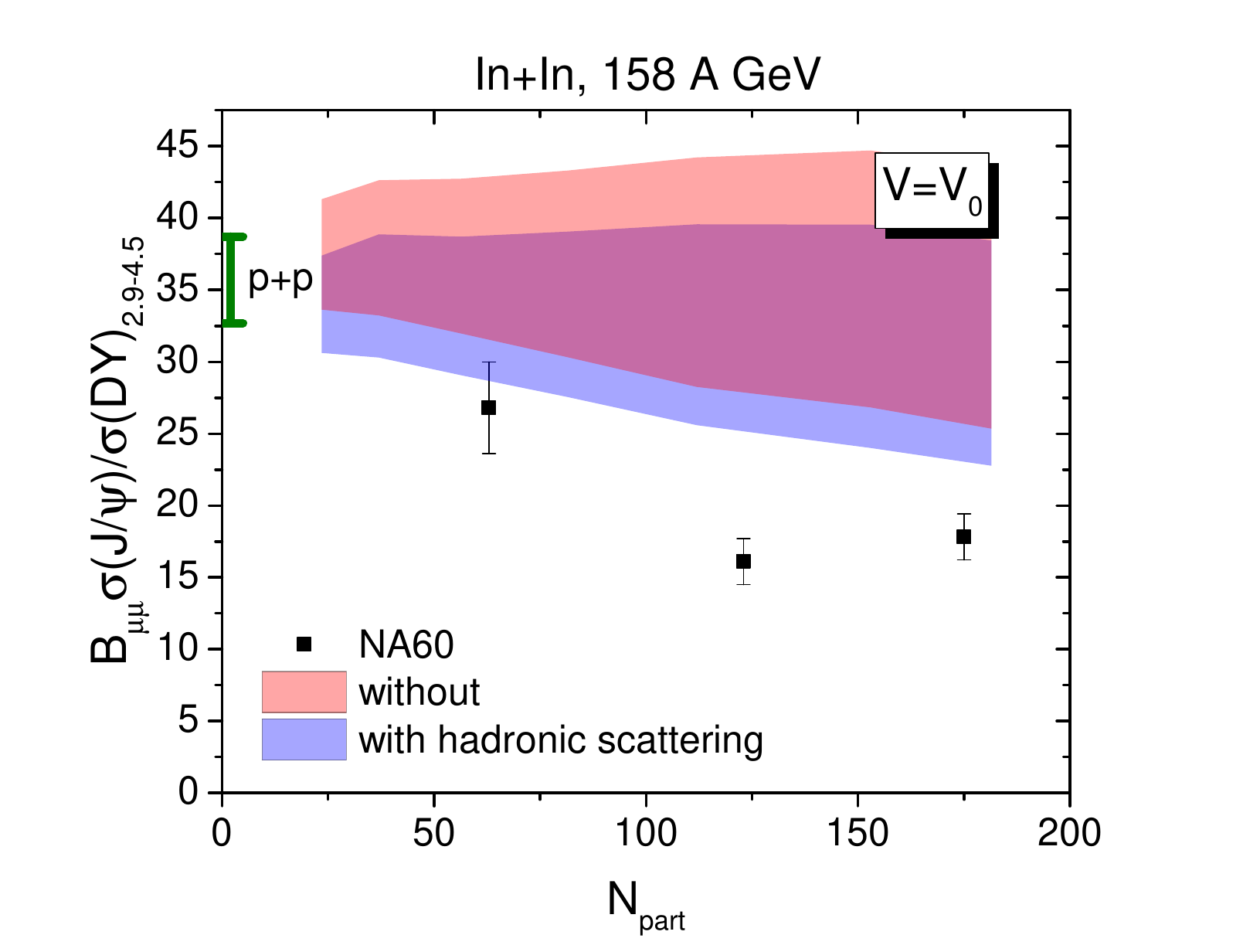}}
\caption{Ratio of dimuons from $J/\psi$ to those from the Drell-Yan process in the invariant mass range 2.9-4.5 GeV as a function of the number of participant nucleons in (upper) Pb+Pb and (lower) In+In collisions at E/A=158 GeV. Experimental data from the NA50 and NA60 Collaborations~\cite{NA50:2004sgj,NA60:2006ncq} are compared with PHSD results with (blue bands) and without (red bands) hadronic interactions for the temperature-independent heavy-quark potential.}
\label{npart}
\end{figure}

Fig.~\ref{npart} shows the ratio of dimuons from $J/\psi$ decay to dimuons from the Drell-Yan process with an invariant mass between 2.9 and 4.5 GeV as a function of the number of participant nucleons in Pb+Pb and In+In collisions at E/A= 158 GeV.
In p+p collisions ($N_{part}=$2), this ratio is measured to be 35.7$\pm$3.0~\cite{NA60:2006ncq}. 
$V_0$ in Fig.~\ref{npart} denotes that in heavy-ion collisions the same radii of charmonia as in p+p and p+A collisions are used for Wigner projection which does not depend on temperature. 

The red bands represent $J/\psi$ production at $T_c$, excluding hadonic interactions.
The upper limit of the bands corresponds to the case where charmonium interactions in the QGP are neglected, whereas the lower limit assumes that the charmonium interaction rate equals twice the charm-quark interaction rate in the QGP.
In other words, the upper limit originates solely from the initial Wigner projection, as in p+p collisions. Due to antishadowing effects, the resulting $R_{AA}$ is slightly above the ratio observed in p+p collisions.
The error bands also account for the uncertainty in the ratio of dimuons from $J/\psi$ to those from the Drell-Yan process in p+p collisions.
The upper limit of the colored bands assumes a ratio of 38.7 (=35.7+3.0), while the lower limit assumes a ratio of 32.7 (=35.7-3.0).
One can see that the upper and lower limits of the red bands coincide with the upper and lower error bars of the p+p data at small $N_{part}$ in Fig.~\ref{npart}.
As discussed in Ref.~~\cite{Song:2023zma}, charm quark scattering in the QGP tends to reduce $J/\psi$ production because the charm and anticharm quarks become increasingly separated in space over time, which reduces the Wigner projection probability.

The blue bands show the results including hadronic interactions after $T_c$, which include the nuclear absorption, comover interactions (e.g., $J/\psi+meson\rightarrow D+\bar{D}$) and their reverse reactions (e.g., $D+\bar{D} \rightarrow J/\psi+meson$) as discussed in Sec.~\ref{hadronic-int}. 
One can see that the effect of hadronic interactions on $J/\psi$ suppression is relatively weak compared with that in p+A collisions, partly because the presence of a partonic phase delays the onset of hadronic interactions.
However, the suppression resulting from both partonic and hadronic interactions is still insufficient to reproduce the experimental data reported by the NA50 and NA60 Collaborations.

\begin{figure}[h]
\centerline{
\includegraphics[width=9 cm]{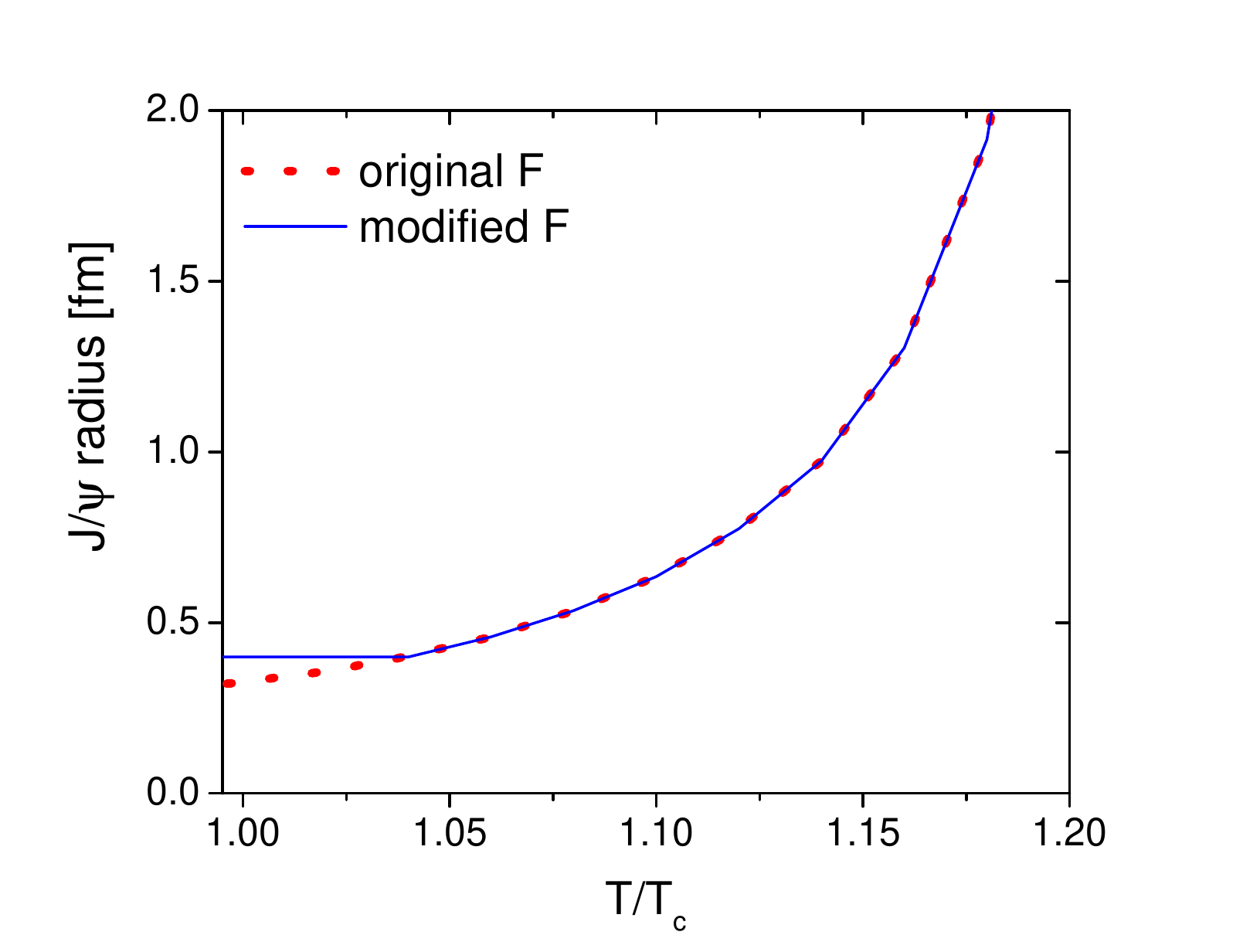}}
\caption{Radius of $J/\psi$ as a function of temperature from the free energy of a heavy-quark pair in the QGP (blue solid line)~\cite{Kaczmarek:2003ph,Burnier:2014ssa,Gubler:2020hft} and its modification near $T_c$ (red dotted line).}
\label{radii}
\end{figure}

We therefore introduce thermal effects on the heavy-quark potential.
The red dotted line in Fig.~\ref{radii} represents the $J/\psi$ radius obtained by solving the Schr\"odinger equation with the free-energy heavy-quark potential~\cite{Kaczmarek:2003ph,Burnier:2014ssa,Gubler:2020hft} as a function of temperature.
In this approach, the $J/\psi$ survives up to approximately 1.2 $T_c$ in the QGP.
However, near the dissociation temperature, the $J/\psi$ radius becomes extremely large, implying that the Wigner projection at this moment is strongly suppressed, because a change of $\sqrt{\langle r^2 \rangle}$ implies a change of $\sigma$ and hence a modification of the $\exp[-\sigma^2 p^2]$ term in Eqs.~(\ref{wigner1}) and (\ref{wigner2}).
Therefore, we wait until the Wigner projection reaches its maximum value as temperature decreases and the charmonium radius correspondingly shrinks.
This prescription was also adopted in our previous study of bottomonum formation in Pb+Pb collisions at LHC energies~\cite{Song:2023zma}.
We note that the actual formation temperature of $J/\psi$, at which the initial Wigner projection is maximized at SPS energies, is around 1.15 $T_c$.

In addition, the $J/\psi$ radius obtaind from the free energy potential below 1.04 $T_c$ becomes smaller than 0.4 fm, which is the $J/\psi$ radius in p+p collision, as shown in Fig.~\ref{radii}.
We therefore slightly modify the radius so that it saturates at 0.4 fm as the temperature approaches $T_c$.
This adjustment helps the $R_{AA}$ of $J/\psi$ in heavy-ion collisions converges to 1.0 at small numbers of participants, assuming the absence of hadronic interactions.
For the excited states $\chi_c$ and $\psi^\prime$, the dissociation temperatures obtained from the free energy heavy quark potential are slightly below $T_c$.
Nevertheless, for the same reason discussed above, we assume that they dissolve at $T_c$ and adopt radii of 0.42 and 0.75 fm, respectively, as in p+p collisions.

\begin{figure}[h]
\centerline{
\includegraphics[width=9 cm]{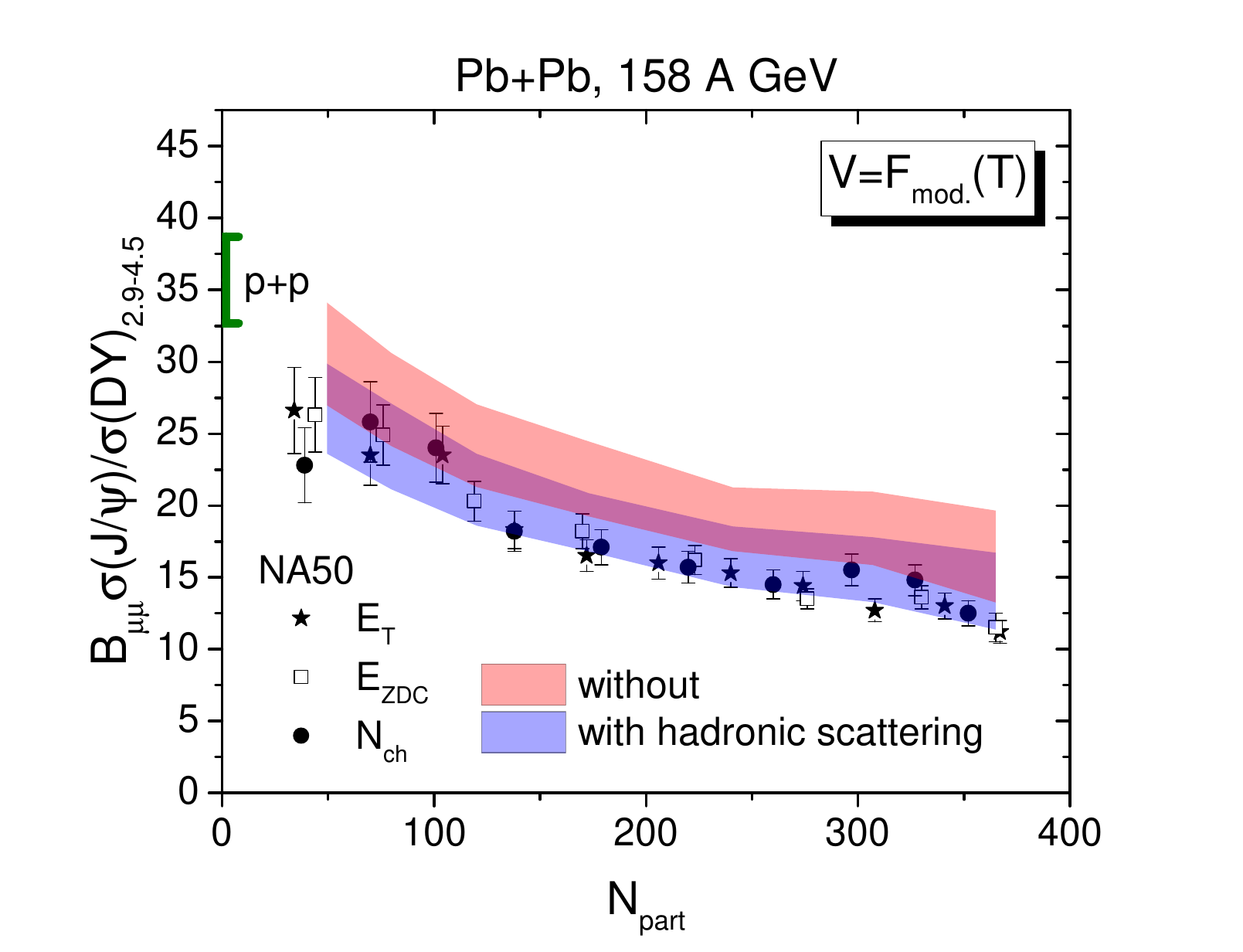}}
\centerline{
\includegraphics[width=9 cm]{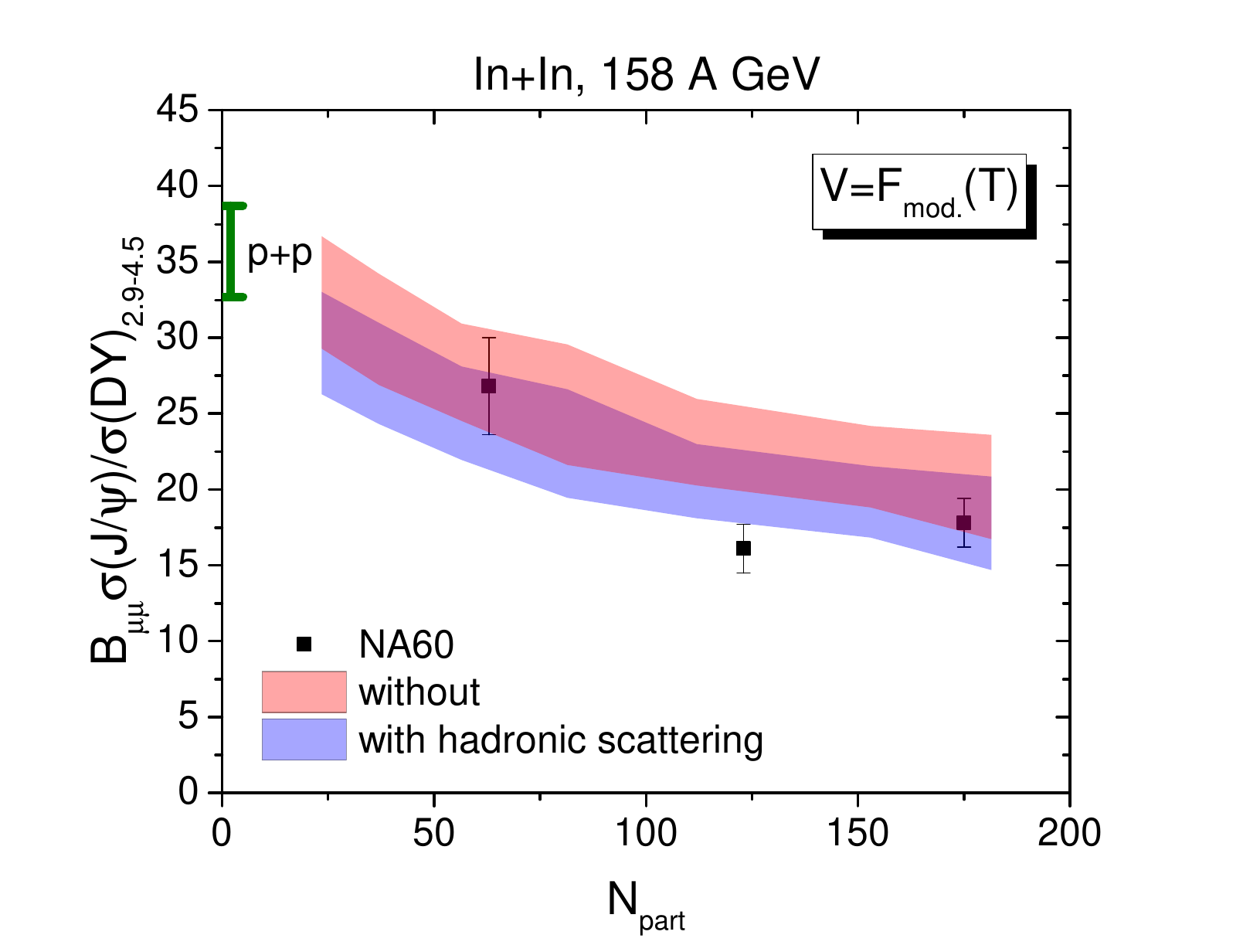}}
\caption{Same as Fig.~\ref{npart}, but for the temperature-dependent heavy-quark potential derived from the free energy of a heavy quark pair in the QGP.}
\label{npartF}
\end{figure}

The results obtained with the temperature-dependent heavy-quark potential are presented in Fig.~\ref{npartF}. 
The PHSD calculations reproduce the measured $J/\psi$ production in both Pb+Pb and In+In collisions at E/A=158 GeV.
The main reason for the additional suppression relative to Fig.~\ref{npart} is that the initial projection occurs later, when the spatial separation between the charm and antiquark quarks is larger.
Because the Wigner projection is updated only within the narrow temperature interval between 1.2 $T_c$ (effectively 1.15 $T_c$) and 1.0 $T_c$, charm-quark scattering in the QGP has little effect on the final $J/\psi$ yield. 
This contrasts with the wide bands in Fig.~\ref{npart}, where the Wigner projection is updated continously from the beginning of the evolution down to 1.0 $T_c$.

Although the bands in Fig.~\ref{npartF} are narrow, the experimental data, especially those from the NA50 Collaboration, appears to favor a large scattering cross section of $J/\psi$ in the QGP when comparing the upper limit ($\sigma(J/\psi)\approx 0$) and the lower limit ($\sigma(J/\psi)\approx 2\sigma(c)$) of the blue band, where $\sigma(c)$ denotes the charm scattering cross section either with light (anti)quark or gluon.
\com{Using the lattice-QCD free energy as the heavy-quark potential corresponds
to a relatively weak-binding scenario, which is also supported by QCD
sum-rule studies~\cite{Lee:2013dca}. In the quasi-free limit, the $c$ and
$\bar c$ can be approximately resolved independently, motivating
$\sigma_{J/\psi}\simeq 2\sigma_c$. We therefore regard this value as an
upper-limit estimate for the $J/\psi$ dissociation cross section.}

The blue band in Fig.~\ref{npartF} shows the PHSD results including hadronic interactions of hidden-charm mesons with comoving hadrons. These interactions include charmonium absorption by baryons, with a cross section of 10 mb at a beam energy of 158 AGeV, as well as charmonium interactions with mesons, as described in Sec.~II.D.

We emphasize that hadronic interactions are particularly important at SPS energies because of the large hadronic corona surrounding the QGP fireball, cf. Fig.~\ref{eden}. As shown in Fig.~\ref{npartF}, the inclusion of these interactions leads to additional charmonium suppression and is essential for reproducing the experimental data from the NA50 and NA60 Collaborations~\cite{NA50:2004sgj,NA60:2006ncq}.

\begin{figure}[h]
\centerline{
\includegraphics[width=9 cm]{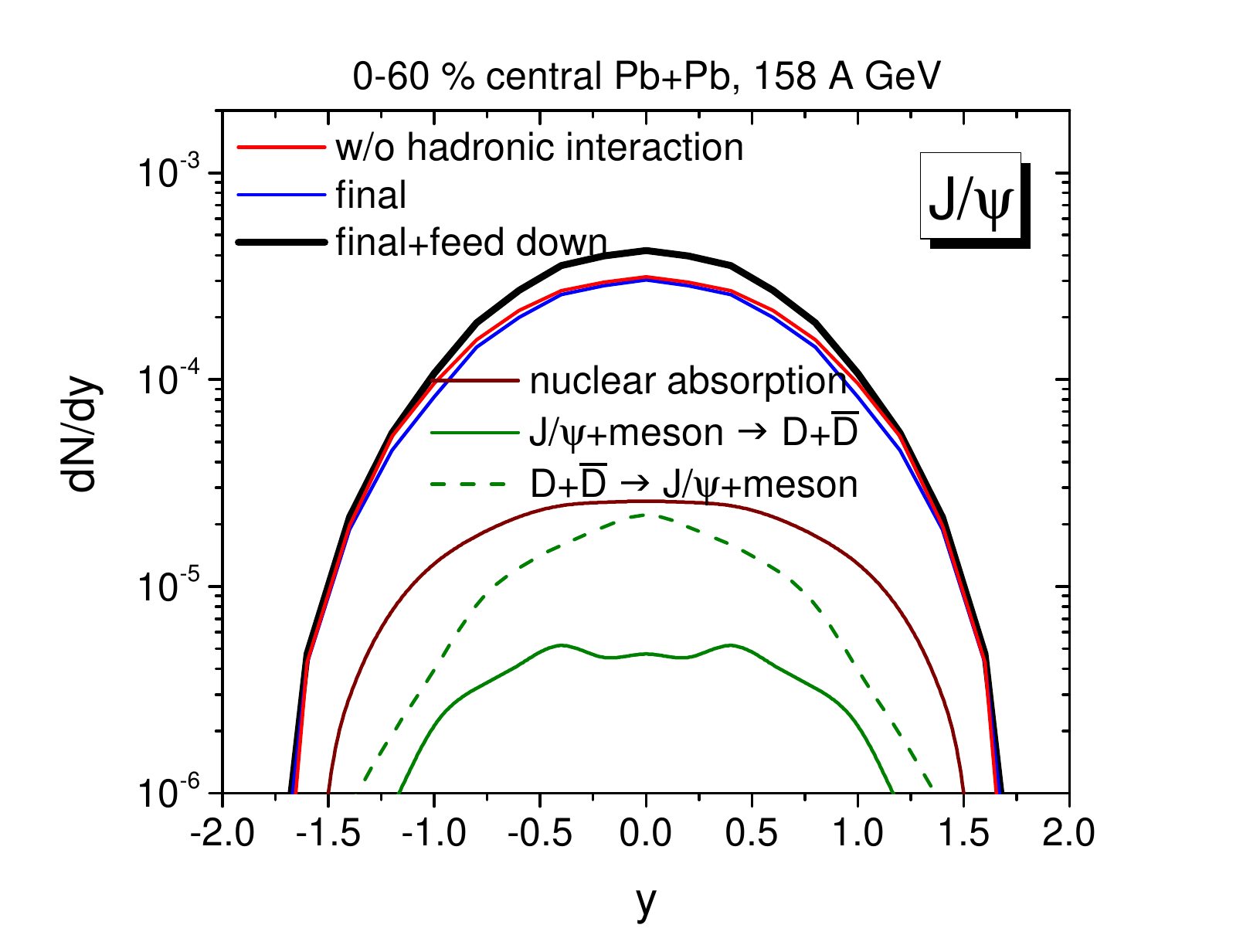}}
\centerline{
\includegraphics[width=9 cm]{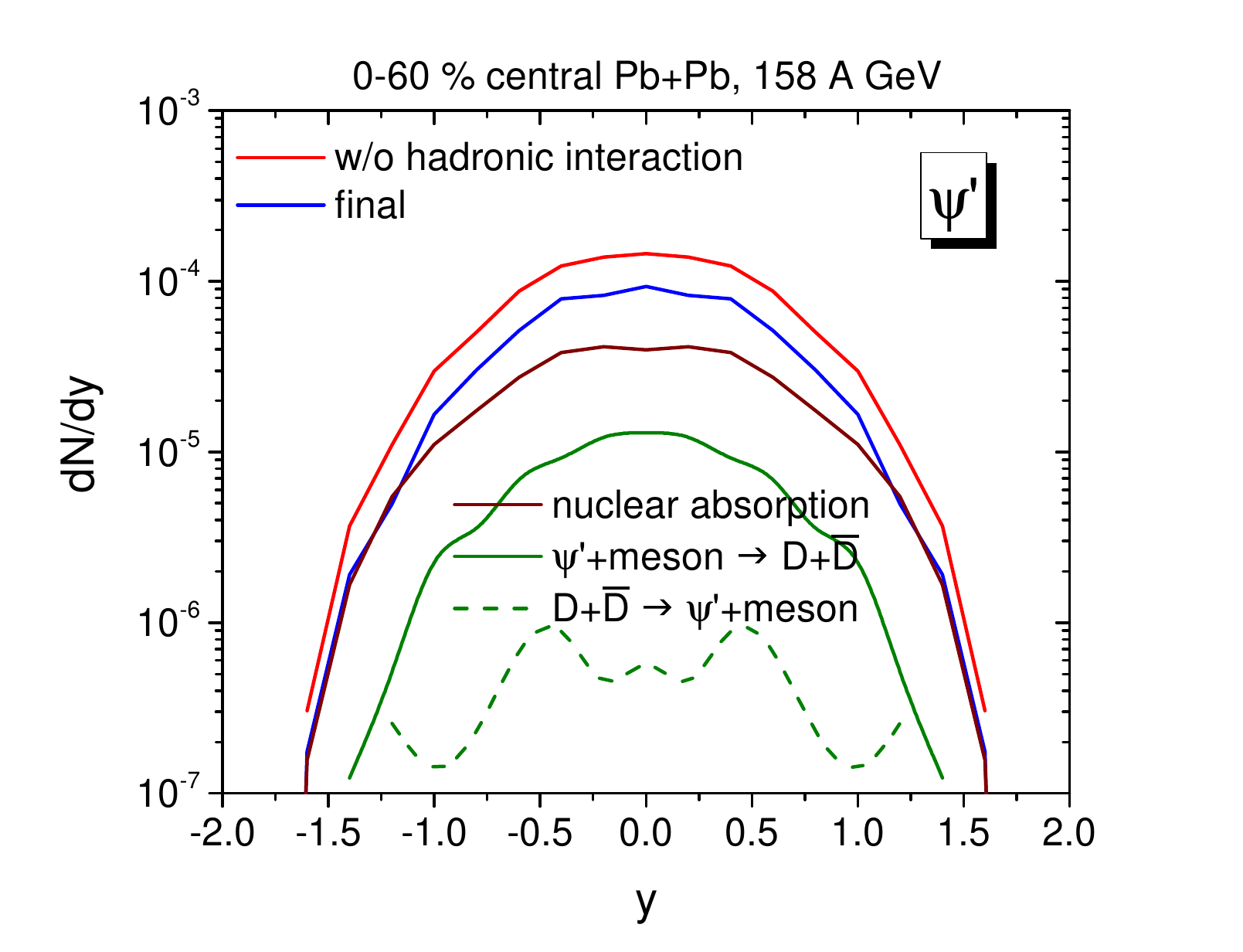}}
\caption{Rapidity distributions of (upper) $J/\psi$ and (lower) $\psi'$ in 0-60 \% central Pb+Pb collisions at E/A=158 GeV without and with nuclear reactions after production.}
\label{dndy}
\end{figure}

For a better understanding of the results shown in Fig.~\ref{npartF}, we present in Fig.~\ref{dndy} the channel decomposition of the rapidity distributions of $J/\psi$ and $\psi^\prime$ for 0--60\% central Pb+Pb collisions at $E/A=158$ GeV. The red lines correspond to charmonia without final-state hadronic interactions, i.e., to their production at $T_c$ solely within the Remler formalism, whereas the blue lines include subsequent interactions with hadrons.

A comparison of the PHSD calculations with and without hadronic interactions demonstrates that these interactions affect the different charmonium states in very different ways. The directly produced $J/\psi$ component is only moderately modified during the hadronic stage, whereas the $\psi^\prime$ yield is strongly reduced. This difference can be traced back to two main mechanisms. First, the $\psi^\prime$ has a larger nuclear absorption cross section than the $J/\psi$. Moreover, $J/\psi$ dissociation in baryon-induced reactions requires the scattering energy to exceed the thresholds for the $\Lambda_c+N$ or $2D+N$ channels. By contrast, owing to its larger mass and smaller binding energy, the $\psi^\prime$ can be dissociated much more easily in collisions with nucleons. Second, regeneration from open-charm mesons is far more efficient for the $J/\psi$ than for the $\psi^\prime$: the reaction $D+\bar{D}\rightarrow J/\psi+\mathrm{meson}$ is exothermic, while $D+\bar{D}\rightarrow \psi^\prime+\mathrm{meson}$ is endothermic.

The black line in the upper panel denotes the final $J/\psi$ rapidity distribution including feed-down contributions from $~\chi_c^0,~\chi_c^1,~\chi_c^2$ and $\psi^\prime$ decays. Consequently, the additional $J/\psi$ suppression due to hadronic interactions observed in Fig.~\ref{npartF} originates predominantly from the suppression of the excited charmonium states, $~\chi_c^0,~\chi_c^1,~\chi_c^2$ and $\psi^\prime$, rather than from direct $J/\psi$ absorption.
In the calculations we employ an elastic scattering cross section of 5 mb for charmonium interactions with light mesons. This elastic rescattering causes a modest broadening of the $J/\psi$ rapidity distribution toward forward and backward rapidities.

\begin{figure}[h]
\centerline{
\includegraphics[width=9 cm]{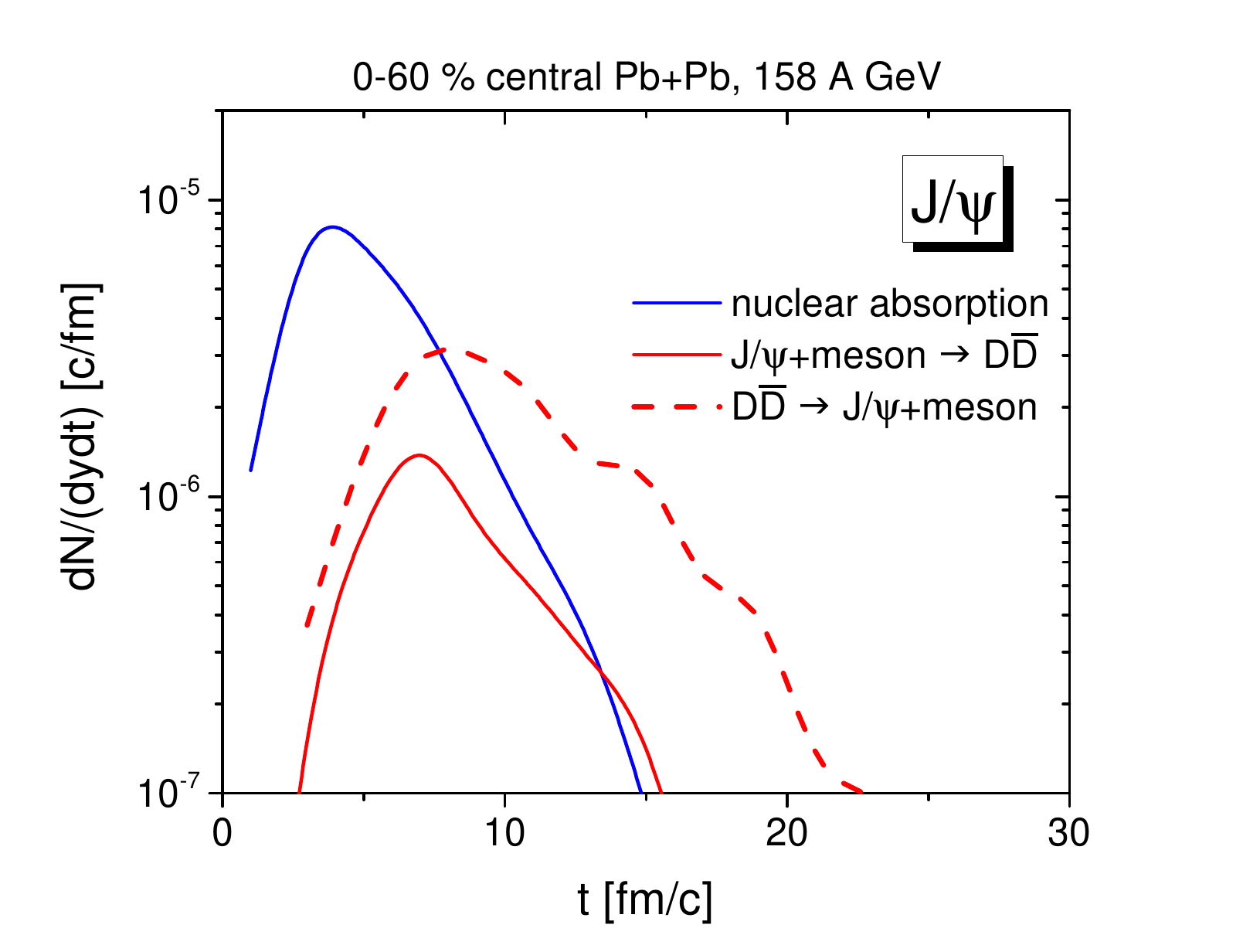}}
\centerline{
\includegraphics[width=9 cm]{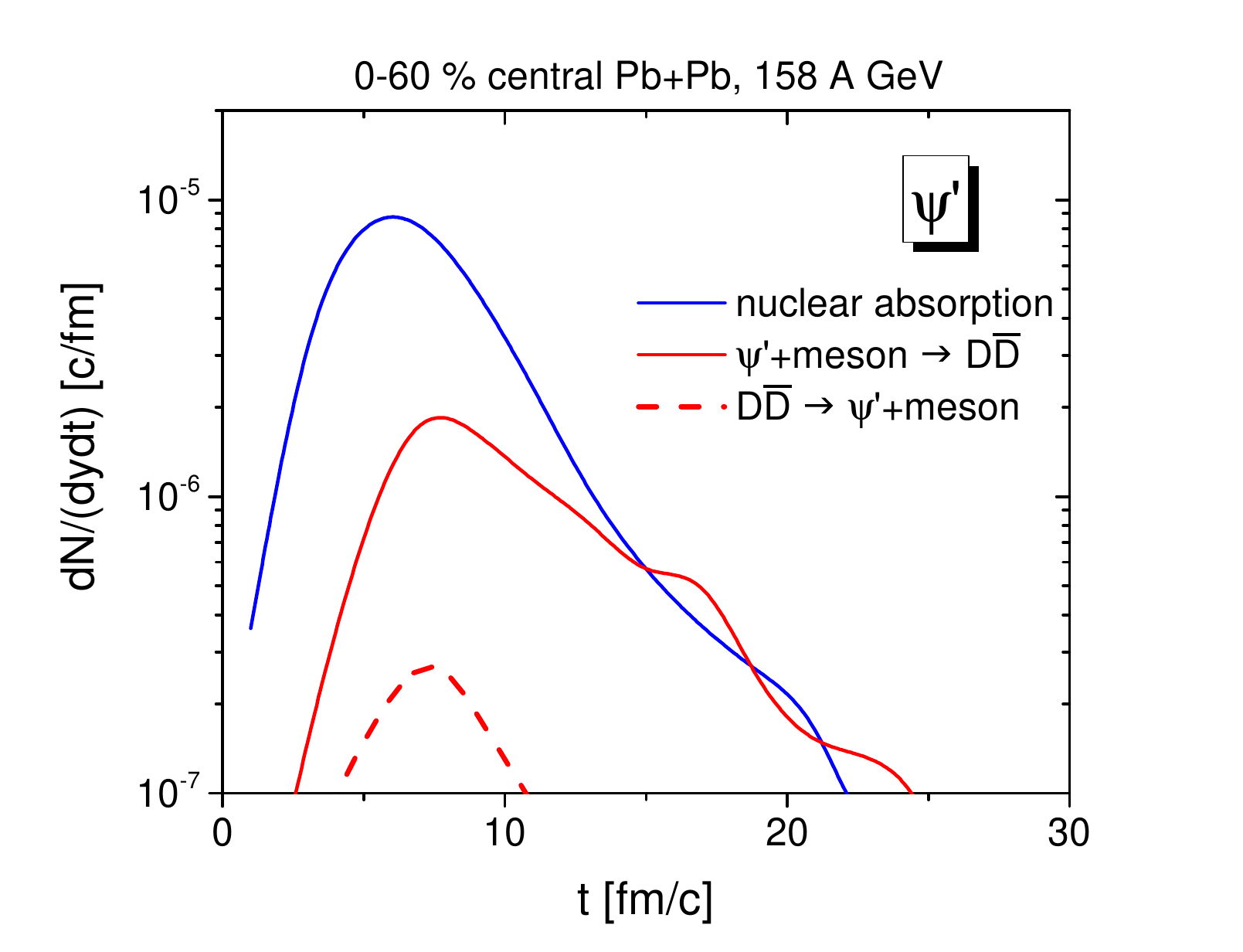}}
\caption{Time evolution of (blue solid) nuclear absorption, (red solid) meson-induced dissociation, and (red dashed) regeneration for (upper) $J/\psi$ and (lower) $\psi'$ in 0-60 \% central Pb+Pb collisions at E/A=158 GeV. 
}
\label{dndt}
\end{figure}

Fig.~\ref{dndt} shows the time evolution of nuclear absorption of $J/\psi$ and $\psi^\prime$, of their dissociation by light mesons, and of their regeneration from $D$ and $\bar{D}$ mesons in 0-60 \% central Pb+Pb collisions at E/A=158 GeV. 
One can see that both $J/\psi$ and $\psi^\prime$ are suppressed predominantly by nuclear absorption and, to less extent, by interactions with light mesons.
In contrast, regeneration from $D$ and $\bar{D}$ mesons is important for $J/\psi$, whereas the regeneration of $\psi^\prime$ is negligible.
The reason is that the former cross section diverges at low scattering energies, whereas the latter vanishes.

Comparing the nuclear absorption of $J/\psi$ and $\psi^\prime$, the absorption of $\psi^\prime$ exhibits a longer tail in time.
This behavior is a consequence of its larger absorption cross section.
A larger cross section implies a larger interaction range, meaning that interactions persist longer in an expanding system such as the hadronic matter produced in heavy-ion collisions. 
For the same reason, one can observe a long tail in the regeneration of $J/\psi$ from $D\bar{D}$ mesons.
We note that most $J/\psi$ regeneration occurs near the threshold energy, where the cross sections diverge, as shown in Fig.~\ref{cs-jpsi}.

\begin{figure}[h]
\centerline{
\includegraphics[width=9 cm]{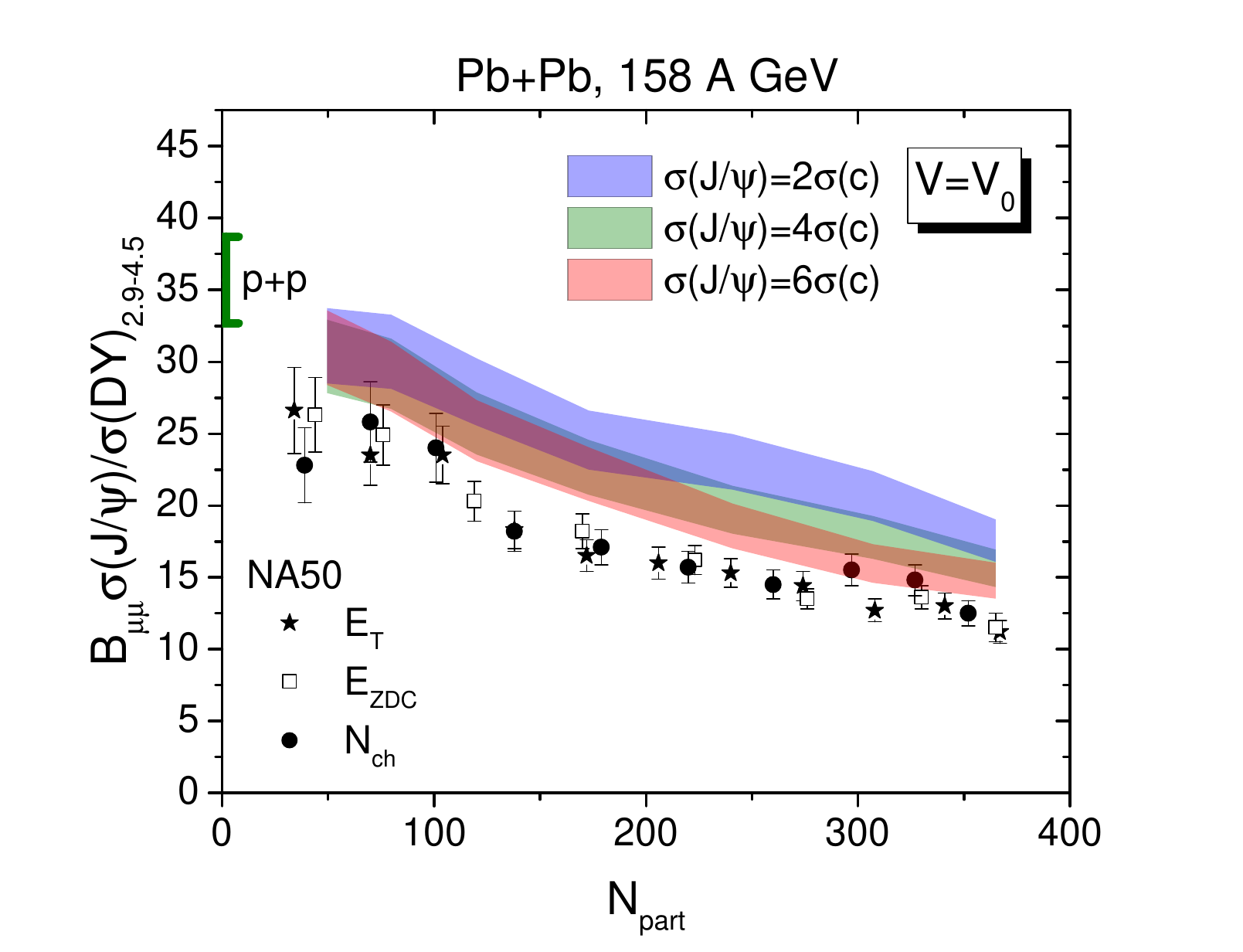}}
\caption{Same as the upper panel of Fig.~\ref{npart} but with 
increasing scattering cross section of charm quark pair in color singlet $\sigma(J/\psi)$, compared to the scattering cross section of charm quark $\sigma(c)$ in QGP.
}
\label{Npart-sigmacc}
\end{figure}

The lattice results for the temperature dependence of the potential between a heavy quark and a heavy antiquark are inconclusive. Some lattice calculations predict an increase, as shown in Fig.~\ref{npart}, whereas others yield an almost temperature-independent potential. 

Therefore we employ another approach, which offers an alternative explanation of the data presented in Fig.~\ref{Npart-sigmacc}. According to the Operator Product Expansion (OPE) method~\cite{Peskin:1979va,Bhanot:1979vb} or pQCD calculations~\cite{Oh:2001rm,Song:2005yd}, the cross sections for charmonium dissociation in a QGP can exceed the simple summation of the charm cross section and that of the anticharm quark. The reason is that an external gluon, which destroys a charmonium, can also interact with the gluon exchanged between charm and anticharm quarks besides with charm and anticharm quarks themselves~\cite{Song:2010ix}.
The large spectral width of charmonium in QGP is also supported by recent lattice calculations~\cite{Bazavov:2023dci}.

The consequences of increasing scattering cross sections for (anti)charm quarks interacting with (anti)quarks or gluons in the QGP are shown in Fig.~\ref{Npart-sigmacc} for Pb+Pb collisions at a kinetic energy of 158 A GeV. We note that the blue band corresponding to $\sigma(c\bar{c})=2\sigma(c)$ is equivalent to the lower limit of the blue band in Fig.~\ref{npart}. On the other hand, $\sigma(c\bar{c})=4\sigma(c)$ and $\sigma(c\bar{c})=6\sigma(c)$ imply that the charm and anticharm quarks forming a charmonium state scatter, respectively, two and three times more frequently.
As one can see, charmonium production decreases with increasing charm-parton scattering cross section. 
However, we need a quite large cross section to reproduce the experimental data from NA50 Collaboration.


\begin{figure}[h]
\centerline{
\includegraphics[width=9 cm]{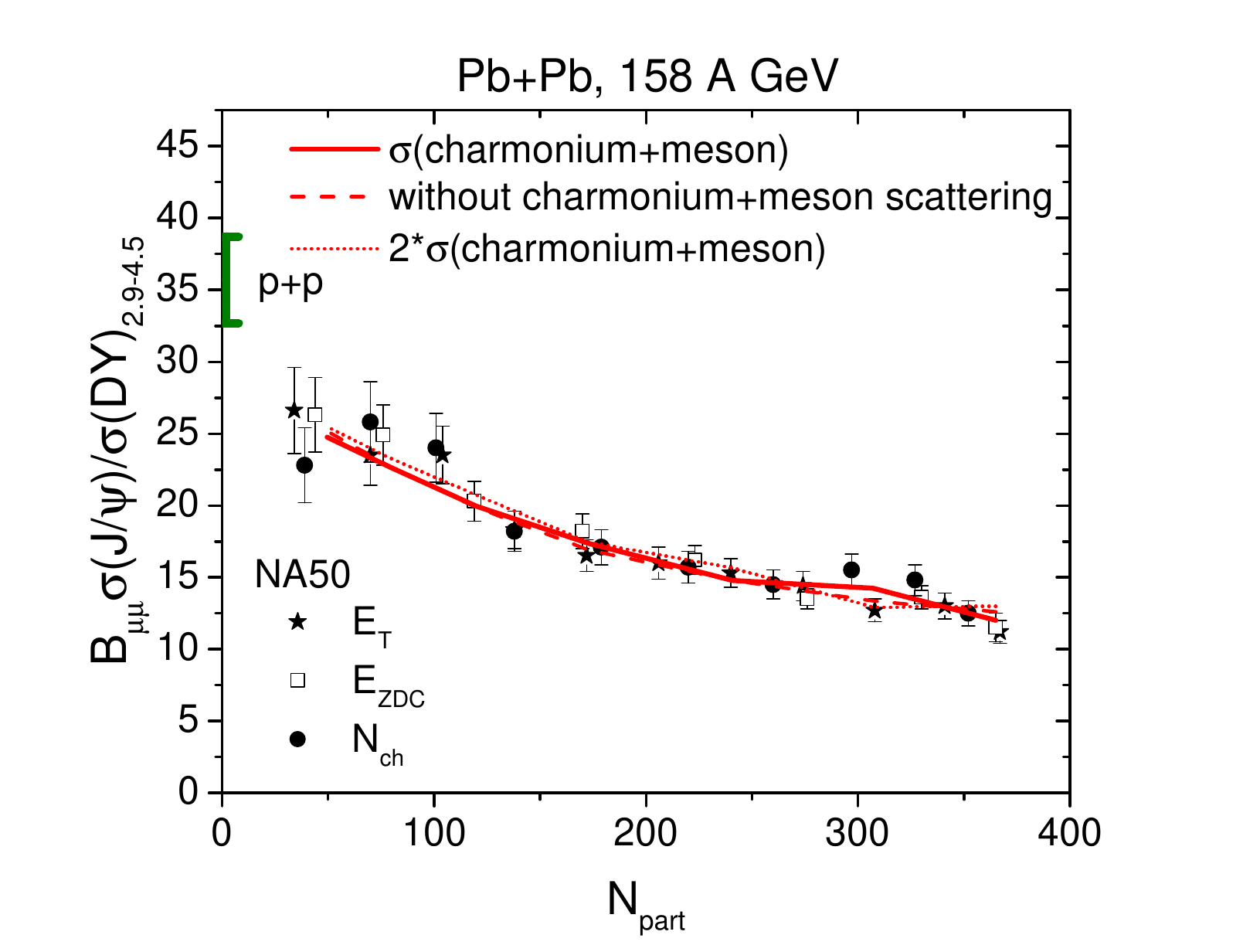}}
\caption{Dependence of dilepton production from $J/\psi$ decay on the scattering cross sections for charmonium interactions with light mesons in Pb+Pb collisions at E/A=158 GeV. The solid line represents the results obtained using the default cross sections; the dashed line shows the results without charmonium and light meson interactions; and the dotted line corresponds to results with scattering cross sections doubled relative to the default.}
\label{Npart-sigma}
\end{figure}

Finally we examine the sensitivity of $J/\psi$ production in heavy-on collisions to the scattering amplitude squared of charmonia with light mesons, $|M_0|^2$, which is tentatively chosen such that the maximum cross section is around 10 mb.
Fig.~\ref{Npart-sigma} shows the ratio of dimuons from $J/\psi$ to those from the Drell-Yan process in Pb+Pb collisions at E/A=158 GeV for three different scenarios.
Since the lower limit of the blue band is close to the experimental data in Fig.~\ref{npartF}, we choose $\sigma(c\bar{c})=2\sigma(c)$ and assume that the ratio of dimuons from $J/\psi$ to those from the Drell-Yan process in p+p collisions is 35.7, ignoring the error bar.
The dashed line represents the result without charmonium-meson interactions, the dotted line corresponds to the case where the cross sections are doubled, and the solid line shows the results obtained with the standard  cross sections corresponding to  
$|M_0|^2=0.11$~mb/GeV$^2$.

The results are only weakly sensitive to the precise values of these cross sections. This can be understood from Fig.~\ref{dndy}, where the contribution of the $J/\psi+\mathrm{meson}$ production and annihilation channels to the total $J/\psi$ yield amounts to only about 4\% at midrapidity. Consequently, moderate variations of the corresponding cross sections have only a limited impact on the final $J/\psi$ yield.

\subsection{Charmonium production at GSI/FAIR}

Motivated by the successful description of charmonium production at SPS energies, discussed in the previous subsection, we now extend the analysis to lower beam energies. This energy regime is of particular interest because it probes strongly interacting matter at higher baryon chemical potential, $\mu_B$, and will be explored experimentally at the upcoming Facility for Antiproton and Ion Research (FAIR).

\begin{figure}[b!]
\centerline{
\includegraphics[width=9 cm]{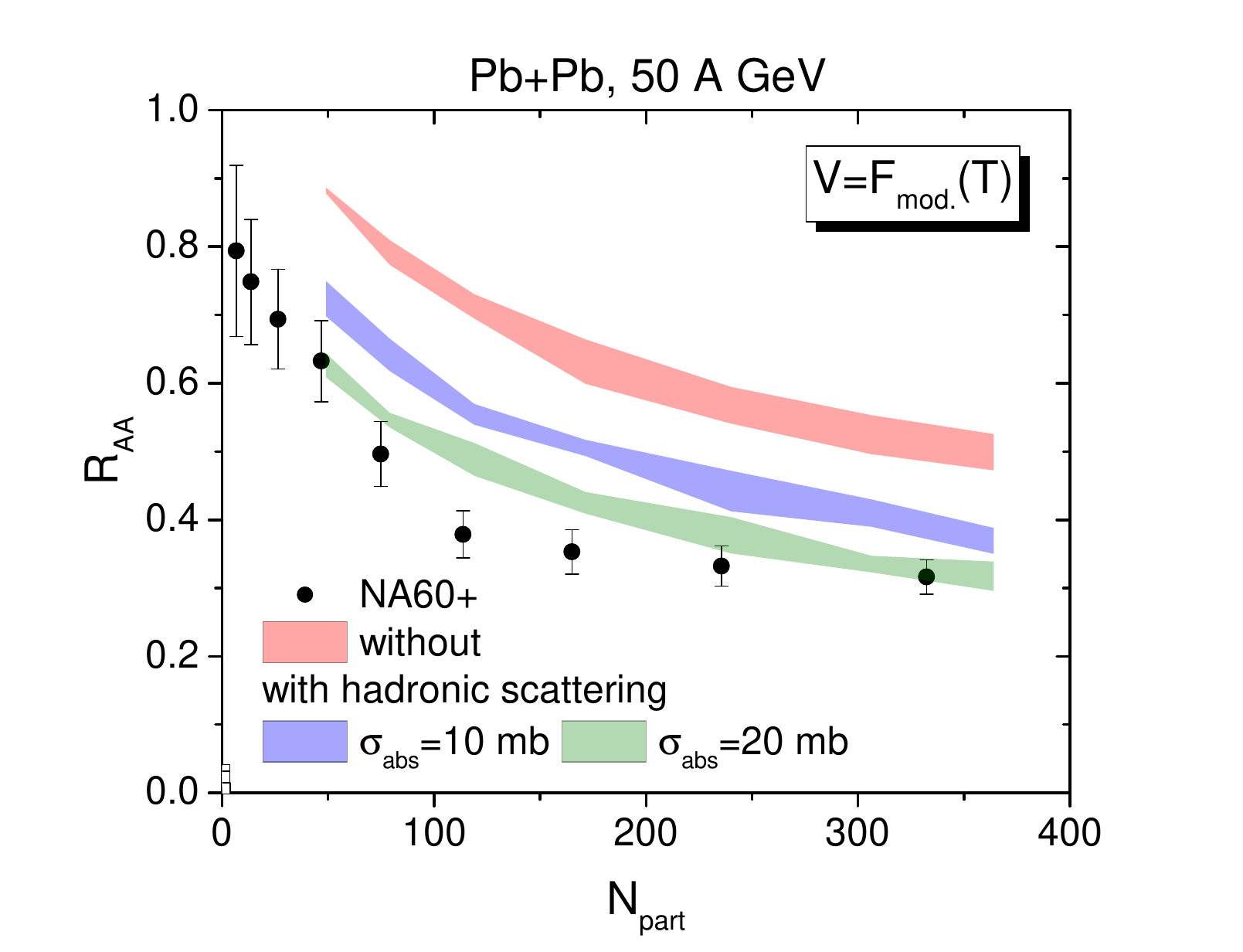}}
\caption{$R_{\rm AA}$ of $J/\psi$ as a function of participant number in Pb+Pb collisions at E/A=50 GeV with and without hadronic scattering. The nuclear absorption cross section is taken to be 10 mb or 20 mb.}
\label{PbPb-NA60+}
\end{figure}

Before turning to FAIR energies, we first examine $J/\psi$ production in Pb+Pb collisions at 50 A GeV, which is planned to be measured by the NA60+ Collaboration~\cite{Arnaldi:2023zlh}. This system provides an important link between the earlier SPS measurements and the future FAIR program. Since no corresponding p+A data are available at the same beam energy, the nuclear absorption cross section cannot be fixed independently. We therefore consider two representative values, $\sigma_{\rm abs}^{J/\psi}=10$ mb and 20 mb \com{($\sigma_{\rm abs}^{\psi^\prime}=20$ mb and 40 mb)}~\cite{Lourenco:2008sk}, and show the resulting calculations in Fig.~\ref{PbPb-NA60+}.

\com{
The free energy from lattice QCD used in this study is calculated at $\mu_B=$0.
However, the heavy-quark potential is expected to be modified in baryon-rich matter ($\mu_B\ne 0$).
There are a few lattice QCD studies that extract the free energy at finite $\mu_B$ up to $\mu_B/T\approx 1$~\cite{Doring:2005ih,Takahashi:2013mja}, as well as a phenomenological study for heavy-ion collisions~\cite{Zhao:2022cvl}.
However, $\mu_B/T$ at SPS energies is around 1.5-2.0~\cite{Braun-Munzinger:1999hun,Moreau:2019vhw} and is expected to be larger at FAIR energies, indicating that the currently available lattice QCD results are not yet sufficient to quantitatively describe the relevant baryon-rich regime. 
Nevertheless, these studies show that the Debye screening mass increases with increasing $\mu_B$ at fixed temperature, which is consistent with pQCD predictions. If the free energy at finite $\mu_B$ is used as the heavy-quark potential, charmoniun states will begin to form at lower temperatures than at $\mu_B=0$, and the $J/\psi$ will be more strongly suppressed than in the present results.  
We leave a detailed investigation of this effect for future work.
}

\begin{figure}[h]
\centerline{
\includegraphics[width=9 cm]{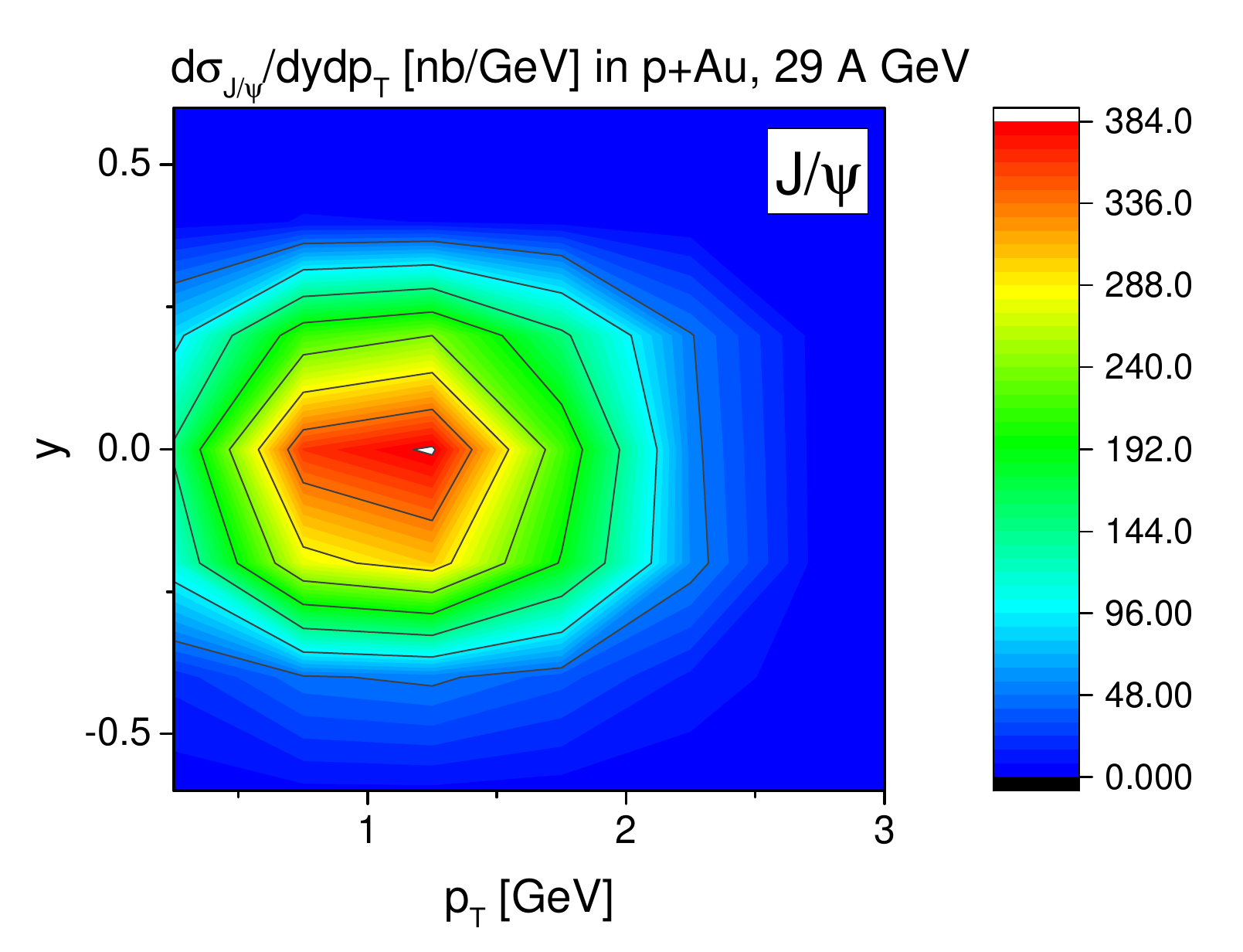}}
\centerline{
\includegraphics[width=9 cm]{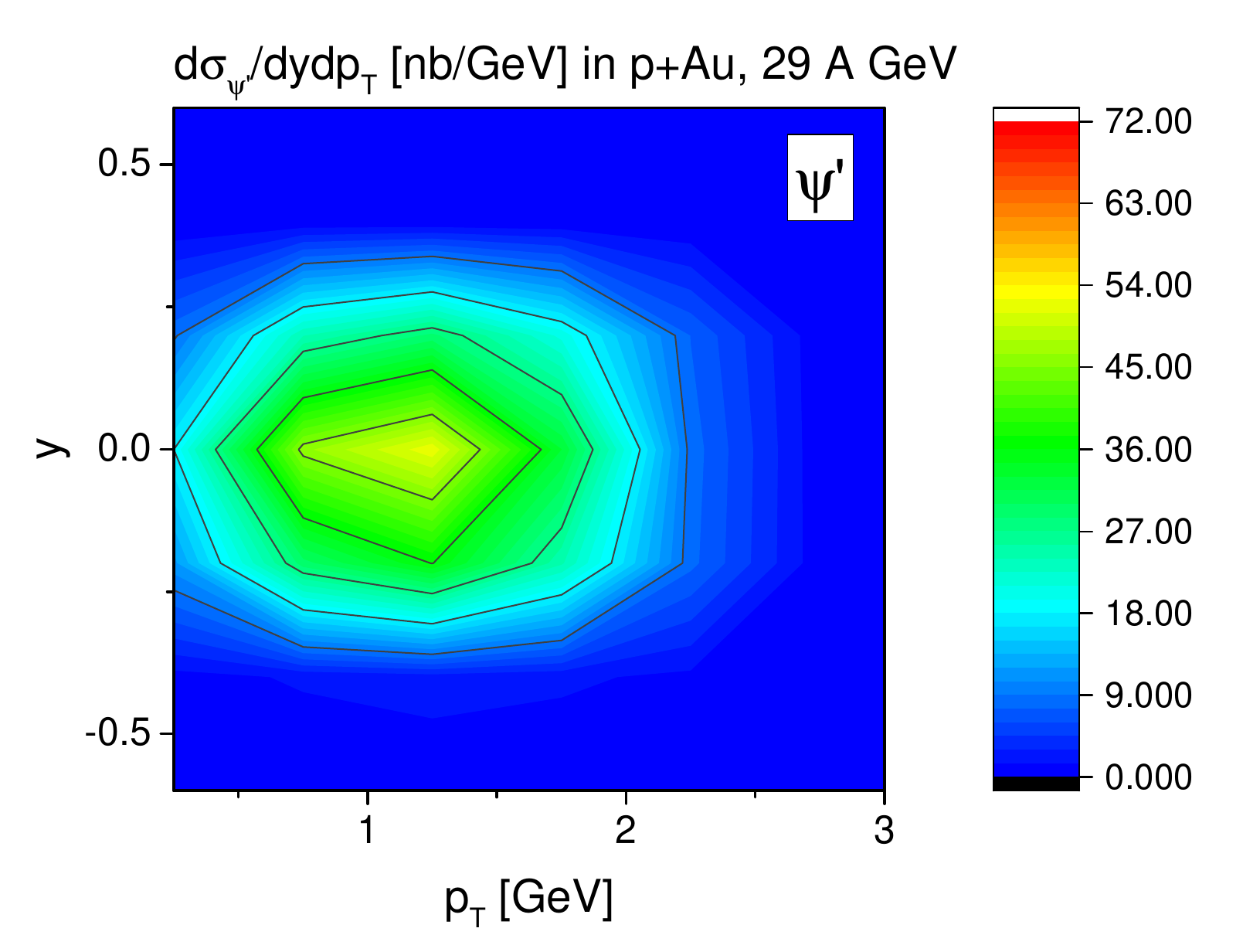}}
\caption{Differential cross sections for the production of (upper) $J/\psi$ and (lower) $\psi^\prime$ as functions of rapidity and transverse momentum in p+Au collisions at E/A=29 GeV, assuming absorption cross sections of 10 mb for $J/\psi$ and $\chi_c$, and 20 mb for $\psi^\prime$.}
\label{pAu}
\end{figure}

Fig.~\ref{pAu} shows the differential cross sections for $J/\psi$ and $\psi^\prime$ production as a function of rapidity and transverse momentum in p+Au collisions at 29 GeV.
The nuclear absorption cross sections are assumed to be 10 mb for $J/\psi$ and 20 mb for $\psi^\prime$, as in p+A collisions at $E_{kin}=$158 A GeV.
One can see that the production are mainly concentrated around midrapidity ($|y|<0.3$) and at low transverse momentum ($p_T<2$ GeV). The integrated production cross sections are 310 $nb$ for $J/\psi$ and 37 $nb$ for $\psi^\prime$.
If the absorption cross section of $J/\psi$ and $\psi^\prime$ increases to 20 mb and 40 mb, 
respectively, the production cross sections decrease to 210 $nb$ for $J/\psi$ and 22 nb for $\psi^\prime$.

\begin{figure}[h]
\centerline{
\includegraphics[width=9 cm]{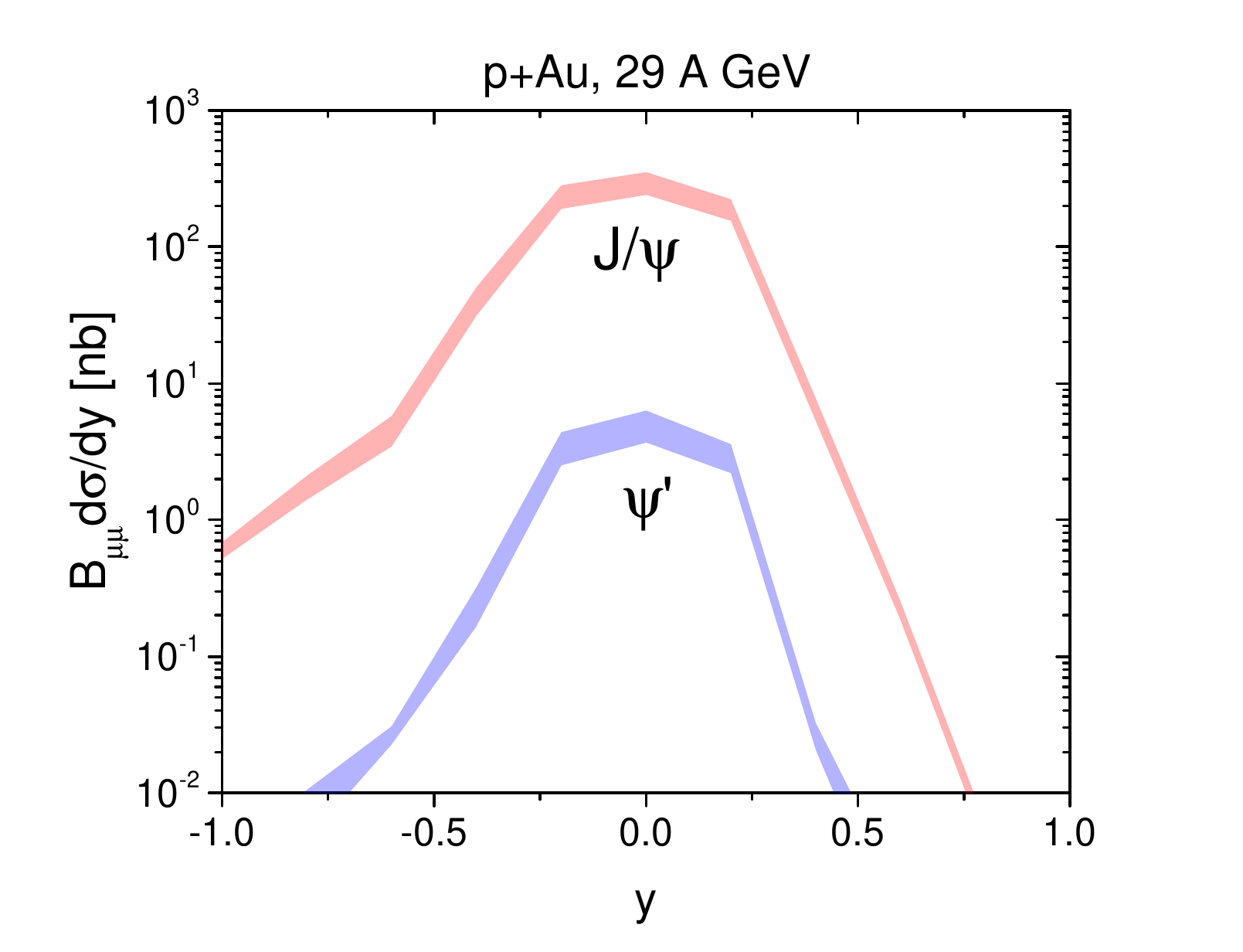}}
\centerline{
\includegraphics[width=9 cm]{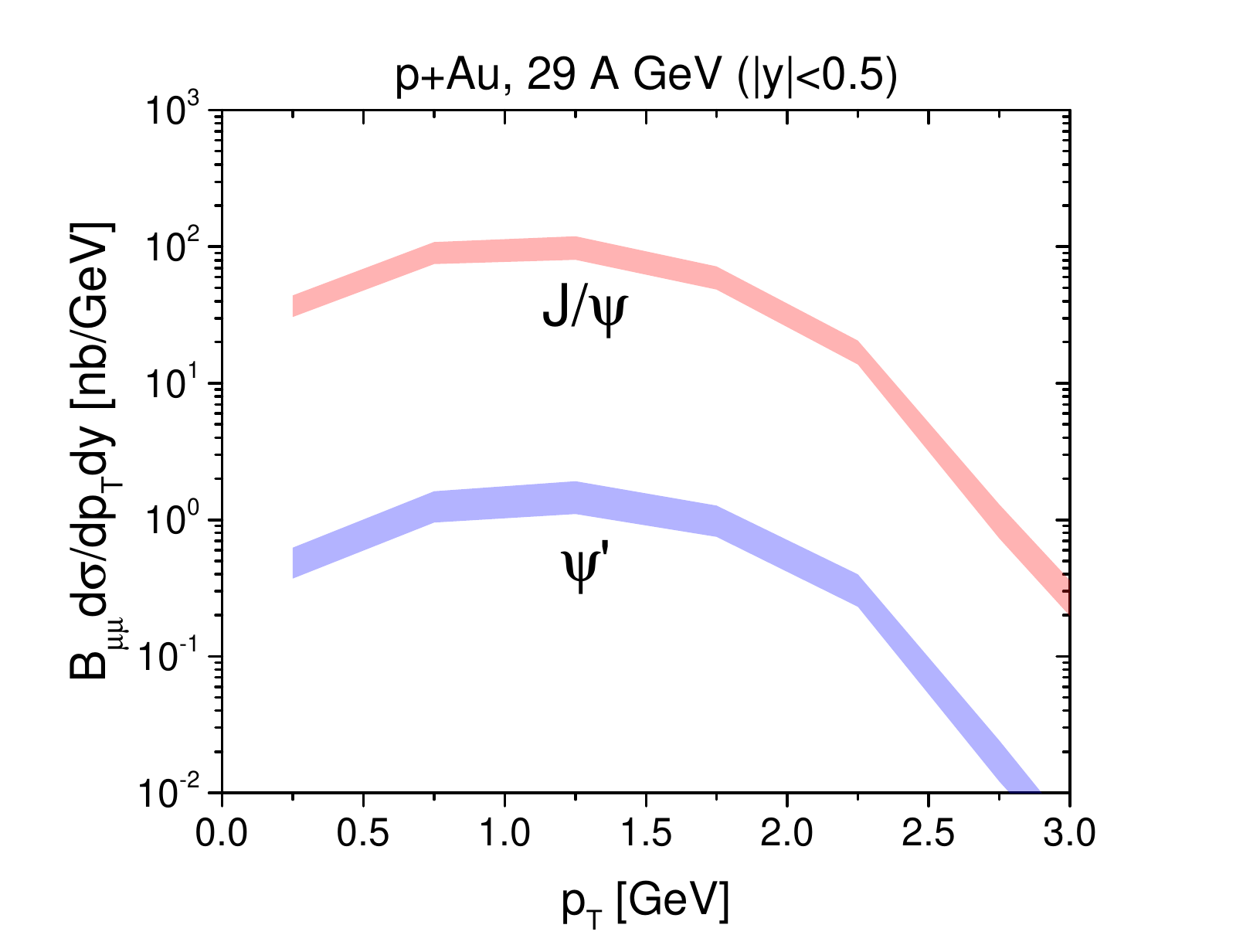}}
\caption{Production cross sections of $J/\psi$ and $\psi^\prime$ multiplied by the branching ratio to dimuon as a function of rapidity and transverse momentum in p+Au collisions at E=29 GeV, assuming absorption cross sections of 10-20 mb for $J/\psi$ and $\chi_c$, and 20-40 mb for $\psi^\prime$.}
\label{pAu2}
\end{figure}

Fig.~\ref{pAu2} shows the production cross sections of $J/\psi$ and $\psi^\prime$ multiplied by the branching ratio to dimuon as a function of rapidity and transverse momentum at midrapidity ($|y|<0.5$) in p+Au collisions at E=29 GeV, assuming absorption cross sections of 10-20 mb for $J/\psi$ and $\chi_c$, and 20-40 mb for $\psi^\prime$.
We note that the cross sections are not scaled by the mass number of target nucleus, different from the lower panels in Fig.~\ref{sigma-pp}.
As discussed in Fig. ~\ref{RpA400}, the nuclear absorption cross section in principle depends on charmonium rapidity and transverse momentum, and the results can be improved when more information on absorption cross section is available.

\begin{figure}[h]
\centerline{
\includegraphics[width=9 cm]{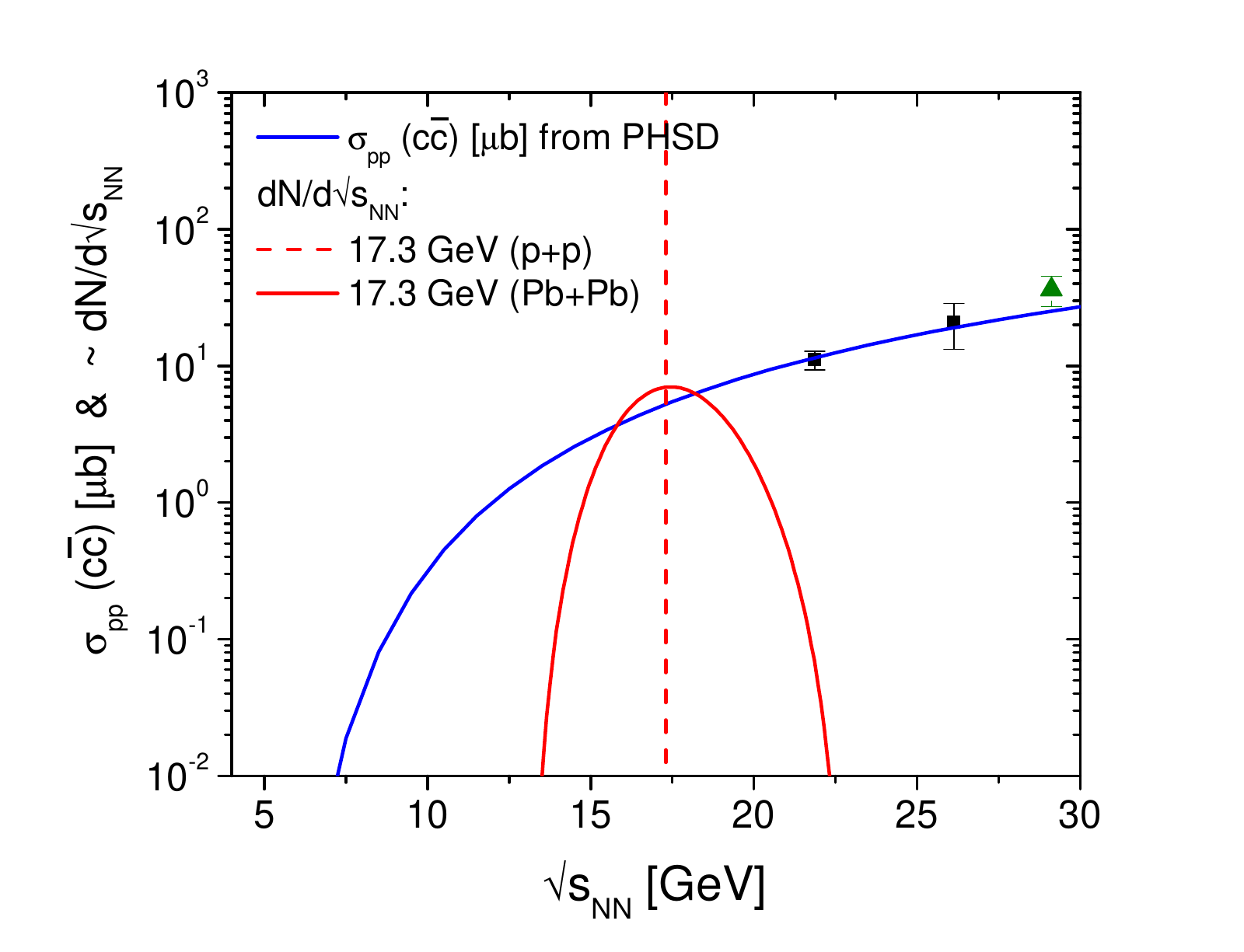}}
\centerline{
\includegraphics[width=9 cm]{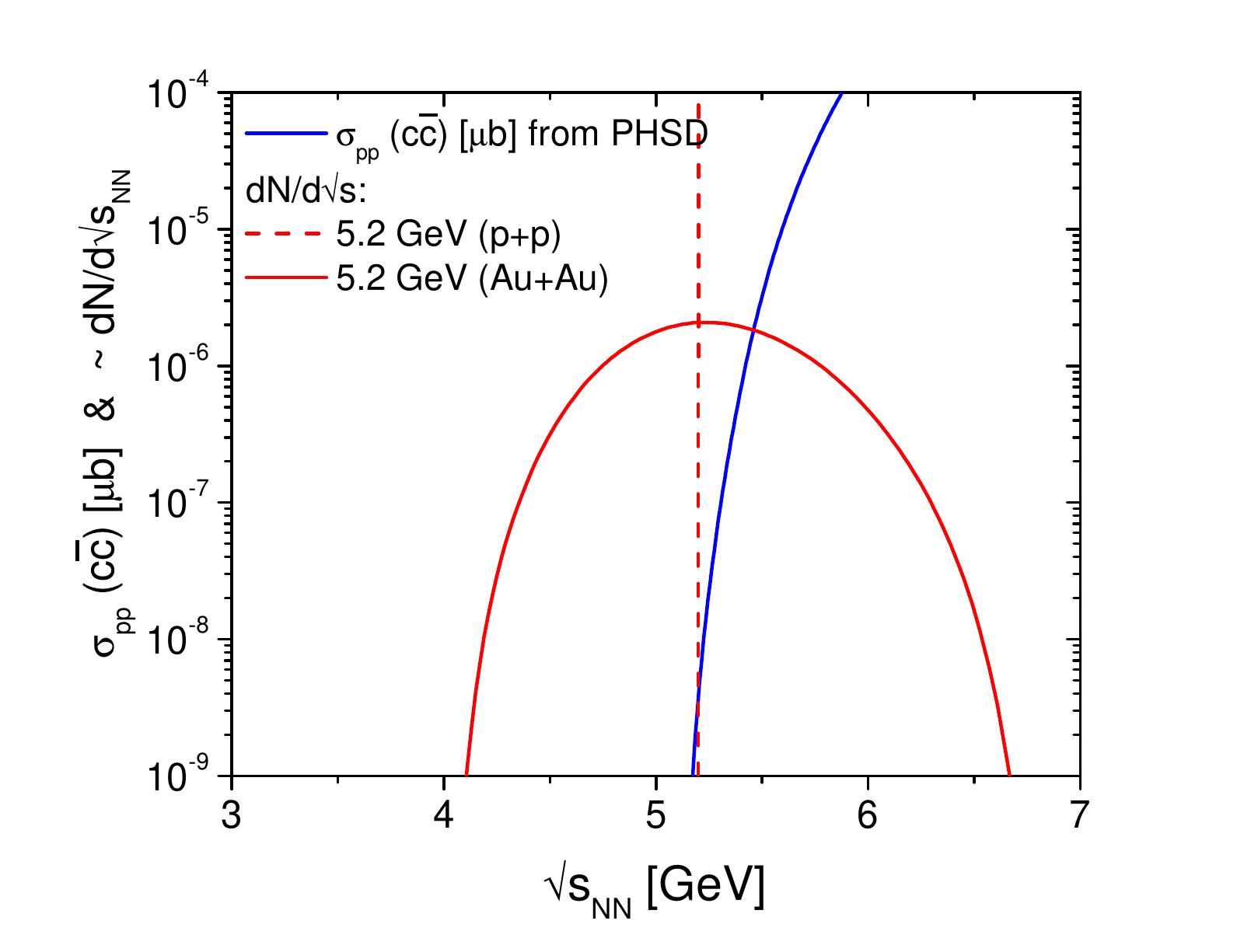}}
\caption{Distribution of nucleon-nucleon scattering energies in (upper) p+p and Pb+Pb collisions at $\sqrt{s_{NN}}=$ 17.3 GeV and (lower) p+p and Au+Au collisions at $\sqrt{s_{NN}}=$ 5.2 GeV, together with the charm production cross section in p+p collisions.}
\label{cs-ccbar}
\end{figure}

\com{
The maximal collision energy foreseen for heavy-ion collisions at GSI/FAIR is approximately $\sqrt{s_{NN}}=5.2$ GeV, only slightly above the threshold for charm-pair production in reactions such as $NN\rightarrow D\Lambda_c$ and $NN\rightarrow J/\psi NN$. The blue curves in Fig.~\ref{cs-ccbar} show fits to the charm-production cross sections, parameterized using the experimental data presented in Fig.~\ref{sigma}. The upper panel corresponds to Pb+Pb collisions at $E/A=158$ GeV ($\sqrt{s_{NN}}=17.3$ GeV), while the lower panel shows Au+Au collisions at $\sqrt{s_{NN}}=5.2$ GeV, representative of the SPS and FAIR energy regimes, respectively.

The red dashed lines indicate the nominal nucleon-nucleon collision energies, whereas the red solid curves show the distributions of actual nucleon-nucleon center-of-mass energies obtained from PHSD simulations. These distributions are broadened by the Fermi motion of nucleons inside the colliding nuclei. Although the Fermi momentum in the nuclear rest frame is typically below $0.2$ GeV, its longitudinal component is strongly amplified by the Lorentz boost in relativistic heavy-ion collisions.

At the nominal collision energies, the charm-production cross section per nucleon-nucleon collision amounts to about $5.26~\mu$b at SPS energy and only $0.0038$ pb at FAIR energy, according to the parameterization shown by the blue curves. After folding the production cross section with the distribution of nucleon-nucleon collision energies generated by Fermi motion, the corresponding average cross sections increase to approximately $5.7~\mu$b at SPS and $3$ pb at FAIR. Thus, Fermi motion enhances charm production at FAIR energies by nearly three orders of magnitude, corresponding to a factor of about $800$. This dramatic enhancement arises because $\sqrt{s_{NN}}=5.2$ GeV lies very close to the charm-production threshold, where the production cross section rises steeply with energy. In contrast, at SPS energies, where the nominal collision energy is well above threshold, the enhancement amounts to only about 8\%.

Estimating charmonium production in heavy-ion collisions at FAIR energies remains challenging. In particular, the momentum distributions of charm and anticharm quarks at such low collision energies, which are required as input to the Remler formalism, are presently not sufficiently constrained and are highly sensitive to the treatment of the nucleon Fermi motion. Nevertheless, the results above indicate that charmonium production in heavy-ion collisions can be substantially enhanced relative to a naive estimate based solely on elementary $p+p$ collisions, owing to the broadening of the effective nucleon-nucleon collision energies by Fermi motion. In addition, short-range nucleon-nucleon correlations may further enhance the high-energy tail of this distribution and thereby increase the charm-production probability~\cite{Reichert:2025egt}.
}


\section{summary}\label{summary}

We applied the Remler formalism to study charmonium production, first at SPS energies, in order to extract the charmonium interaction rate in the QGP as well as the scattering cross sections of charmonium in the hadron environment.
These parameters were then used to estimate charmonium production at GSI/FAIR energies.

In an initial model study, shown in Fig. 11 of Sec.~\ref{HIC}, we assumed that the properties of charmonium do not change in the QGP by employing temperature-independent charmonium radii identical to those in p+p collisions. 
Within the Remler formalism, a large interaction rate of charmonium in the QGP suppresses $J/\psi$ production in heavy-ion collisions because it delays the formation of charmonium.
However, even the largest interaction rate - taken to be twice the interaction rate of charm quark - fails to reproduce the experimental data in Pb+Pb and In+In collisions at E/A=158 GeV measured by the NA50 and NA60 Collaborations.
This failure clearly indicates the necessity of including an in-medium heavy-quark potential.

We therefore considered subsequently the free energy of a heavy-quark pair as the effective heavy-quark potential. The rms radius of the wave function obtained for this potential depends on temperature and diverges near the dissociation temperature. 
Since the diverging radius 
indicates that the bound state disappears, the initial projection is performed when it reaches its maximum value, which occurs at approximately 1.15 $T_c$, although the actual formation temperature of $J/\psi$ is around 1.2 $T_c$.
The experimental data on $J/\psi$ production in both Pb+Pb and In+In collisions at SPS energies are then successfully reproduced.
This result demonstrates the importance of the in-medium heavy-quark potential and suggests that the dissociation temperature of $J/\psi$ is close to $T_c$.
The results also depend on the scattering cross section of $J/\psi$ in the QGP.
Although it is not fully conclusive, the results suggest that the scattering cross section is approximately twice the charm-quark scattering cross section in the QGP and large as expected for a weakly bound system with a large rms radius, 

For the hadronic phase, we obtained from p+A collisions a $J/\psi$ nuclear absorption cross section of 7 mb at $E_{kin}$= 400 GeV and of 10 mb at $E_{kin}$= 158 GeV, while the absorption cross section of $\psi^\prime$ is approximately twice that of $J/\psi$.
The scattering cross sections of charmonium with light meson were calculated under the assumption of a universal constant scattering amplitude, tuned such that the maximum cross section is around 10 mb.
The reverse reaction, the regeneration of charmonium from $D$ and $\bar{D}$ mesons, was determined through detailed balance.
We find that the results for $J/\psi$ production at SPS energies are not very sensitive to the precise value of the scattering amplitude.

Since the NA60+ Collaboration will provide experimental data on $J/\psi$ production in Pb+Pb collisions at energies between those of the earlier NA50 and NA60 experiments and that of the upcoming FAIR experiment, 
we simulated $J/\psi$ production in Pb+Pb collisions at E/A=50 GeV with the nuclear absorption cross section of 10-20 mb.
Finally, using the parameters extracted from SPS energies, we estimated charmonium production at GSI/FAIR.
As a results, the cross section for $J/\psi$ production in p+Au collisions at 29  GeV is expected to be $210-310 ~nb$, while that for $\psi^\prime$ is about $22-37 ~nb$, depending on the assumed nuclear absorption cross section (10-20 mb).
Our results indicate that charm and charmonium production in Au+Au collisions may be significantly enhanced compared to p+p collisions at the same nominal energy due to the Fermi motion of nucleons. While the Fermi momentum itself is modest, its effect becomes particularly important near the threshold energy of charm production.

\section*{Acknowledgements}
We are grateful for useful discussions with  R. Arnaldi, W. Cassing, I. Grishmanovskii, S. H. Lee, E. Scomparin,  O. Soloveva, I. Vassiliev and R. Vogt. We thank A.W.R. Jorge for assistance with Figure 3.
We acknowledge support by the Deutsche Forschungsgemeinschaft (DFG, German Research Foundation) through the grant CRC-TR 211 'Strong-interaction matter under extreme conditions' - Project number 315477589 - TRR 211. 
The computational resources have been provided by the LOEWE-Center for Scientific Computing and the "Green Cube" at GSI, Darmstadt and by the Center for Scientific Computing (CSC) of the Goethe University.
The publication is funded by the Open Access Publishing Fund of GSI Helmholtzzentrum fuer Schwerionenforschung.

\bibliography{main}

\end{document}